\documentclass[iop,apj]{emulateapj}
\usepackage{amsmath,dsnr,epsfig,ifthen,natbib,soul}
\usepackage[export]{adjustbox}
\usepackage[caption=false]{subfig}
\usepackage[pdftex,backref,breaklinks,colorlinks,citecolor=blue]{hyperref}
\usepackage{booktabs}
\usepackage[all]{hypcap}

\newcommand{\hst}{{\it HST}}
\newcommand{\zw}{I~Zw~1}
\newcommand{\fohfive}{F05189$-$2524}
\newcommand{\fohseven}{F07599$+$6508}
\newcommand{\mrk}{Mrk~231}
\newcommand{\fonethreetwo}{F13218$+$0552}
\newcommand{\fonethreethree}{F13342$+$3932}
\newcommand{\pgonefour}{PG1411$+$442}
\newcommand{\pgonesix}{PG1613$+$658}
\newcommand{\pgoneseven}{PG1700$+$518}
\newcommand{\ftwoone}{F21219$-$1757}
\newcommand{\vfifty}{\ifmmode v_{50\%}\else $v_{50\%}$\fi}
\newcommand{\vfiftyave}{\ifmmode \langle v_{50\%}\rangle \else $\langle
  v_{50\%}\rangle$\fi}
\newcommand{\vtsigr}{\ifmmode v_{02\%}\else $v_{02\%}$\fi}
\newcommand{\vtsig}{\ifmmode v_{98\%}\else $v_{98\%}$\fi}
\newcommand{\vtsigave}{\ifmmode \langle v_{98\%}\rangle \else $\langle
  v_{98\%}\rangle$\fi}

\shorttitle{Feedback in Nearby Type 1 Quasars}
\shortauthors{Rupke et al.}

\begin{document}

\slugcomment{Resubmitted to ApJ 13 Sep 2017}

\title{Quasar-Mode Feedback in Nearby Type 1 Quasars: Ubiquitous
  Kiloparsec-Scale Outflows and Correlations with Black Hole
  Properties}

\author{David S. N. Rupke} \affil{Department of Physics, Rhodes
  College, Memphis, TN 38112, USA; Research School of Astronomy and
  Astrophysics, The Australian National University, Canberra, ACT
  2611, Australia} \email{drupke@gmail.com}

\author{Kayhan G\"{u}ltekin} \affil{Department of Astronomy,
  University of Michigan, Ann Arbor, MI 48109, USA}

\author{Sylvain Veilleux} \affil{Department of Astronomy and Joint
  Space-Science Institute, University of Maryland, College Park, MD
  20742, USA}

\begin{abstract}
  The prevalence and properties of kiloparsec-scale outflows in nearby
  Type 1 quasars have been the subject of little previous
  attention. This work presents Gemini integral field spectroscopy of
  ten Type 1 radio-quiet quasars at $z<0.3$. The excellent image
  quality, coupled with a new technique to remove the point spread
  function using spectral information, allow the fitting of the
  underlying host on a spaxel-by-spaxel basis. Fits to stars,
  line-emitting gas, and interstellar absorption show that 100\%\ of
  the sample host warm ionized and/or cool neutral outflows with
  spatially-averaged velocities
  ($\langle\vtsig\rangle \equiv \langle v+2\sigma\rangle$) of
  200--1300~\kms\ and peak velocities (maximum \vtsig) of
  500--2600~\kms. These minor-axis outflows are powered primarily by
  the central AGN, reach scales of 3--12~kpc, and often fill the field
  of view. Including molecular data and Type 2 quasar measurements,
  nearby quasars show a wide range in mass outflow rates ($dM/dt = 1$
  to $>$1000 \smpy) and momentum boosts
  [$(c~dp/dt)/L_\mathrm{AGN} = 0.01-20$]. After extending the mass
  scale to Seyferts, $dM/dt$ and $dE/dt$ correlate with black hole
  mass ($dM/dt \sim M_\mathrm{BH}^{0.7\pm0.3}$ and
  $dE/dt \sim M_\mathrm{BH}^{1.3\pm0.5}$). Thus, the most massive
  black holes in the local universe power the most massive and
  energetic quasar-mode winds.
\end{abstract}



\section{INTRODUCTION} \label{sec:introduction}

An increasingly large body of evidence points to the presence of
wide-angle, galaxy-scale outflows of gas and dust in the host galaxies
of many nearby Type 2 quasars
\citep{westmoquette12a,rupke13b,rupke13a,
  veilleux13a,cicone14a,harrison14a,mcelroy15a,bae17a,gonzalezalfonso17a,
  wylezalek17a}. These galactic winds, driven by the radiative and
mechanical luminosity of the central engine, expel gas from the
nucleus, and thereby from the sites of ongoing star formation and
black hole accretion. Increasingly, this evidence comes from 3D
observations at optical, near infrared, and millimeter
wavelengths. Such data are ideal for distinguishing the wind from the
host galaxy and constraining its structure and power source.

Quasar-mode feedback is an essential element of modern simulations and
semi-analytic models of galaxy formation and evolution
\citep[e.g.,][]{silk98a,king03a,dimatteo05a,somerville08a,booth09a,richardson16a}. It
may quench galaxies \citep{zubovas12a,wylezalek16a,pontzen17a}, shape
the stellar mass function
\citep{somerville08a,schaye15a,taylor15a,kaviraj17a}, shape galaxy
morphology \citep{dubois16a}, impact the distribution of the
Ly$\alpha$ forest \citep{gurvich17a}, and regulate BH accretion
\citep{hopkins16a,volonteri16a}. Though such outflows are harder to
detect at high $z$ than at low $z$, observations are beginning to
suggest that quasar-driven winds could be common at the epochs of peak
star formation and BH accretion
\citep{perna15a,brusa15a,carniani15a,carniani16a,leung17a}. At high
$z$ but low AGN luminosity, the impact of winds may be less, though
the data are sparse \citep{yesuf17a}.

\setlength{\tabcolsep}{3pt}
\capstartfalse\begin{deluxetable*}{cclccccc}
  \tablecaption{Sample\label{tab:sample}}
  \tabletypesize{\scriptsize}
  \tablewidth{\textwidth}

  \tablehead{\colhead{Galaxy} & \colhead{Other name} & \colhead{$z$} &
    \colhead{log($L_\mathrm{bol}/\lsun$)} & \colhead{AGN Frac.} &
    \colhead{log($M_\mathrm{BH}/\msun$)} & \colhead{$\sigma$ (\kms)} &
    \colhead{log[$P_\mathrm{radio}/(W~Hz^{-1})$])} \\
    \colhead{(1)} & \colhead{(2)} & \colhead{(3)} & \colhead{(4)} &
    \colhead{(5)} & \colhead{(6)} & \colhead{(7)} & \colhead{(8)}}

  \startdata
  \zw             & PG0050$+$124, F00509$+$1225 & 0.0608  & 12.07 & 0.93$_{-0.10}^{+0.07}$ & 8.53$_{-0.5}^{+0.5}$ & 188$\pm$36 & 22.7 \\
  \fohfive        & \nodata                     & 0.04288 & 12.22 & 0.71$_{-0.21}^{+0.29}$ & 8.32$_{-0.5}^{+0.5}$ & 201$\pm$64 & 23.1 \\
  \fohseven       & \nodata                     & 0.1483  & 12.58 & 0.75$_{-0.39}^{+0.25}$ & 8.59$_{-0.5}^{+0.5}$ & \nodata    & 24.4 \\
  \mrk            & F12540$+$5708               & 0.0422  & 12.60 & 0.71$_{-0.07}^{+0.07}$ & 8.58$_{-0.5}^{+0.5}$ & 233$\pm$113 & 24.1 \\
  \fonethreetwo   & \nodata                     & 0.2047  & 12.68 & 0.83$_{-0.21}^{+0.17}$ & 8.55$_{-0.5}^{+0.5}$ & \nodata    & 23.8 \\
  \fonethreethree & \nodata                     & 0.1797  & 12.49 & 0.69$_{-0.24}^{+0.23}$ & 9.12$_{-0.5}^{+0.5}$ & \nodata    & 23.8 \\
  \pgonefour      & \nodata                     & 0.0898  & 11.78 & 1.00$_{-0.00}^{+0.00}$ & 8.54$_{-0.48}^{+0.46}$ & 216$\pm$31 & 22.3 \\
  \pgonesix       & Mrk~876                     & 0.12925 & 12.29 & 0.82$_{-0.09}^{+0.11}$ & 8.34$_{-0.52}^{+0.47}$ & \nodata & 23.1 \\

  \pgoneseven     & F17002$+$5153               & 0.2902  & 13.12 & 0.80$_{-0.19}^{+0.16}$ & 8.79$_{-0.45}^{+0.45}$ & \nodata & 24.8 \\
  \ftwoone        & \nodata                     & 0.1127  & 12.17 & 0.78$_{-0.13}^{+0.15}$ & 8.61$_{-0.5}^{+0.5}$ & 121$\pm$11 & 23.7
  \enddata

  \tablecomments{Column 3: Redshift, meausred from the current
    data. Column 4: Galaxy bolometric luminosity
    \citep{veilleux09a}. Column 5: ``AGN fraction'', or fraction of
    the bolometric luminosity of the galaxy due to an AGN
    \citep{veilleux09a}. The error bars reflect the full range of
    values derived from six mid-infrared diagnostics. Column 6: Black
    hole mass, taken from reverberation mapping \citep{bentz15a} where
    possible, or \hst\ photometric measurements otherwise
    \citep{veilleux09b}. Errors from reverberation mapping have the
    uncertainty in the virial coefficient $f$ added in quadrature
    ($\delta f = 0.44$; \citealt{woo10a}). Photometric error bars are
    from the scatter in the $M_\mathrm{BH}$--$L_{H}$ relationship from
    which the masses were determined \citep{marconi03a}. Column 7:
    Stellar velocity dispersions
    \citep{dasyra06a,dasyra07a,rothberg13a,grier13a}. The large error
    bars for \fohfive\ and \mrk\ are due to conflicting optical and
    infrared measurements \citep{dasyra06a,rothberg13a}; the mean of
    the two measurements is shown, and the error bar is half the
    difference between the measurements. Column 9: 1.4 or 1.5 GHz
    radio power; from \citet{barvainis97a} [\pgonefour],
    \citet{carilli98a} [\mrk], \citet{condon90a} [\fohfive],
    \citet{condon98a} [\fohseven, \fonethreetwo, \fonethreethree,
    \ftwoone], or \citet{kukula98a} [\zw, \pgonesix, \pgoneseven].}
  
\end{deluxetable*}\capstarttrue
\setlength{\tabcolsep}{6pt}

\mrk, the nearest Type 1 quasar, has become a poster-child for
quasar-mode feedback
\citep[e.g.,][]{fischer10a,feruglio10a,rupke11a,fiore15a,morganti16a}. However,
larger studies of kpc-scale outflows in Type 1 quasars are few. At
low redshift ($z < 0.3$), one study of Type 1 quasars using integral
field spectroscopy (IFS) has found extended narrow line regions (NLRs)
photoionized by the central active galactic nucleus (AGN) but a
paucity of galaxy-scale ionized outflows
\citep{husemann13a,husemann14a}. IFS studies of nearby Type 2 quasars,
however, point to a very high incidence (as high as 80--100\%) of
ionized gas outflows with sizes of order 1--10~kpc
\citep{rupke13a,harrison14a,mcelroy15a,bae17a}. This is a puzzling
discrepancy in light of the AGN unification scheme, which suggests
that the presence and properties of large-scale outflows should be
identical in Type 1 and 2 quasars.

Results at somewhat higher redshifts ($z\sim0.6$) provide a more
consistent picture. Large-scale ionized gas outflows are powerful and
ubiquitous in both Types 1 and 2 \citep{liu13a,liu13b,liu14a}. The
properties of these higher-redshift systems depend on the details of
the removal of unresolved emission \citep{husemann16a}, though clearly
outflows are present in many cases.

The properties of quasar-mode AGN outflows (velocity, mass outflow
rate) correlate with bolometric AGN luminosity
\citep{sturm11a,spoon13a,veilleux13a,rupke13a,cicone14a,stone16a,
  gonzalezalfonso17a,sun17a,fiore17a}. The gas depletion timescales
are short in the most luminous AGN, suggesting that feedback is most
effective in these systems
\citep{cicone14a,gonzalezalfonso17a}. Direct observational evidence
for quasar-mode AGN feedback on star formation in galaxies is still
inconclusive, though suggestive correlations exist
\citep[e.g.,][]{farrah12a,wylezalek16a,baron17a}.

\begin{figure}[h]
  \includegraphics[width=0.9\columnwidth]{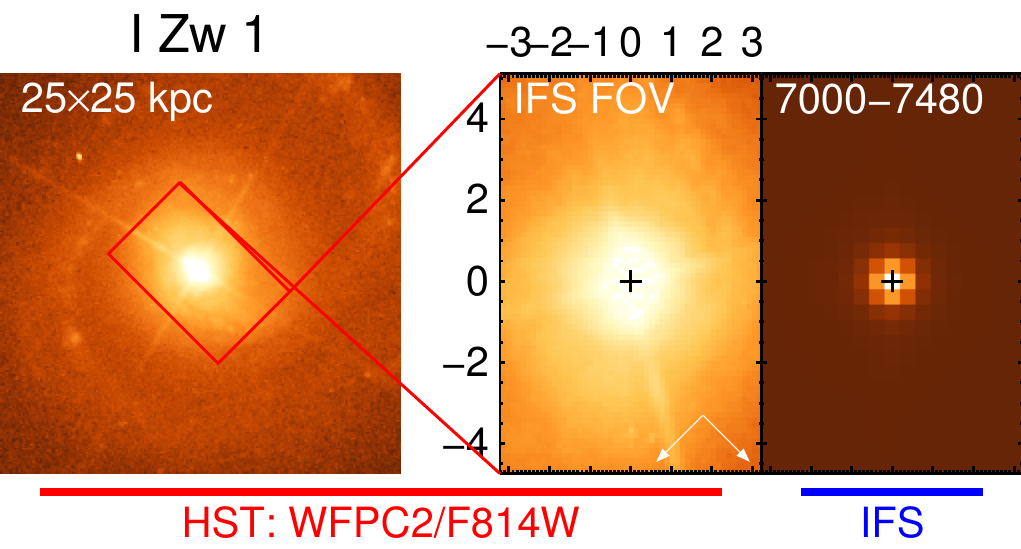}
  \caption{Images of each galaxy. (left) \hst\ image showing the
    galaxy's large-scale structure, labeled with the instrument and
    filter. The GMOS FOV is indicated by a red box. (middle) \hst\
    image cutout matching the GMOS FOV. When the nuclear point source
    is not saturated, the next panel to the right shows the same image
    smoothed with a Gaussian kernel to match the ground-based seeing
    and rebinned to the GMOS spaxel size. North and East are indicated
    by the white arrows. The black cross in each panel is the \hst\
    point used to co-register the \hst\ and IFS images. This point can
    differ by up to 0\farcs03 from the fitted centers of the smoothed
    \hst\ or GMOS data cubes, which are shown as red
    crosses. Differences result from the host contribution to the
    ground-based PSF (in \fohfive, \fonethreetwo, and \fonethreethree;
    \S\,\ref{sec:psfmodel}). (right) GMOS image, summed over the range
    (in \AA, listed in the panel) that overlaps the \hst\ filter
    transmission profile. The black cross is the location of the
    quasar, determined from unsaturated \hst\ images (\fohfive,
    \fonethreetwo, and \fonethreethree) or from the quasar PSF image
    extracted from the IFS fitting (in the other cases; see
    Figure~\ref{fig:cont_radprof}). The axes give galactocentric
    coordinates in kpc. The unsmoothed \hst\ scaling is asinh
    \citep{lupton99a}. The smoothed \hst\ and GMOS scales are
    linear. (The complete figure set (10 images) is available.)}
  \label{fig:cont}
\end{figure}

This study presents detailed observations of a nearby ($z<0.3$) sample
of Type 1 quasars observed with Gemini IFS. In particular, this sample
has considerable ancillary data (\S\,\ref{sec:sample}), including
black hole mass measurements, by which to study the dependence of
outflow properties on black hole properties. Studies of large-scale
outflows in Type 1 quasars have been neglected because of the impact
on the data of the bright nuclear point source. \S\,\ref{sec:red-anl}
details a new technique for spectral removal of the quasar point
spread function (PSF), which allows fitting of stellar host properties
on a spaxel-by-spaxel basis across the field of view (FOV). The
subsequent fitting of strong emission lines and the \nad\ absorption
feature are described, and maps of the quasar and host galaxy, strong
emission line properties, and absorption line properties are shown for
each quasar. The ubiquitous presence and basic properties of the
detected outflows are described in \S\,\ref{sec:of_presence}; global
properties are computed and compared to black hole properties in
\S\S\,\ref{sec:ofprop}--\ref{sec:bhprop}. The wind power source is
analyzed in \S\,\ref{sec:powersource}, and the implications for wind
models and AGN feedback are discussed in
\S\S\,\ref{sec:models}--\,\ref{sec:feedback}. \S\,\ref{sec:summary}
summarizes the paper.

\section{THE DATA} \label{sec:data}

\capstartfalse\begin{deluxetable*}{cccccccrc}
  \tablecaption{Observations\label{tab:obs}}
  \tabletypesize{\footnotesize}
  \tablewidth{\textwidth}

  \tablehead{\colhead{Galaxy} & \colhead{PID} & \colhead{Dates} &
    \colhead{Grating} & \colhead{$t_\mathrm{exp}$} &
    \colhead{PSF} & \colhead{Range} & \colhead{PA} & \colhead{FOV}\\
    \colhead{(1)} & \colhead{(2)} & \colhead{(3)} & \colhead{(4)} &
    \colhead{(5)} & \colhead{(6)} & \colhead{(7)} & \colhead{(8)} &
    \colhead{(9)}}

  \startdata
  \zw             & GN-2015A-DD-9  & 10jul2015             & B600 & 4$\times$1800s & 0\farcs6 & 4610--7490~\AA & 225$\arcdeg$ & 5\farcs1$\times$7\farcs8 \\
  \fohfive        & GS-2011B-Q-64  & 03dec,13dec,31dec2011 & B600 & 6$\times$1800s & 0\farcs6 & 4560--7430~\AA &   0$\arcdeg$ & 5\farcs2$\times$5\farcs0 \\
  \fohseven       & GN-2007A-Q-12  & 08apr2007             & R831 & 3$\times$1440s & 0\farcs5 & 6520--8640~\AA & 180$\arcdeg$ & 3\farcs0$\times$4\farcs5 \\
  \mrk            & GN-2013A-Q-51  & 16jun,17jun2013       & B600 & 5$\times$1800s & 0\farcs8 & 5620--6950~\AA & 215$\arcdeg$ & 6\farcs3$\times$7\farcs5 \\
  \fonethreetwo   & GN-2012A-Q-15  & 02apr,08apr2012       & B600 & 8$\times$1800s & 0\farcs8 & 5340--8230~\AA & 315$\arcdeg$ & 3\farcs9$\times$4\farcs5 \\
  \fonethreethree & GN-2012A-Q-15  & 17may2012             & B600 & 8$\times$1800s & 0\farcs6 & 5340--8230~\AA & 313$\arcdeg$ & 4\farcs2$\times$5\farcs1 \\
  \pgonefour      & GN-2015A-DD-9  & 22jul,23jul2015       & B600 & 4$\times$1800s & 0\farcs8 & 4610--7480~\AA & 335$\arcdeg$ & 3\farcs9$\times$4\farcs5 \\
  \pgonesix       & GN-2012A-Q-15  & 22apr,24apr2012       & B600 & 6$\times$1800s & 0\farcs9 & 5340--8220~\AA & 135$\arcdeg$ & 5\farcs1$\times$7\farcs8 \\
  \pgoneseven     & GN-2003B-C-5   & 23sep2003             & B600 & 2$\times$1800s & 0\farcs7 & 4320--7190~\AA &  14$\arcdeg$ & 3\farcs0$\times$4\farcs2 \\
  \ftwoone        & GS-2005B-Q-50  & 09sep2005             & B600 & 2$\times$1800s & 0\farcs9 & 4060--6860~\AA &   0$\arcdeg$ & 2\farcs7$\times$4\farcs5
  \enddata

  \tablecomments{Column 2: Program ID. Column 3: UT dates of
    observations (in DDmmmYYYY format). Column 4: Grating. Column 5:
    Number of exposures $\times$ length of each exposure. Column 6:
    FWHM of quasar PSF in the data cube, except for \fohfive, in which
    the number is an estimate of the seeing from observing
    logs. Column 7: Observed wavelength range of combined data
    cube. Column 8: E of N position angle of IFU during
    observations. Column 9: Size of combined FOV.}
\end{deluxetable*}\capstarttrue

\subsection{Sample} \label{sec:sample}

The present low-redshift sample ($z < 0.3$) was selected from the
larger QUEST sample (Quasar and ULIRG Evolution Study;
\citealt{veilleux06a,schweitzer06a,netzer07a,veilleux09b,veilleux09a}). The
QUEST sample consists of local ultraluminous infrared galaxies
(ULIRGs) from the 1~Jy sample \citep{kim98a} and optically-selected
Palomar-Green quasars \citep{schmidt83a}. Properties of the Gemini
subsample of QUEST are listed in Table \ref{tab:sample}. The
heterogeneous selection of this subsample was due to the combination
of three archival datasets with seven observations designed to meet
the current science goals. For the sources observed for this program,
selection was based on a combination of observability, proximity (to
maximize spatial resolution), and diversity in galaxy properties.

All of the quasars in the sample are traditional Type 1s except for
F05189$-$2524. However, this quasar has a broad line region (BLR)
observed in the near infrared \citep{veilleux99b}; thus a portion of
the BLR is unobstructed by the obscuring medium. Three are
broad-absorption-line (BAL) quasars: \fohseven, \mrk, and \pgoneseven.
All three are also low-ionization BALs (LoBALs), with broad absorption
observed in \ion{Mg}{2} \citep{smith95a,turnshek97a,veilleux16a} and
\nad\ \citep{boroson92a,rupke05c}.

\begin{figure*}
  \includegraphics[width=0.9\textwidth,center]{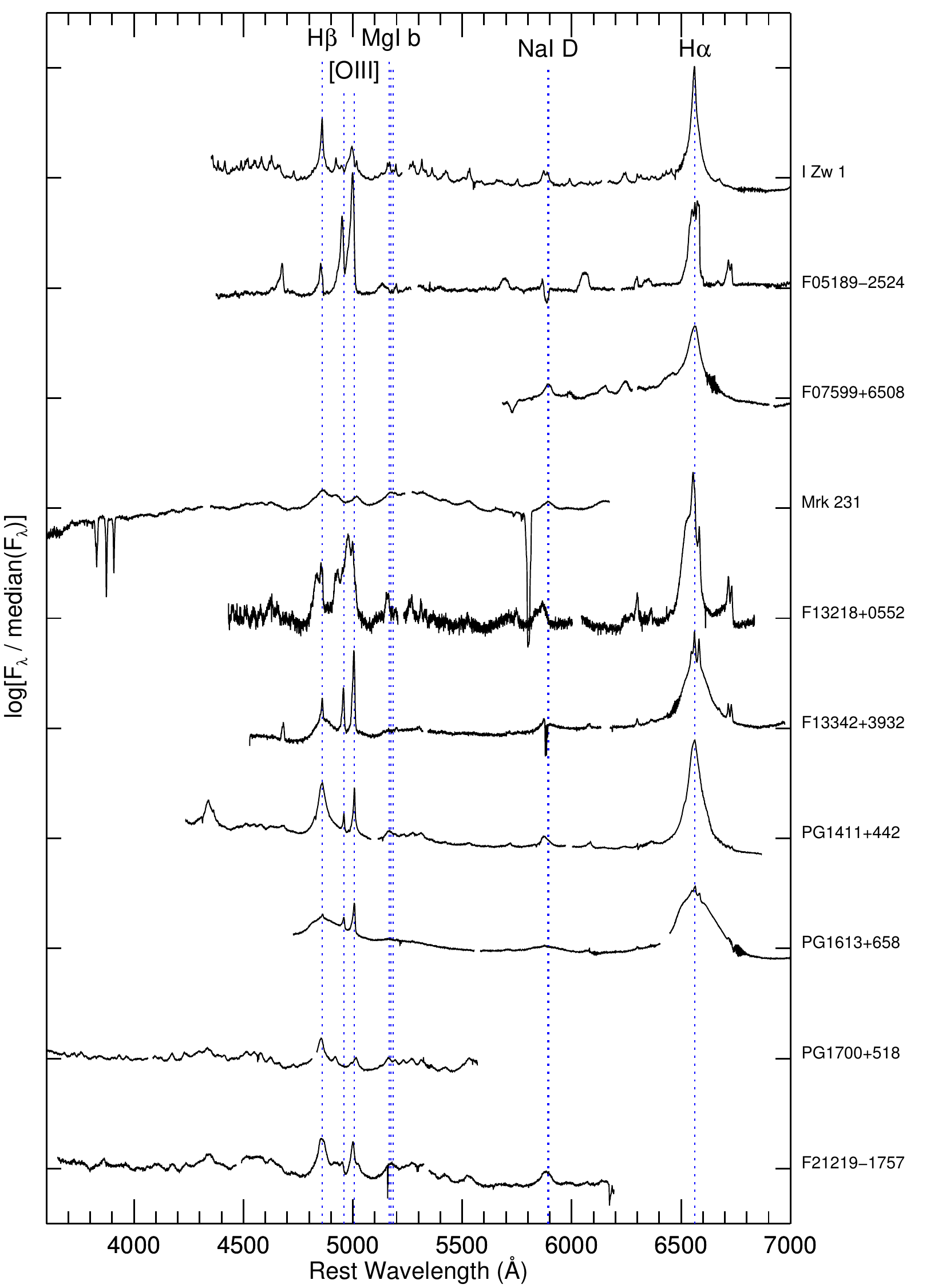}
  \caption{Quasar spectra, derived from the central spaxel in the
    0\farcs1 spaxel data cubes. These are used as fit inputs, and are
    represented by $I_\mathrm{quasar}^{0}$ in \S\,\ref{sec:psfmodel}
    and equation 2.}
  \label{fig:psfspec}
\end{figure*}

The average value of $L_H/L_H^*$ in the sample is 2.8
\citep{veilleux06a,veilleux09a}, which corresponds to a halo mass of
log($M_\mathrm{halo}/\msun$) $\sim$ 12.8 based on abundance matching
\citep{moster13a} and a baryon fraction $f_b = 0.17$. Thus, the sample
probes well above the halo mass at which the stellar-to-halo mass
fraction peaks; i.e., the region in which AGN feedback is thought to
be necessary in quenching star formation
\citep[e.g.,][]{somerville08a}.

\subsection{Observations and Data Reduction} \label{sec:obs}

The ten quasars were observed with the Integral Field Unit in the
Gemini Multi-Object Spectrograph (GMOS;
\citealt{allington-smith02a,hook04a}) on the Gemini North and South
telescopes. Details of each observation are listed in Table
\ref{tab:obs}. All observations used the 1-slit mode, and 9/10 used
the B600 grating; this setup yields an approximately constant
resolution element of full width at half-maximum (FWHM) 1.5~\AA. All
observations were dithered to increase the FOV. The resulting total
FOV is centered on the quasar in each case except for \pgoneseven. The
wavelength range for the typical setup included most strong emission
lines (\ha, \hb, \othll, \ntll, \sutl, \ooll). For two archival setups
(\pgoneseven\ and \ftwoone) and a new \mrk\ observation, the grating
was tilted blueward, including only \otl\ through \othll. All cubes
but one (\pgoneseven) also included the \nadl\ absorption
doublet. Image quality of the reduced data cubes, as determined from a
PSF fit (\S\,\ref{sec:psfmodel}), was 0\farcs5--0\farcs9. (Separate
exposures for a given quasar were convolved to match the worst seeing
in the data.)

The Gemini IRAF package (v1.12) was used to reduce the data,
supplemented by revision 114 of the IFUDR GMOS package and IFSRED
\citep{rupke14b}. Data cubes were visualized with QFitsView
\citep{ott12a}. In addition to the steps described in \citet{rupke13a}
and \citet{rupke15a}, scattered light was removed from the flat fields
and the data prior to extraction. Because of the strong nuclear point
source in most cubes, scattered light removal was crucial for accurate
continuum and emission-line fitting.

Scattered light could not be completely removed from two systems: \zw\
and \pgonefour. The dithering strategy adopted in these cases put the
quasar near the edge of the field of view along the short axis of the
IFU. Consequently, the quasar light was concentrated near either the
edge or center of fiber blocks in the raw data. Because scattered
light subtraction relies on extrapolating from interblock regions, the
off-centered quasar light in the blocks meant that this extrapolation
was imperfect. The result is an under-subtraction of the scattered
light, and is visible as a vertical stripe emerging from the point
source in the reconstructed host galaxy image (\S\ref{sec:psfmodel}).

Each final data cube was rebinned to a spaxel size of
0\farcs3$\times$0\farcs3, which balanced sensitivity and sampling of
the seeing disk (the sampling was Nyquist or better in almost every
case).

\subsection{Data Analysis} \label{sec:red-anl}

The spectrum in each spaxel was modeled with IFSFIT \citep{rupke14a},
which utilizes PPXF \citep{cappellari12a} for starlight fitting and
MPFIT \citep{markwardt12a} for other continuum and emission line
fitting. In brief, IFSFIT masks emission line regions, fits the
continuum, and then simultaneously fits all emission lines in the
continuum-subtracted spectrum. Subsequently, interstellar absorption
lines are modeled.

\subsubsection{Separation of Quasar and Host Galaxy
  Light} \label{sec:psfmodel}

In most of this sample, the luminous AGN, or quasar, outshines the
host galaxy. \hst\ $H$-band data show that the quasar light fraction
is 0.45--0.89 \citep{veilleux06a,veilleux09b}. Heavy nuclear
obscuration in \fohfive, and to a lesser degree in \fonethreetwo\ and
\fonethreethree, lower this fraction in the optical. At \hst\
resolution, the quasar contributes 45\%\ of the $H$-band light of
\fohfive\ but $<$1\%\ of its $I$-band light \citep{kim13a}, consistent
with the appearance of a BLR only in the near infrared
\citep{veilleux99b}. On visual inspection of optical \hst\ images
(Figure~\ref{fig:cont}) \fonethreetwo\ and \fonethreethree\ show a
small quasar light fraction, while six other quasars dominate their
hosts. The final quasar, \ftwoone, has no optical \hst\ image.

To visually compare \hst\ and IFS data, the cubes were summed over
wavelengths that overlap with the \hst\ filter range
(Figure~\ref{fig:cont}). For the three systems with small quasar light
fraction in the optical, the \hst\ images were also Gaussian-smoothed
to match the seeing.

\begin{figure*}[h]
  \centering
  \includegraphics[width=0.9\textwidth,center]{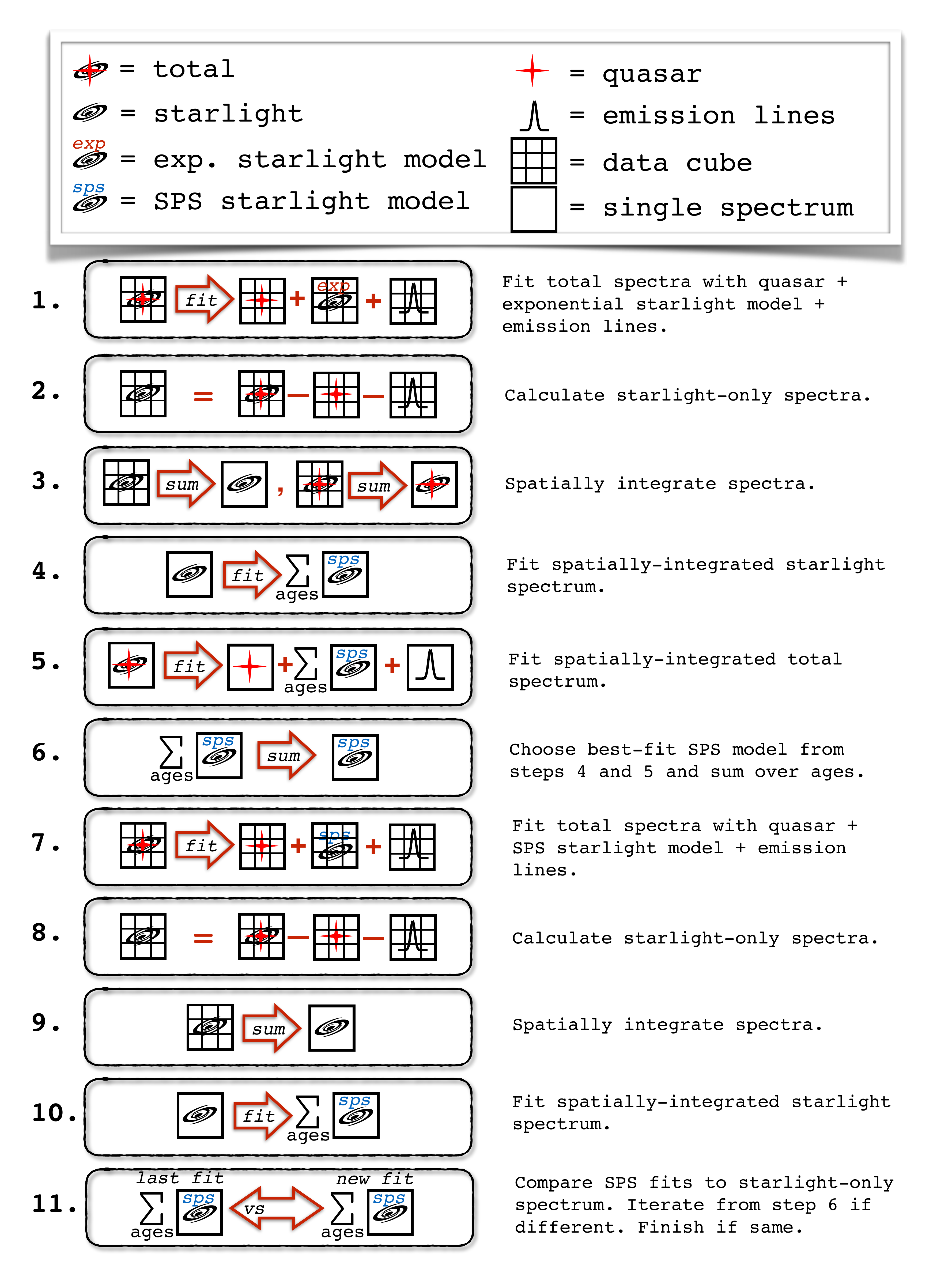}
  \caption{Illustration of the procedure for separating the quasar and
    host galaxy light (\S\,\ref{sec:psfmodel}).}
  \label{fig:flowchart}
\end{figure*}

In 9/10 cases, the peak IFS spaxel in 0\farcs1 sampling represents a
pure quasar spectrum to very good approximation (Figure
\ref{fig:psfspec}). For \fohfive, the peak spaxel traces an unresolved
point source at optical ground-based resolution (\citealt[][47\%\
contribution to total light]{veilleux02a}), but it is a combination of
quasar light and starlight. It has very strong, high-ionization
emission lines from the NLR that are either spatially unresolved or
dominated by an unresolved component (including \othll, \ion{He}{2}
4686~\AA, [\ion{Fe}{7}] 5159~\AA, [\ion{Fe}{7}] 5721~\AA,
[\ion{Fe}{7}]~6087~\AA, and [\ion{Fe}{10}] 6375~\AA;
\citealt{farrah05a}).


A multi-step process is adopted to fit the quasar and host galaxy in
each spaxel. This iterative approach is necessary for three reasons.

First, subtracting the starlight accurately is crucial for properly
constraining the underlying interstellar component of the \nad\
absorption feature. It also produces more accurate recombination-line
fluxes by accounting properly for Balmer absorption. In dusty,
infrared-luminous galaxies where \nad\ is strong, the stellar
continuum subtraction is much less important because interstellar
absorption dominates the feature. However, in many of these (less
dusty) quasars, the stellar and interstellar \nad\ absorption are
comparable in strength.

Second, simultaneously fitting the quasar, starlight, and emission
lines was technically infeasible. The starlight must be fit in PPXF
(to allow the velocity and dispersion of the stellar spectrum to be
free parameters). In principle the quasar spectrum could also be input
into PPXF as a sky spectrum, since the sky spectrum remains
unconvolved with the fitted stellar velocity dispersion. However, PPXF
uses Legendre polynomials to correct the stellar fit for imperfectly
calibrated data.  Linear combinations of polynomials are too flexible
because they have the freedom to produce negative continuum
values. This is especially a problem in spaxels where the host galaxy
contribution is uncertain due to domination by quasar light or the
total flux is very low.

Third, there is a range of sensitivity in this dataset. This is partly
by design -- galaxies were chosen over a range of redshifts. It is
also by accident -- galaxies that were targeted for this program were
observed $\sim$2$\times$ longer than galaxies observed for different
purposes and retrieved from the archive. There is also intrinsic
variation in the strength of the host galaxy starlight compared to the
quasar. Galaxies with very high signal-to-noise (S/N) in the stellar
continuum (e.g., \fohfive\ and \mrk) can be fit using multiple stellar
templates. However, the composition of the stellar population could
not be constrained on a spaxel-by-spaxel basis in most of the
sample. In one system (\fohseven), the stellar population could not be
constrained even in the spatially-integrated spectrum.

\begin{figure*}
  \includegraphics[width=0.9\textwidth,center]{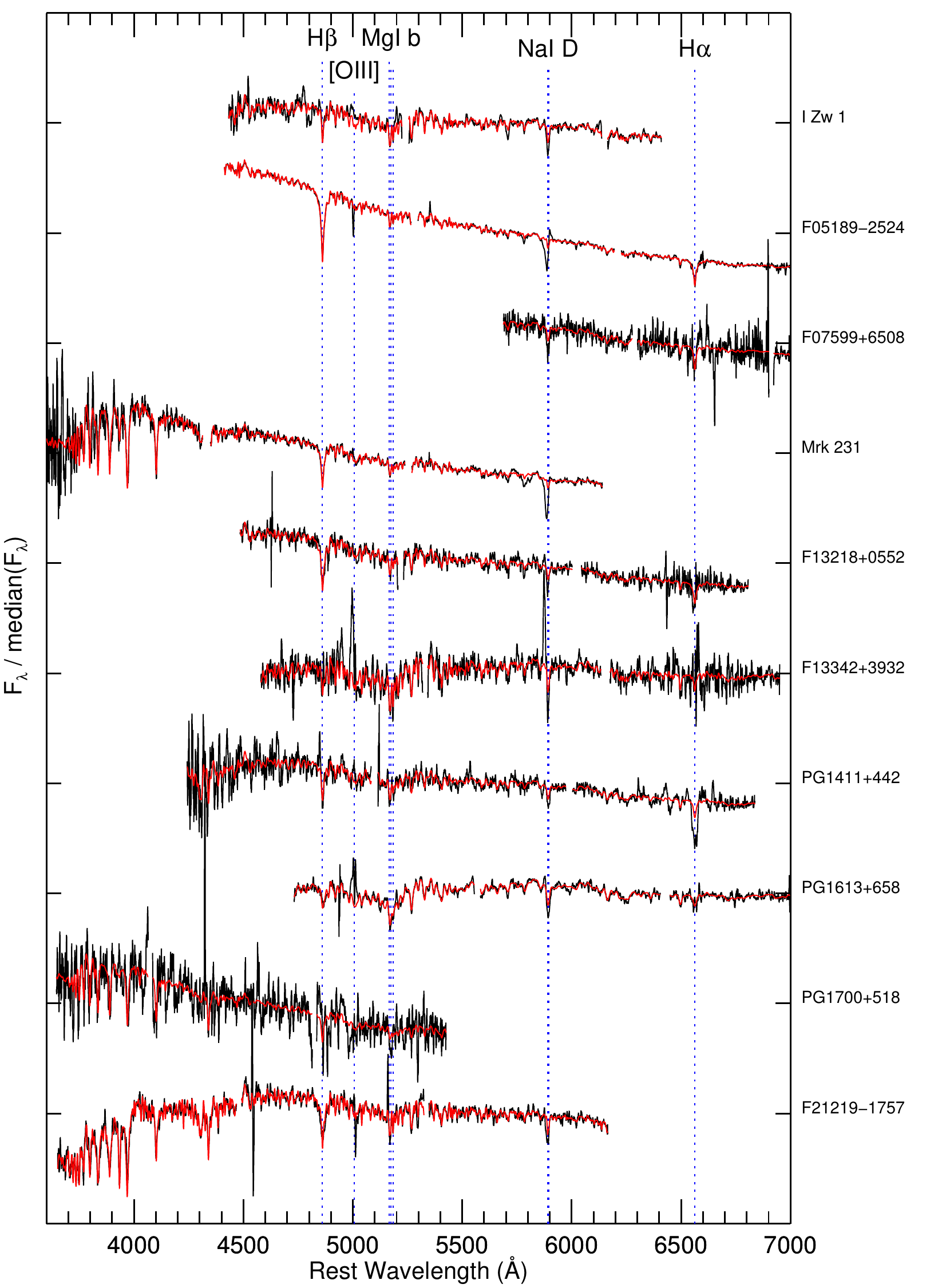}
  \caption{Stellar population synthesis fits (red) to starlight
    spectra (black) that have been spatially-integrated over the IFS
    FOV. The quasar light and emission lines have been removed
    (\S\,\ref{sec:psfmodel}). The data is smoothed by a 10-pixel
    boxcar to facilitate comparison with the fits.}
  \label{fig:spsfits}
\end{figure*}

A description of the fitting process follows, and it is illustrated in
Figure \ref{fig:flowchart}. As discussed above, all fitting is
performed with IFSFIT. IFSFIT calls PPXF when fitting starlight
spectra with stellar population synthesis (SPS) models. In the
discussion below, ``total'' refers to a complete spectrum with all
physical components.
\begin{enumerate}
\item{\ul{Fit total spectra with quasar $+$ exponential
      starlight model $+$ emission lines.} Each spaxel $n$ is fit with
    the sum of a scaled quasar spectrum, a starlight model, and
    emission lines:
    \begin{equation}
      I_\mathrm{total}^{n} = I_\mathrm{quasar}^{n} + I_\mathrm{starlight,~exp. model}^{n} + I_\mathrm{emission}^{n}.
    \end{equation}
    Emission lines are first masked; the quasar and starlight
    continuum components are fit simultaneously using MPFIT; the
    continuum is subtracted from the data; and finally the emission
    lines are fit with MPFIT. The starlight model and quasar scaling
    in each spaxel are each the sum of four exponentials:
    \begin{equation}
      I_\mathrm{total}^{n} = \sum_\mathrm{i=1}^{4}I_\mathrm{i}^{n}I_\mathrm{quasar}^{0} + \sum_\mathrm{j=1}^{4}I_\mathrm{j}^{n} + I_\mathrm{emission}^{n},
    \end{equation}
    where $I_\mathrm{quasar}^{0}$ is the nuclear quasar spectrum
    (Figure~\ref{fig:psfspec}); the exponentials are
    \begin{alignat*}{2}
      & I_{1}^{n}  && = a_{1}^{n}e^{-b_{1}^{n} \bar{\lambda}}\\
      & I_{2}^{n}  && = a_{2}^{n}e^{-b_{2}^{n}(1-\bar{\lambda})},\\
      & I_{3}^{n}  && = a_{3}^{n}(1-e^{-b_{3}^{n}\bar{\lambda}}),~\mathrm{and}\\
      & I_{4}^{n}  && = a_{4}(1-e^{-b_{4}^{n}(1-\bar{\lambda})});
    \end{alignat*}
    $a_\mathrm{i}^{n}\geq0$; $b_\mathrm{i}^{n}\geq0$;
    $\bar{\lambda} \equiv \frac{\lambda -
      \lambda_\mathrm{min}}{\lambda_\mathrm{max} -
      \lambda_\mathrm{min}}$;
    and [$\lambda_\mathrm{min}$,$\lambda_\mathrm{max}$] is the fit
    range. Exponentials are used because they are monotonic and can be
    made positive-definite. The four exponentials allow for all
    combinations of concave/convex and monotonically
    increasing/decreasing. Physically, the quasar multipliers correct
    for spatially-varying errors due to imperfect scattered light
    removal (or other calibration effects) and for the
    $\lambda^{-0.2}$ dependence of PSF size on atmospheric seeing. For
    a Gaussian PSF, the latter effect produces an exponential of the
    form $e^{-C\lambda^{0.8}}$, which is similar to $I_1$.}
\item{\ul{Calculate starlight-only spectra.} The quasar and
    emission-line components are subtracted from the total spectrum in
    each spaxel. The result is a data spectrum including only
    starlight:
    \begin{equation}
      I_\mathrm{starlight}^{n} = I_\mathrm{total}^{n} - I_\mathrm{quasar}^{n} - I_\mathrm{emission}^{n}.
    \end{equation}}
\item{\ul{Spatially integrate spectra.} The starlight and total
    spectra are spatially integrated:
    \begin{alignat*}{2}
      & I_\mathrm{starlight} && = \sum_{n}I_\mathrm{starlight}^{n}~\mathrm{and}\\
      & I_\mathrm{total} && = \sum_{n}I_\mathrm{total}^{n}.
    \end{alignat*}}
\item{\ul{Fit spatially-integrated starlight spectrum.} The
    spatially-integrated starlight spectrum is fit with the
    \citealt{gonzalezdelgado05a} high-resolution SPS models and
    Legendre polynomials $P_l$:
      \begin{alignat*}{2}
        & I_\mathrm{starlight} && = I_\mathrm{starlight,~SPS~model} + I_\mathrm{Leg.~poly.} \\ \nonumber
        & ~ && = \sum_t A^t I_\mathrm{starlight,~SPS~model}^{t} + \sum_\mathrm{l=0}^\mathrm{l_\mathrm{max}}B_\mathrm{l}P_\mathrm{l}.
      \end{alignat*}
      The models vary in age, $t$, but assume Solar metallicity
      \citep{rupke08a}. A Legendre polynomial of order
      $l_\mathrm{max}=20-50$ (small compared to $\sim$6200 pixels in
      wavelength) is used in order to account for imperfections in
      calibration (e.g., scattered light subtraction) or in the PSF
      subtraction. These imperfections appear as very shallow, broad
      ripples in the continuum or as higher-frequency edge effects.}
  \item{\ul{Fit spatially-integrated total spectrum.} This fit
      proceeds as in steps 1, 2, and 4, but involves only a single
      invocation of IFSFIT. After masking emission lines, IFSFIT fits
      the continuum component of the spatially-integrated total
      spectrum with a scaled quasar plus exponential model for
      starlight as in step 1:
      \begin{equation}
        I_\mathrm{total,cont.} = I_\mathrm{quasar} + I_\mathrm{starlight,~exp.model}.
      \end{equation}
      IFSFIT then calculates the starlight spectrum as in step 2:
      \begin{equation}
        I_\mathrm{starlight} = I_\mathrm{total,cont.} - I_\mathrm{quasar},
      \end{equation}
      and fits the result with the sum of SPS and Legendre components
      as in step 4 above. Finally, the emission lines are unmasked and
      fit. The endpoint of this fit is represented by
      \begin{equation}
        I_\mathrm{total} = I_\mathrm{quasar} + I_\mathrm{starlight,~SPS~model} + I_\mathrm{emission}.
      \end{equation}}
  \item{\ul{Choose best-fit SPS model from steps 4 and 5 and
        sum over ages.} The fit (from either of the previous two
      steps) that shows the minimum contribution from a polynomial is
      typically chosen. The best-fit SPS model is summed over ages
      $t$, ignoring the polynomial component:
      \begin{equation}
        I_\mathrm{starlight,~SPS~model}^\mathrm{fixed} =
        \sum_t A^tI_\mathrm{starlight,~SPS~model}^{t}.
      \end{equation}
      The ``fixed'' superscript indicates that this starlight model
      will not vary in subsequent steps.}
  \item{\ul{Fit total spectra with quasar $+$ SPS starlight
        model $+$ emission lines.} Each spaxel is fit as in step 1,
      but with the (fixed) SPS starlight model from step 6 replacing
      the exponential model:
      \begin{equation}
        I_\mathrm{total}^{n} = I_\mathrm{quasar}^{n} + C^{n}I_\mathrm{starlight,~SPS~model}^\mathrm{fixed} + I_\mathrm{emission}^{n}.
      \end{equation}
      As in step 5, IFSFIT is invoked only once, but there are
      intermediate steps (within SPS) before the SPS model is fit. In
      particular, the entire spectrum in each spaxel is first fit with
      the quasar, an exponential starlight model, and emission
      lines. A starlight spectrum is then calculated by removing the
      best-fit quasar and emission components from the total spectrum,
      and the starlight-only spectrum is refit with the SPS model. The
      fixed SPS model,
      $I_\mathrm{starlight,~SPS~model}^\mathrm{fixed}$, is allowed to
      vary from spaxel to spaxel in velocity, dispersion, and
      normalization ($C^{n}$). The stellar population is thus
      implicitly assumed to be unvarying across the relatively small
      GMOS FOV.}
  \item{\ul{Calculate starlight-only spectra.} Repeat step 2,
      using the new fit from step 7. The new starlight-only spectrum
      can differ from the first one because the first fit (step 1)
      does not properly account for stellar absorption lines, which in
      turn affects how the emission lines are removed from the total
      spectrum. The new fit of each spaxel (step 7) accounts for
      stellar absorption lines.}
  \item{\ul{Spatially integrate starlight spectrum.} Repeat
      step 3 using the new starlight spectrum from step 8.}
  \item{\ul{Fit spatially-integrated starlight spectrum.}
      Repeat step 4, but now using the new spatially-integrated
      starlight spectrum from step 9.}
  \item{\ul{Compare SPS fits to starlight-only spectrum.}
      Compare the results of steps 4 and 10 by calculating the
      fractional contribution to the total light of starlight in four
      age groups. If there is a $\geq$10\%\ difference in any group,
      begin again with the new SPS model, starting from step
      6. Iterate until the current SPS model differs from the previous
      by $<$10\%\ in each of the four age groups.}
\end{enumerate}

\capstartfalse\begin{deluxetable*}{crcrrr}
  \tablecolumns{6} \tablecaption{Stellar Population Synthesis
    Fits\label{tab:spsfits}} \tablewidth{\textwidth}

  \tablehead{\colhead{} & \multicolumn{4}{c}{Ages (Gyr)} & \colhead{}\\
    \cmidrule(lr){2-5} \colhead{Galaxy} & \colhead{$\leq$0.01} &
    \colhead{0.01--0.1} & \colhead{0.1--1} & \colhead{$>$1} &
    \colhead{poly} \\
    \colhead{(1)} & \colhead{(2)} & \colhead{(3)} & \colhead{(4)} &
    \colhead{(5)} & \colhead{(6)}}

  \startdata
  \zw             & 38 &  0 &  0 & 23 & 39 \\
  \fohfive        & 38 &  1 & 36 &  3 & 22 \\
  \fohseven       & 46 &  0 &  0 & 19 & 35 \\
  \mrk            & 59 &  0 & 13 & 11 & 17 \\
  \fonethreetwo   & 22 &  0 & 34 & 13 & 31 \\
  \fonethreethree &  0 &  0 &  0 & 39 & 51 \\
  \pgonefour      & 26 &  0 &  1 & 26 & 47 \\
  \pgonesix       &  0 &  0 &  0 & 58 & 42 \\
  \pgoneseven     & 37 &  0 &  0 &  4 & 59 \\
  \ftwoone        &  6 &  0 & 42 & 19 & 33
  \enddata

  \tablecomments{Columns 2--5: Percent contribution to total flux (in
    the wavelength range of the fits to integrated host galaxy
    spectra) from Solar metallicity stellar populations of different
    ages (\S\,\ref{sec:psfmodel}). Column 6: Percent contribution from
    linear combination of Legendre polynomials.}
\end{deluxetable*}\capstarttrue

\begin{figure*}
  \includegraphics[width=0.95\textwidth,center]{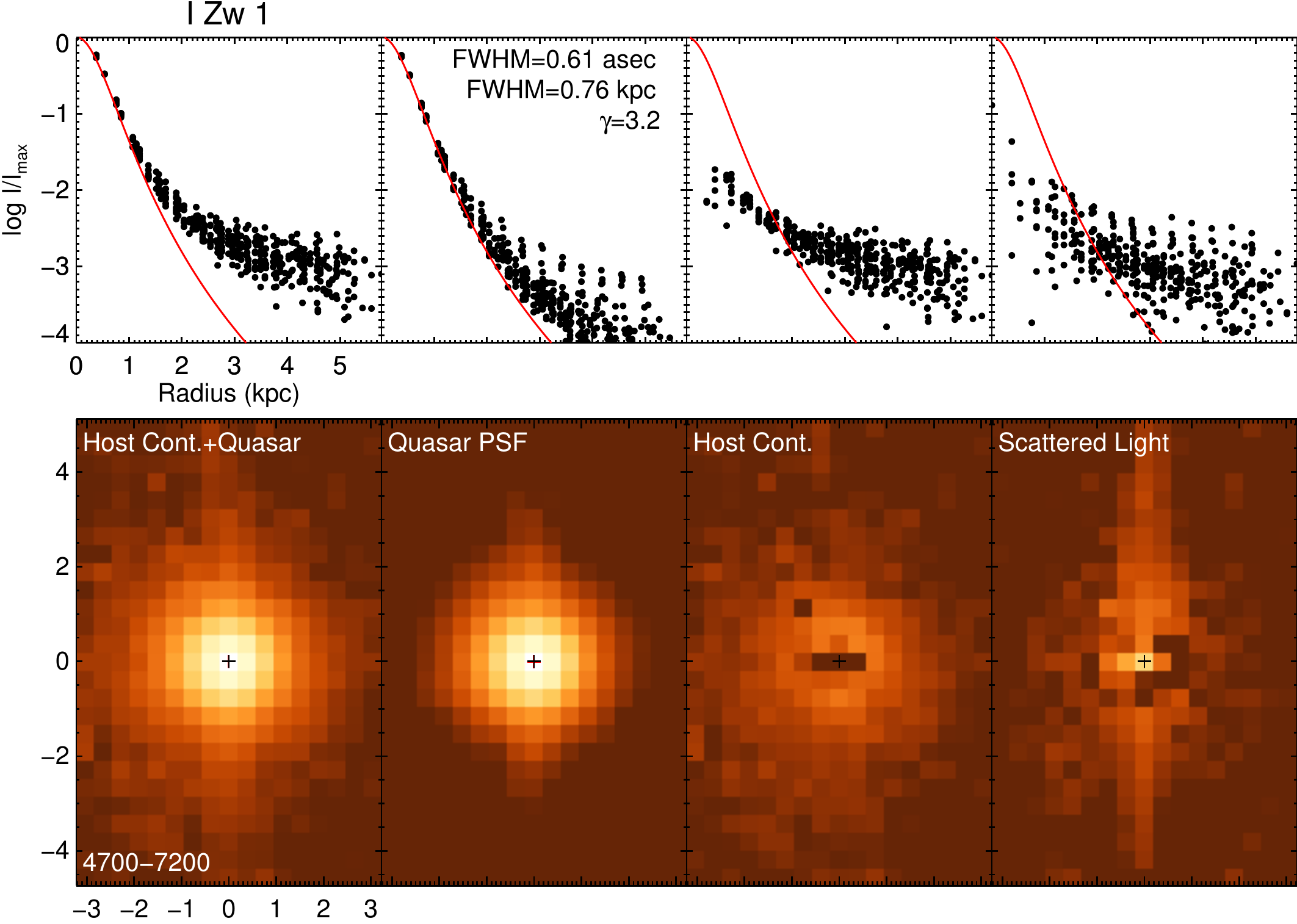}
  \caption{Radial intensity profiles and corresponding images of the
    galaxy continuum and the continuum components extracted from the
    fit (\S\,\ref{sec:psfmodel}). In each column, the top panel shows
    log of normalized intensity vs.\ radius. The bottom panel shows
    the correponding image on a log intensity scale. The left (or
    only) column shows host galaxy continuum plus quasar PSF, summed
    over the fitted wavelength range (listed in the corner of the
    image). For data where an unresolved quasar spectrum is fit, the
    middle column shows the summed quasar-only data. The right column
    then shows the host continuum. A circular 2-D Moffat fit to the
    quasar PSF is in red on the top row, and the best-fit FWHM and
    Moffat index are printed in the upper right corner of the top
    middle panel. In one case, \zw, a fourth column also shows the
    polynomial component separated from the stellar component and
    easily identifiable as strong scattered light by the central
    vertical stripes. In the bottom panels, the black cross locates
    the quasar center, based either on unsaturated \hst\ images
    (\fohfive, \fonethreetwo, and \fonethreethree;
    Figure~\ref{fig:cont}) or fitting of the quasar PSF image (other
    cases). The red cross is the fitted center of each image. The
    black and red crosses differ in the total image (left column) in
    the three cases where the host contributes to the ground-based PSF
    (Figure~\ref{fig:cont} and \S\,\ref{sec:psfmodel}). The fitted and
    assumed QSO centers (middle column) are typically identical but
    differ by 0\farcs04 in \fohfive, reflecting the imperfect PSF
    template due to host galaxy contamination. (The complete figure
    set (10 images) is available.)}
  \label{fig:cont_radprof}
\end{figure*}

For most objects, the first iteration of steps 1--10 resulted in
convergence. For three objects, a second iteration was required.

The final SPS fits to the spatially-integrated starlight spectra are
shown in Figure \ref{fig:spsfits}, and the percent contributions (by
flux in the fitted wavelength range) of the stellar populations in
four age bins are given in Table \ref{tab:spsfits}. The use of the
\citet{gonzalezdelgado05a} templates imposes an upper rest wavelength
limit of 7000~\AA\ to the fits. Reddening of the stellar population,
which is almost certainly present, was not included in the fits
because of the additional degrees of freedom this requires. Inclusion
of reddening would likely result in younger best-fit ages. The
luminosity-weighted age mixtures from these fits vary from a
combination of ages $\la$10~Myr with ages $\ga$100~Myr to populations
dominated by ages $\ga$1~Gyr. The precision of these ages is likely
coarse, given the single metallicity, unreddened populations and the
lack of ages 10--100~Myr in the best-fit populations. However, the
important outcome of these fits is not an accurate set of ages but
rather an accurate fit to the underlying stellar continuum,
particularly in the Balmer lines and \nad, at modest spectral
resolution, which can be produced by more than one combination of
stellar populations.

For the two sources where unsubtracted scattered light was most
problematic (\S\,\ref{sec:obs}; \zw\ and \pgonefour), the starlight
fits also included a separate very broad component in \ha\ and \hb\
which was limited in wavelength but allowed to vary in flux and
width. The scattered light under-subtraction was pronounced in these
two broad emission line features because of their intrinsic strength
in the quasar spectrum and the particular way in which light is
scattered across the 2-D GMOS raw data. Including these free
components, which were discarded after fitting, significantly improved
the fits.

The wavelength-integrated results of the quasar/galaxy decomposition
with IFSFIT are shown in Figure~\ref{fig:cont_radprof}. Moffat fits to
the quasar PSF are also shown in this figure, which illustrate the
effectiveness of the spectral method in recovering a smooth PSF. The
exponential model that multiplies the quasar spectrum in each spaxel
is shaped primarily by the $I_1$ term, likely due to its similarity to
the theoretical dependence of the seeing-only PSF on wavelength. In
every case, the host galaxy (starlight plus emission lines) is visible
as an excess over the PSF in the radial profile. In most cases
specific features in the host galaxy image are easily recognizable in
\hst\ images (Figure~\ref{fig:cont}).

The Legendre polynomial components of the fits represent featureless
continuum contaminants due to incomplete scattered light removal,
imperfect PSF removal, or unfit stellar continuum components. The
percent contributions to the total fitted light are in the range
20--50\%, with a median of 37\%. For the purposes of this analysis,
the polynomial component is assumed to be part of the host galaxy, an
assumption which is borne out by the qualitative comparisons of the
host galaxy reconstructions with \hst\ images. However, in one system
with strong residual scattered light, \zw, the polynomial component is
identifiable as scattered light and is separated from the host
emission (Figure~\ref{fig:cont_radprof}).

\begin{figure*}
  \includegraphics[center]{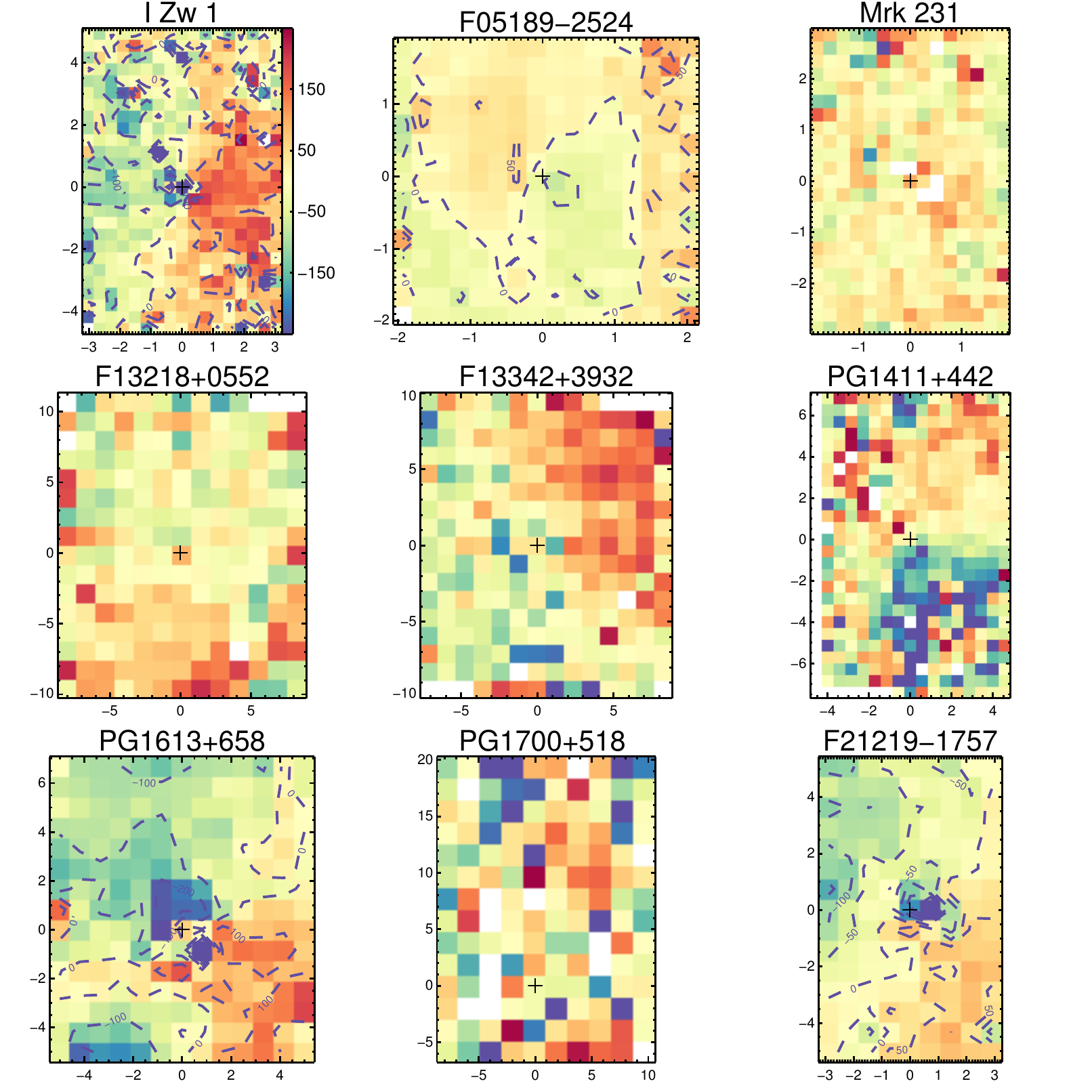}
  \caption{Stellar velocities in each quasar host except \fohseven,
    for which the stellar component could not be fit. Axes are in
    kpc. Contours are shown in cases where the S/N is high enough;
    contour labels are in \kms\ relative to systemic. Rotation is
    evident except in \mrk\ and \pgoneseven; the former may be near
    face-on, while the latter data is shallow. There is remarkable
    spatial coherence within individual maps, despite the quasar PSF
    dominating the light in most data cubes.}
  \label{fig:stelvel}
\end{figure*}

\begin{figure*}
  \includegraphics[width=\textwidth,center]{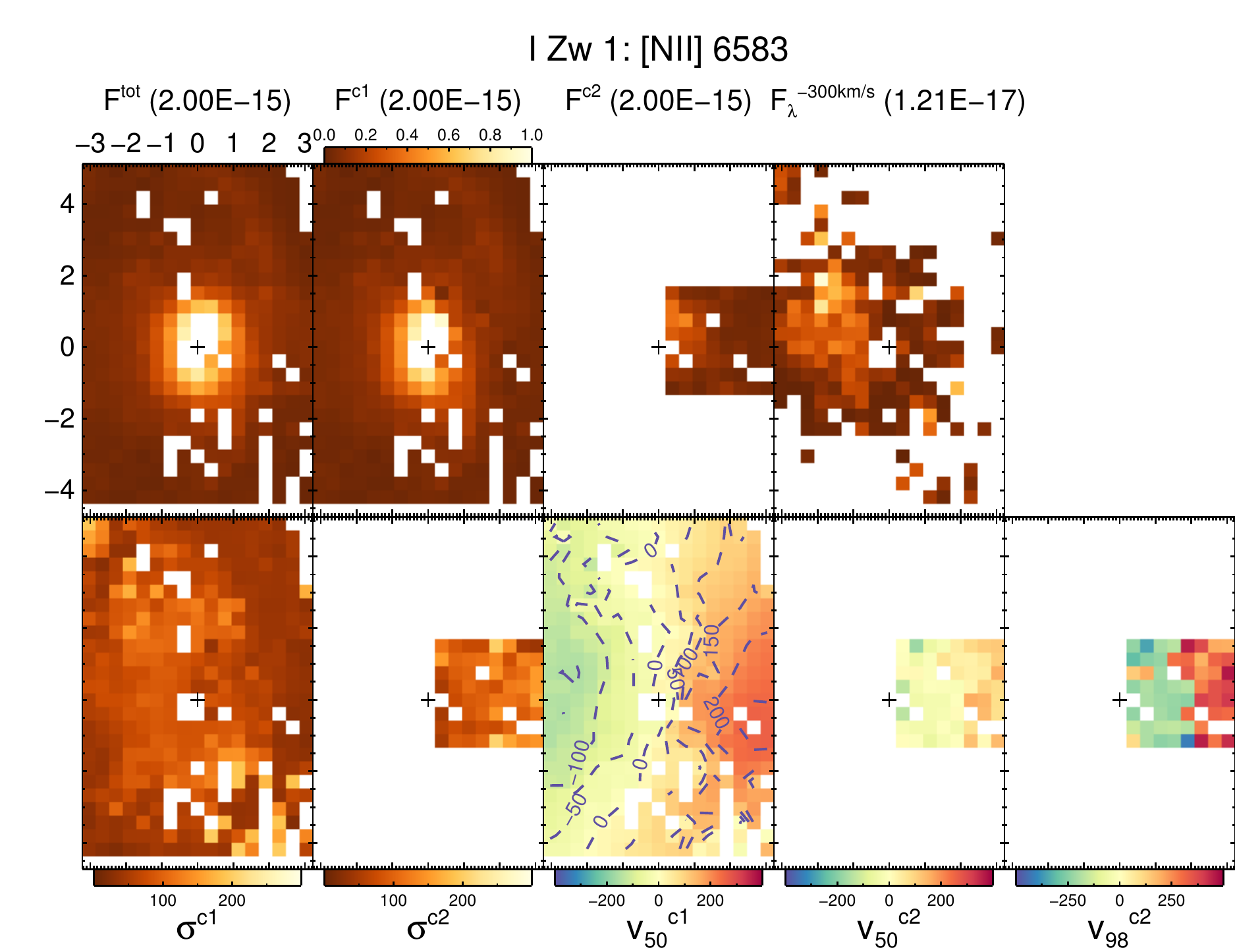}
  \caption{Properties of the \ntl\ or \othl\ emission lines in each
    quasar. Top panels show total flux and flux per component in units
    of erg s$^{-1}$ cm$^{-2}$ arcsec$^{-2}$, and flux density at a
    representative blueshifted velocity (if more than one component is
    fit) in units of erg s$^{-1}$ cm$^{-2}$ \AA$^{-1}$
    arcsec$^{-2}$. The flux scale is from 0 (dark orange) to 1
    (white), and the peak flux is listed in each panel title. Bottom
    panels show velocities: the velocity dispersion $\sigma$ of each
    component, the central velocity (\vfifty) of each component, and
    \vtsig$^\mathrm{c2}$ if a second component is fit. Color bars show
    the velocity scales in \kms. The spatial image scale in kpc is
    shown on the axes of the upper left panel. Velocity contours are
    plotted atop \vfifty$^\mathrm{c1}$ in all cases but \mrk, for
    which the contours are too noisy. (The complete figure set (9
    images) is available.)}
  \label{fig:eml_maps}
\end{figure*}

The stellar velocity maps resulting from the starlight fits are shown
in Figure~\ref{fig:stelvel}.

The fidelity of the host galaxy reconstructions is at a minimum within
1--2 spaxels of the nucleus (i.e., within 0.5--1 spatial resolution
elements). However, even within this radius, some stellar and line
properties can still be extracted accurately (see
Figure~\ref{fig:stelvel} and the emission- and absorption-line maps
presented below), with this accuracy depending on the relative
strength of the PSF, the details of scattered light contamination, and
the strength of the relevant line features.

\subsubsection{Emission Line Fitting} \label{sec:emlfits}

Resolved line emission was detected and fitted in each quasar, though
in \ftwoone\ the emission was too weak for the line properties to be
robustly determined. Every data cube also included unresolved line
emission from the NLR; this unresolved emission was present in the
point source spectrum. The NLR emission was not removed from the point
source spectrum prior to using it as a continuum template, because at
the physical resolutions of the target and instrument combinations,
the NLR emission is dominated by unresolved components. Tests showed
that removing NLR emission from the point source spectrum before
fitting yielded very different fits. This was due to the emission line
fit in many spaxels being dominated by the unresolved point source
component.

\begin{figure*}
  \includegraphics[center]{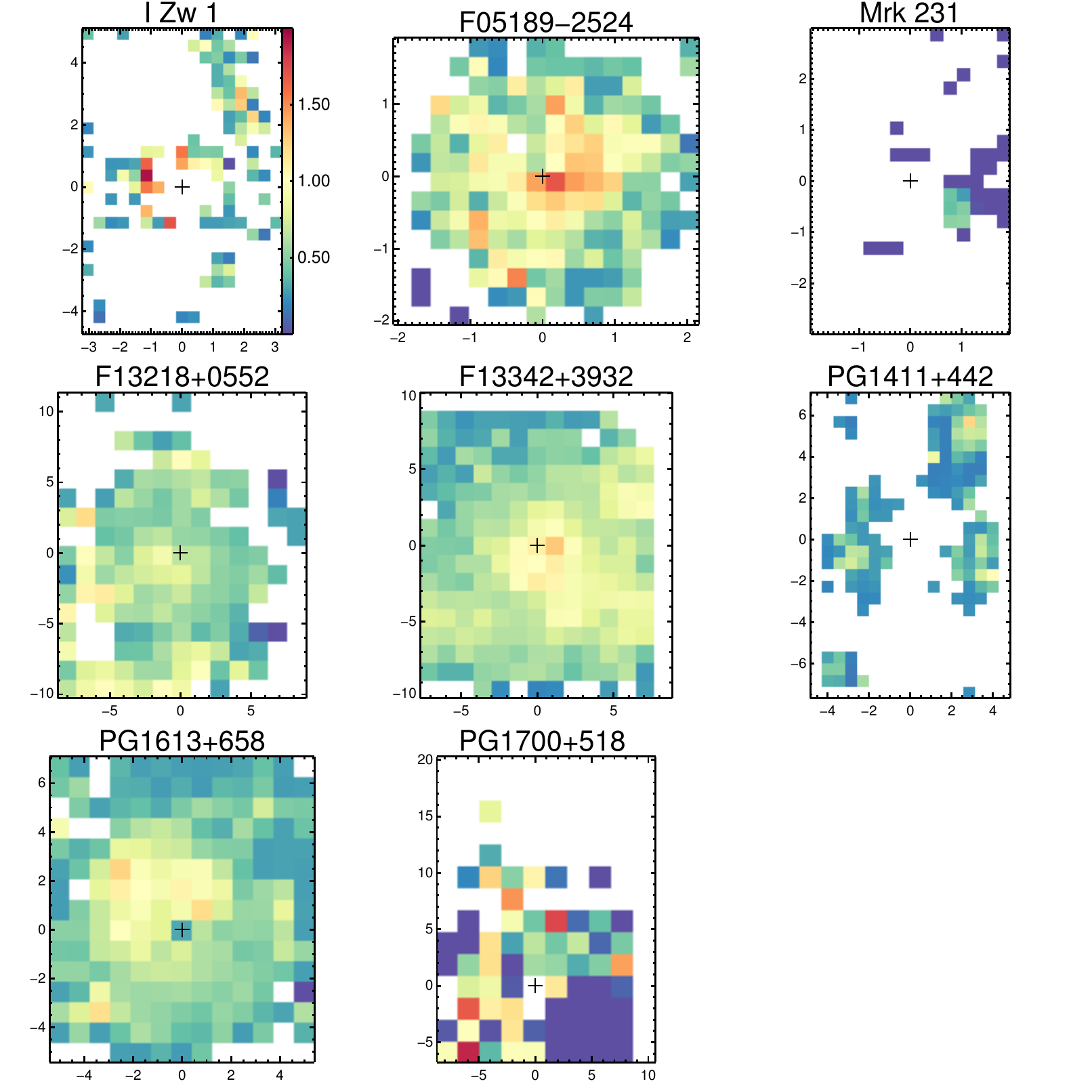}
  \caption{Reddening \ebv\ in quasar hosts with more than one Balmer
    emission line detection, as measured from the Balmer
    decrement. Fluxes are summed over components before \ebv\ is
    calculated. The \citet{cardelli89a} extinction curve and case B at
    10$^{4}$~K are assumed. The \ha/hb\ flux ratio is used in all
    cases but \mrk\ and \pgoneseven, in which the \hb/\hg\ ratio is
    used. No Balmer lines are detected in \ftwoone. \ebv\ is
    calculated only in spaxels where two Balmer lines have 2$\sigma$
    detections.}
  \label{fig:ebv}
\end{figure*}

\begin{figure*}
  \includegraphics[center]{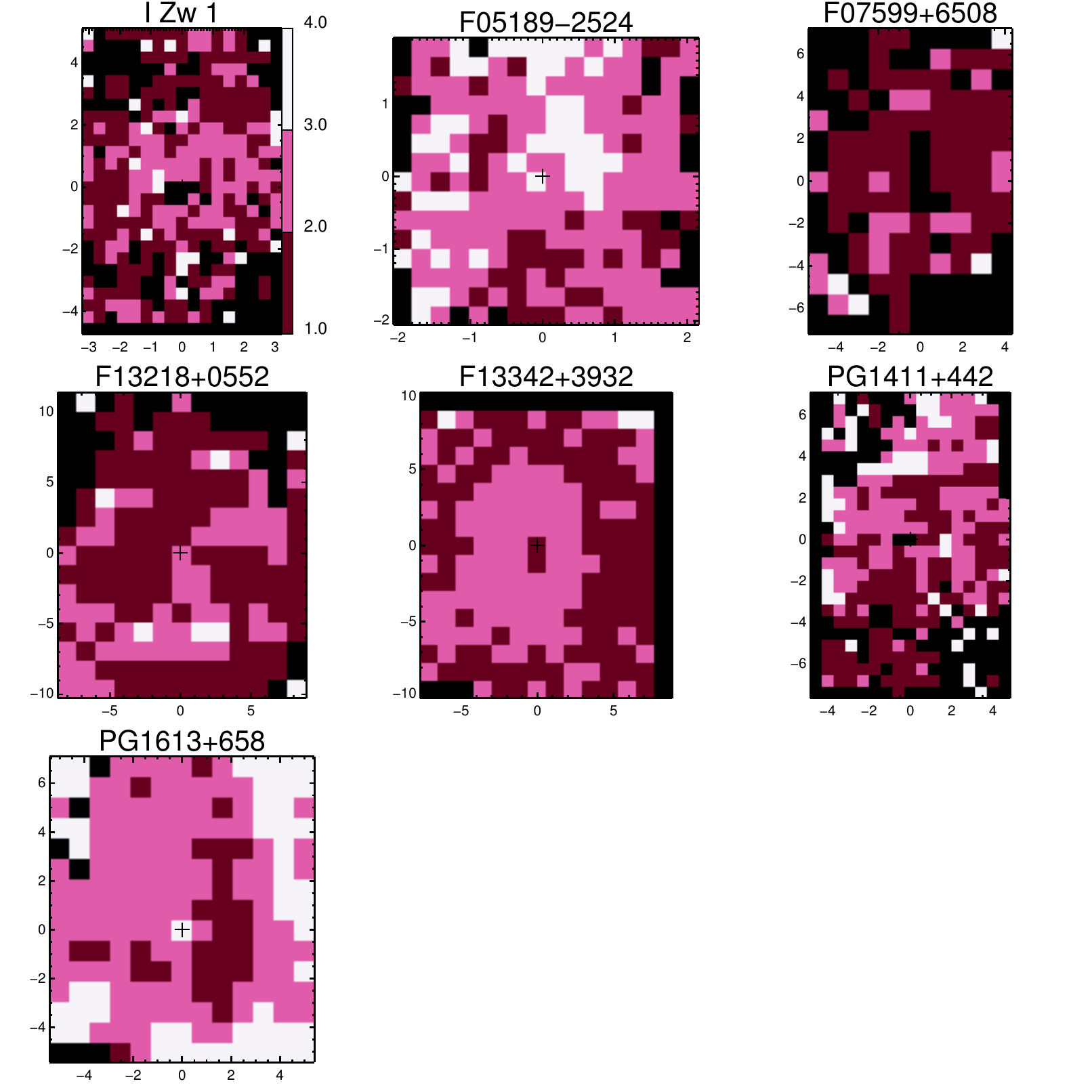}
  \caption{Electron density $n_e$ in quasar hosts with \sutl\
    measurements. Fluxes are summed over components before $n_e$ is
    calculated. Density is calculated from the \sutal/\sutbl\ ratio
    using the parameterization of \citet{sanders16a}. Upper and lower
    sensitivity limits of 10$^1$~cm$^{-2}$ and 10$^4$~cm$^{-2}$ are
    imposed. $n_e$ is calculated only in spaxels where the two lines
    have 2$\sigma$ detections. The colorbar displays ranges of
    log($n_e/\mathrm{cm}^{-2}$).}
  \label{fig:elecden}
\end{figure*}

\begin{figure*}
  \includegraphics[width=0.95\textwidth,center]{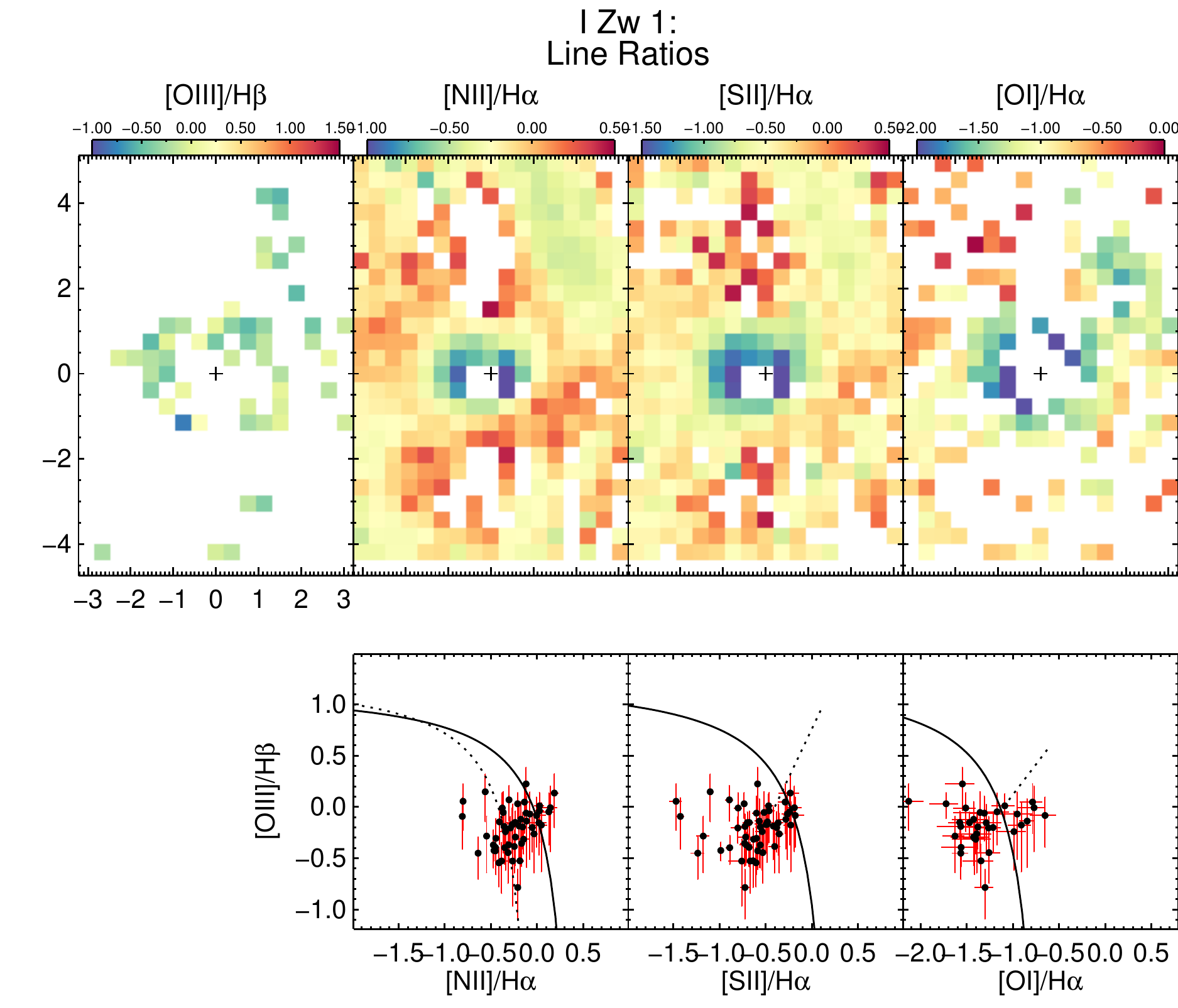}
  \caption{Strong emission line ratios and line ratio diagrams for
    each quasar. Fluxes are summed over components before the ratios
    are calculated. The line ratio scales are logarithmic. In the
    bottom panels, theoretical maximum starburst lines (solid lines;
    \citealt{kewley01a,kewley06a}), an empirical star forming galaxy
    locus (curved dashed line; \citealt{kauffmann03a}); and a line
    separating nuclear line ratios from Seyferts and LINERs (straight
    dashed lines; \citealt{kewley06a}) are shown. The line ratios
    typically reflect a mixing sequence between stellar and AGN
    ionization, with shocks important in some cases
    (\S\,\ref{sec:of_presence}). (The complete figure set (9 images)
    is available.)}
  \label{fig:emlrat_maps}
\end{figure*}

Lines were modeled with 1--2 Gaussian velocity components in each
spaxel. All strong emission lines were fit with the same velocity
components. Only one component was fit in \pgonefour\ and \ftwoone,
and up to two components were fit in the rest of the sample. The
initial guesses for the number of components per spaxel in a given
data cube or region of a cube were set by hand beforehand, and were
tailored for each cube or region to produce better by-eye fits. Model
line profiles were convolved with the spectral resolution before
fitting.  Each component was automatically checked for significance
after the fit and kept only if it exceeded a 3$\sigma$ flux threshold
in at least one strong line. After fitting, each emission line in a
velocity component was checked and set to zero flux if the
significance of that line was below 2$\sigma$.

If two components were fit, the components were sorted into two maps
by wavelength or velocity dispersion. For a given galaxy, the first
component (c1) has either a longer wavelength or a smaller velocity
dispersion than component 2 (c2). The choice of sorting by wavelength
or velocity dispersion was applied to the entire cube. The choice
depended on which method by eye produced the smoothest rotation signal
in c1. If an outflow was present, it was assumed to arise in
c2. However, c2 was not assumed to be an outflow in every case. In
\zw, c2 has very low velocities and does not show velocity gradients
consistent with a coherent outflow. This component is assumed to arise
from other effects (e.g., tidal motions).

This method does not assign c1 or c2 to the galaxy-rotation model or
``other'' model a priori. Rather, c1 and c2 are assigned to
galaxy-rotation and other based on what makes the best galaxy-rotation
model. This predisposes the analysis to a good disk model rather than
having it predisposed toward the other model. When an outflow
component is present in c2, there is thus higher confidence that such
a feature exists.

Ionized outflow is represented by blueshifted emission lines in c2. It
is possible that redshifted emission may in some cases represent the
back side of the outflow; however, in most cases c2 is dominated by
high-velocity blueshifted motions. The only clear-cut cases of
redshifted outflowing emission are \pgonesix\ and \pgoneseven; in
these cases the outflow is taken as the entirety of c2.

Maps of strong emission lines for each galaxy in which these lines are
detected are shown in Figure~\ref{fig:eml_maps}. \ntl\ (or \othl\ in
the cases in which \ntl\ is not present in the wavelength range) was
chosen as a strong representative line. Each figure displays maps of
the total line flux; the flux in each component; the velocity
dispersion of each component, $\sigma^\mathrm{c1}$ and
$\sigma^\mathrm{c2}$; and the central velocity of each component,
\vfifty$^\mathrm{c1}$ and \vfifty$^\mathrm{c2}$.

For galaxies in which more than one emission line component was
fitted, two other panels appear: a map of the flux density at a
representative blueshifted velocity, and a map of the velocity that
encompasses 98\%\ of the cumulative velocity distribution in each
spaxel, \vtsig$^\mathrm{c2}$. This convention matches that used in
some previous IFS studies \citep{rupke11a,rupke13b,rupke13a}. The
cumulative percentages are derived in blueshifted spaxels by
integrating from the red side of the velocity distribution, such that
\vtsig$^\mathrm{c2}$ is representative of a ``maximum'' blueshifted
velocity. In redshifted spaxels, \vtsig$^\mathrm{c2}$ is
correspondingly calculated by integrating from the blue side of the
line.

\capstartfalse
\begin{figure*}
  \centering
  \includegraphics[width=0.8\textwidth]{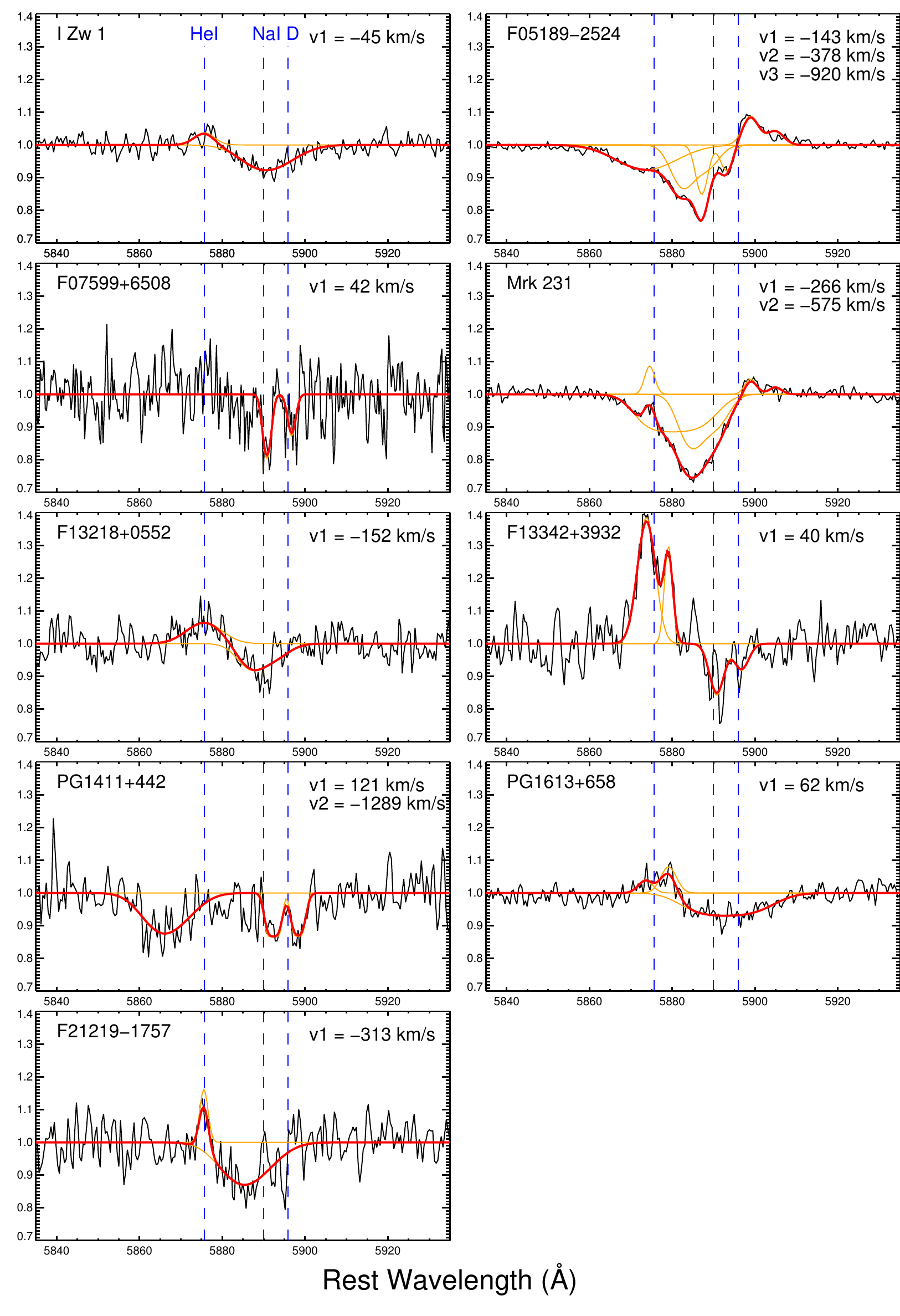}
  \caption{Fits to \nad\ absorption and emission and
    \ion{He}{1}~5876~\AA\ emission in starlight spectra integrated
    over the IFS FOV. The quasar light has been removed from each
    spectrum (\S\,\ref{sec:psfmodel}). Velocities (\vfifty) of the
    absorption components are listed in each panel. Vertical dashed
    lines are line locations based on the galaxy's systemic velocity.}
  \label{fig:nadnucfits}
\end{figure*}
\capstarttrue

Because they are forbidden lines, \nt\ and \oth\ may reflect outflow
or AGN-related physics more than the recombination line maps. The
Balmer lines are also more susceptible to errors in stellar or quasar
subtraction, since Balmer lines feature strongly in these
templates. Nonetheless, maps of the other lines are consistent with
the maps shown.

Besides the fluxes and kinematics, other quantities calculated from
the emission lines on a spaxel-by-spaxel basis are reddening \ebv
(Fig.~\ref{fig:ebv}), electron density $n_e$ (Fig.~\ref{fig:elecden}),
and line ratios (Fig.~\ref{fig:emlrat_maps}). These three quantities
are calculated after fluxes have been summed over both components
(c1$+$c2). The measurements for individual components are noisier than
for the flux or velocity measurements, and summing over both
components produces more spatially-coherent maps. However, the maps of
\ebv, $n_e$, and line ratios for individual components (c1 or c2) are
generally consistent with the c1$+$c2 maps.

\ebv\ is calculated from the Balmer decrement of the c1$+$c2 line flux
(Fig.~\ref{fig:ebv}). Case B at $10^4$~K and the \citet{cardelli89a}
extinction curve are assumed. These reddening maps are not used to
correct the line maps in Figures~\ref{fig:eml_maps} and
\ref{fig:emlrat_maps} because in many spaxels the reddening is too
noisy to be useful (these spaxels are not shown in
Figure~\ref{fig:ebv}). This is largely due to limited sensitivity to
H$\beta$ in some data cubes.

Electron densities are calculated from the c1$+$c2 fluxes and the
\sutal/\sutbl\ ratio (Fig.~\ref{fig:elecden}). The fit of
\sutal/\sutbl\ vs. $n_e$ from \citet{sanders16a} is used, but it is
assumed that the ratio is not sensitive to $n_e$ below 10~cm$^{-2}$
and above 10,000 cm$^{-2}$ as it asymptotes, and calculated densities
outside of this range are set to the lower or upper
limit. Spatially-averaged values of $n_e$ in each quasar range from 50
to 410~cm$^{-2}$, with a median of 150~cm$^{-2}$.

Flux differences among lines are displayed in standard line ratio
plots in Figure~\ref{fig:emlrat_maps}. Resolved line ratio maps are
typically used to distinguish the dominant excitation mechanism
(stellar photoionization, shock ionization, or AGN photoionization;
\citealt{veilleux87a,kewley06a}), to determine the fraction of each
mechanism that is ionizing the gas in each spaxel
\citep[e.g.,]{rich11a,davies16b}, and to measure physical parameters
of the ISM and radiation field by comparison to theoretical models
\citep{dopita95b,groves04b,dopita06a}. These plots are discussed
further in \S\,\ref{sec:of_presence}.

\capstartfalse
\begin{figure*}
  \includegraphics[width=0.95\textwidth,center]{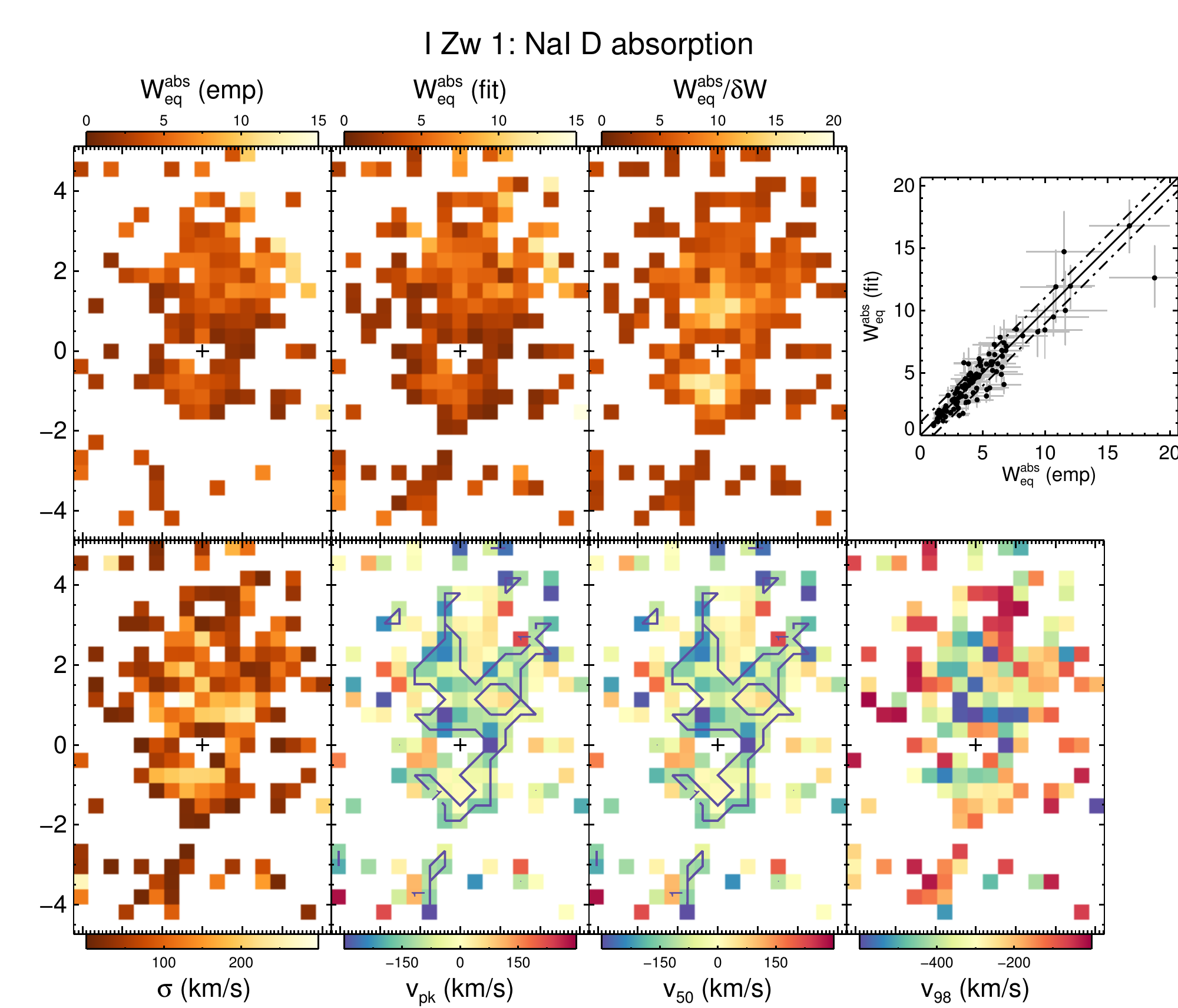}
  \caption{{\scriptsize \nad\ absorption-line properties in each
      quasar host in which resolved absorption is observed. Top panels
      show empirical and fitted observed-frame equivalent widths
      $W_\mathrm{eq}$ (in units of \AA), fitted $W_\mathrm{eq}$
      signal-to-noise, and the correlation between empirical and
      fitted measurements. Empirical equivalent widths are calculated
      from simple integration, while fitted equivalent widths are
      calculated from line fits. The empirical estimates are a sanity
      check on the fitted equivalent widths. Bottom panels show
      velocity maps: velocity dispersion $\sigma$, peak velocity (in
      optical depth space) $v_{pk}$, \vfifty, and \vtsig. In the cases
      where only one component is fit, $v_{pk} = \vfifty$. In the
      three cases where two components are fit (\fohfive, \mrk, and
      \pgonefour), the components are combined in a cumulative
      distribution based on optical depth and the velocity maps
      reflect this (\S\,\ref{sec:absfits}). Zero-velocity contours
      divide blueshifted and redshifted spaxels. (The complete figure
      set (9 images) is available.)}}
  \label{fig:nad_maps}
\end{figure*}
\capstarttrue

\subsubsection{Absorption Line Fitting} \label{sec:absfits}

Half of the quasars in this sample show blueshifted resonant
absorption from the \nad\ doublet in the PSF template spectra (Figure
\ref{fig:psfspec}). \fohseven\ and \mrk\ show broad absorption
\citep{boroson92a,veilleux13b,veilleux16a}. \fohfive\
\citep{rupke15a}, \fonethreethree, and \pgonefour\ (Hamann et
al. 2017, in prep.) show narrow absorption. Many also show possible
resonant broad emission, which requires high column densities, though
it is difficult to distinguish it from nearby \ion{He}{1}~5876~\AA\
emission without careful fitting \citep{thompson91a}.

In this work, the unresolved absorption is fit out as part of the PSF
removal and discarded. The goal is instead to measure the resolved,
extended absorption and emission that appears as part of the host
galaxy spectrum.

The \nad\ doublet and \ion{He}{1} emission line were fit using the
methods of \citet[][for \nad\ absorption]{rupke05a} and \citet[][if
\nad\ emission is present]{rupke15a}. These techniques robustly
separate optical depth and covering factor in absorption, even for a
blended doublet and multiple velocity components, by assuming a
Gaussian in optical depth. They also handle cases of combined resonant
absorption and emission. In all cases but one, the \nad\ and
\ion{He}{1} lines were fit simultaneously, but the \ion{He}{1} was
constrained where needed by fits to other emission lines. In one case
with very high S/N, \fohfive, the \ion{He}{1} line was fit with other
emission lines rather than as part of the \nad\ fitting.

In both the integrated and the spaxel-by-spaxel fits, one \nad\
component was fit in all cases but \pgonefour\ (see below), \mrk, and
\fohfive. The latter two are the nearest galaxies in this sample and
have deep, high-S/N integrations. The \fohfive\ fits are similar to
those previously published \citep{rupke05c,rupke15a}, though here the
binning is coarser and the spatially unresolved continuum has been
removed to match the treatment of the rest of the sample. The
integrated spectrum requires 3 components in absorption, while the
spaxel-by-spaxel fit uses 2 components. The \mrk\ fits use deeper data
than previously published \citep{rupke11a}, which enables 2-component
fitting.

Two cases of the strongest and highest-S/N absorption (\fohfive\ and
\mrk) also show redshifted \nad\ emission. The \fohfive\ emission is
analyzed in \citet{rupke15a}; the deep \mrk\ data reveals the new
\nad\ emission, but it is not discussed further here.

For all quasar hosts but one, the host continuum for these fits is
assumed to be a combination of the stellar and polynomial
components. In one case with strong scattered light, \zw, the
polynomial component was successfully identified as scattered light
and the continuum is assumed to be the purely stellar component. (The
effect of removing the polynomial component is to increase the
resulting values of the covering factor $C_\mathrm{f}$, and
consequently to increase the global flow rates; \S\,\ref{sec:ofprop}.)
 
Fits to \nad\ in the integrated quasar host spectra
(Figure~\ref{fig:nadnucfits}) reveal extended interstellar absorption
in every case except \pgoneseven\ (for which the \nad\ feature is not
in the observed wavelength range).

In the spaxel-by-spaxel fits, spaxels with fitted equivalent width
S/N$>$5 were considered significant detections. Of the nine quasar
hosts with extended interstellar absorption, eight show this
absorption on a spaxel-by-spaxel basis
(Figure~\ref{fig:nad_maps}). The maximum extent of this absorption
ranges from 1--10 kpc, and the structure of the absorption is
varied. Wind structure is discussed further in the next section. Most
equivalent widths are in the range 1--10~\AA.

Figure~\ref{fig:nad_maps} also shows the velocity maps of the neutral
outflows. In cases where two outflowing absorption line components are
fit, the lines in each spaxel are first combined into a cumulative
velocity distribution weighted by optical depth; \vfifty\ and \vtsig\
are calculated with respect to this total distribution. The maps also
show the velocity of the peak optical depth, which differs from
\vfifty\ for two or more components. Finally,
Figure~\ref{fig:nad_maps} compares the fitted equivalent width with an
empirical estimate from simple integration \citep{rupke15a}; this is a
sanity check on the quality of the fits. Generally, the fits tend to
recover more signal and are a better estimate of the line properties
\citep{rupke15a}.

\pgonefour\ shows a weak but significant high-velocity extended
component ($-$1300~\kms). Tests of integrated spectra of varying
apertures show that this absorption has a maximum extent of 1--2 kpc
from the quasar (in projection). The low-velocity absorption extends
to larger radii. Thus, two components are fit within five spaxels of
the nucleus, and one component outside of this.

For \pgonesix, the spaxel-by-spaxel accounting for the line does not
reveal that it extends in every direction around the nucleus. The
extended absorption that is detected on a spaxel-by-spaxel basis is a
strong velocity component that arises in front of a small satellite
galaxy that is $\sim$5~kpc from the quasar in projection (Figure
\ref{fig:cont}), and is redshifted at a velocity similar to the
ionized gas disk of the quasar host at that location.

The detected absorption is generally blueshifted; spaxels with
$\vfifty<0$~\kms\ are assumed to be outflowing. There is some
absorption near systemic or slightly redshifted (up to
$\sim$200~\kms). In three systems (\fohseven, \pgonefour, and
\pgonesix) these make up a substantial fraction of the spaxels with
detected absorption. In \pgonesix, this is due to the nearby
companion. The same may be true in \pgonefour. In \fohseven, the
origin of the redshifted absorption is unclear.

\section{RESULTS} \label{sec:results}

A key result from this study is the set of maps of the properties of
emission and absorption lines, as well as the continuum maps, for each
galaxy. These maps can be used to constrain properties of the AGN and
its host galaxy. However, the scientific focus of this paper is the
large-scale outflows in these systems. This section outlines how these
maps constrain the properties of ionized and neutral outflows. In it,
the outflow properties in the current sample are also compared to host
galaxy properties, black hole properties, and to the outflows in other
galaxies.
 
Though a detailed physical interpretation of the disk kinematics of
these quasars is not essential to constraining the outflow properties
of these data, it is important to assess the presence of disks because
of their dominant imprint on the spectra and because their geometry
can relate to the outflow orientation. Every quasar in this sample
does show evidence for a rotating ionized gas or stellar disk. 9/10
quasars show an ionized gas disk, and in the 10th system (\ftwoone)
the surface brightness of the ionized gas is too low (and the S/N too
low) to determine its properties. 7/10 quasars show a stellar disk
(Figure~\ref{fig:stelvel}); in two of the other cases (\fohseven\ and
\pgoneseven) the S/N is again too low to robustly constrain the
stellar velocities, while in the 10th case (\mrk) the stellar disk is
either slowly rotating or nearly face-on, similar to the gas disk. In
each case these disks extend to the edges of the FOV. Finally, 3/10
quasars have low-luminosity companions within or at the edge of the
FOV (\pgonefour, \pgonesix, and \pgoneseven). In at least one case,
\pgoneseven, this companion may have a significant effect on emission
line component c1.

\begin{figure*}
  \includegraphics[center,width=0.95\textwidth]{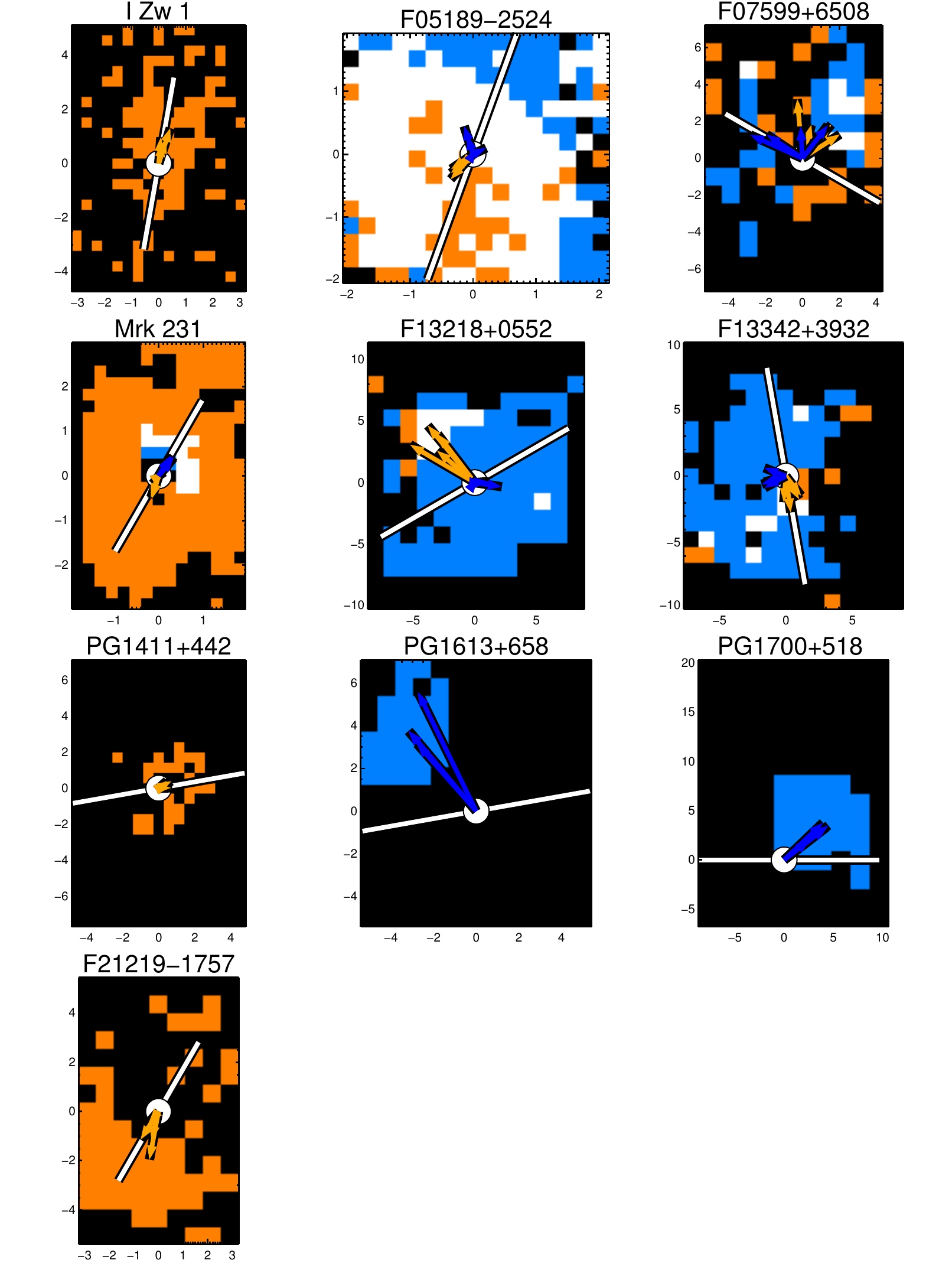}
  \caption{\scriptsize The outflow geometry of each detected outflow
    (\S\,\ref{sec:of_presence}). The neutral outflow is shown in
    orange, the blueshifted part of the ionized outflow in blue, and
    the overlap in white. The galaxy nucleus and minor axis are shown
    as a black-outlined white circle and line, respectively. The
    average outflow positions are means over all outflow
    spaxels. Arrows are drawn from the nucleus to this average
    position. Four averages, each with a different positional weight,
    are calculated and plotted for each gas phase to illustrate the
    uncertainty in the outflow position: unweighted, weighted by
    \vfifty$^2$, weighted by \vtsig$^2$, and weighted by
    $W_\mathrm{eq}$ or flux. Orange (blue) arrows denote the mean
    neutral (ionized) outflow position. Most outflows are oriented
    along or near the minor axis in projection, with some exceptions
    (the neutral outflow in \fohseven, the outflows in \fonethreetwo,
    the ionization cone in \pgonesix, and the jet-driven outflow in
    \pgoneseven\ are not along the projected minor axis). It is
    possible that the outflows in \fohseven\ and \fonethreetwo\ are
    along the actual minor axis if the minor axis is near the line of
    sight; some cases like this are expected. The outflows in
    \fonethreethree\ might appear to be exceptions to the minor-axis
    orientation, but are not (see \S\,\ref{sec:of_presence}).}
  \label{fig:ofgeometry}
\end{figure*}

\subsection{The Presence, Orientation, and Ionization of Large-Scale
  Outflows} \label{sec:of_presence}

Of the radio-quiet Type 1 quasars in this sample, all contain
large-scale (maximum sizes 3--12~kpc) outflows of ionized and/or
neutral gas, as revealed by the presence of broad, blueshifted
absorption lines and broad, blueshifted emission lines. If we break
the detections down by phase, 70\%\ have ionized outflows, 80\%\ have
neutral outflows, and 50\%\ have both. This detection rate contrasts
with the $\sim$10\%\ detection rate of ionized outflows in a previous
IFS study of Type 1 quasars \citep[][hereafter H13]{husemann13a}, but
is consistent with the very high detection rate in nearby Type 2s
\citep{harrison14a,mcelroy15a,bae17a}.

For the two objects in common between H13 and this work (\zw\ and
\pgoneseven), there is agreement on the ionized outflow
properties. The significant disagreement on the detection rate could
thus result from differences in sample selection, as H13 select
lower-luminosity AGN. Part, though not all, of the discrepancy could
also result from our choice to probe both neutral and ionized gas
outflows, rather than just ionized outflows, which could increase the
detection rate.

A third possibility is that the different detection rate between our
study and H13 results from data quality differences. The data in H13
were taken with the Potsdam Multi-Aperture Spectrophotometer (PMAS) on
the 3.5m Calar Alto telescope. The quoted emission-line surface
brightness limit of H13 (at lower spatial and spectral resolution) is
a factor of 3--4 worse than the best values achieved in the current
study ($\sim$5$\times10^{-17}$ erg s$^{-1}$ cm$^{-2}$ arcsec$^{-2}$;
Figure~\ref{fig:eml_maps}). The 0\farcs5 sampling of PMAS is larger
than the 0\farcs2 GMOS hexagonal spaxels, and the typical seeing of
H13 was about twice the current study. The spectral resolution in H13
varied from 2 to 4$\times$ coarser than the current study. These
factors make separating a second (typically broad, faint) outflowing
emission line component from a disk component more challenging. The
exception is when the outflow dominates the line emission. In
\pgoneseven, the outflow completely dominates the line emission in
\oth, and both studies classify it as a large-scale outflow.

The approximate projected outflow boundaries and orientations are
shown in Figure~\ref{fig:ofgeometry}. The average outflow position in
each phase is calculated in projected Cartesian coordinates; the
position weighted by \vfifty$^2$, \vtsig$^2$, and $W_\mathrm{eq}$ or
flux is also calculated. These different weightings give a sense of
the uncertainty in computing the true orientation of the outflow.

Loosely-collimated outflows along their host galaxy's minor axes are
common on kpc scales in many starburst and active galaxies observed
with IFS \citep[e.g.,][]{rupke13a}. The observed outflows in these
quasars are also typically oriented near their hosts' minor
axes. Clear exceptions to this are the outflows in \fonethreetwo, the
neutral outflow in \fohseven, and the ionized outflows in \pgonesix\
and \pgoneseven. One expects some exceptions due to projection
effects, which will enhance misalignments for systems that are nearly
face-on. \pgonesix\ and \pgoneseven\ have outflows misaligned with the
minor axis for reasons described below. The outflows in
\fonethreethree\ would at first glance appear to be exceptions as
well. However, the highest velocity neutral gas lies right along the
minor axis at 1--2~kpc scales, and the largest ionized fluxes are
found along the minor axis on the same scales in the opposite
direction. This behavior is not captured in the outflow positional
averages, as the outflow is spatially dominated by an apparently less
collimated flow at larger scales.

The opening angle of an outflow (between 0$^\circ$ and 180$^\circ$)
quantifies its collimation; loosely-collimated outflows have large
opening angles. In the bipolar cone model, the opening angle is simply
the projected two-dimensional angle between the sides of each
cone. The broad footprints of the outflows in this sample
(Figure~\ref{fig:ofgeometry}) suggest loosely-collimated flows with
large opening angles (perhaps approaching 180$^\circ$).

In \pgoneseven, the outflow lies along a known radio jet. This quasar
is the most distant and radio-luminous in the sample with $z = 0.2905$
and log$(P_\mathrm{1.4~GHz}) = 24.8$~W~Hz$^{-1}$. This places it at
the knee of the radio luminosity function of nearby AGN
\citep{condon02a}, but two orders of magnitude below the knee of the
$z\sim0.3$ quasar 1.4~GHz luminosity function \citep{condon13a}. It is
one of two quasars in the sample with a known jet
\citep{kukula98a,yang12a}. (The other is \mrk; see
\citealt{morganti16a} and references therein.)

In \pgonesix, there is an extended NLR or ionization cone, presumably
aligned with and collimated by the AGN torus, that is roughly along
the galaxy major axis and traced by velocity component 2. It has very
high \oth/\hb\ values (Figure~\ref{fig:emlrat_maps}), indicative of
AGN photoionization. The flux and bulk velocity of c2 peak about 2 kpc
from the nucleus. Though the projected bulk motions (as traced by
\vfifty$^{c2}$) of this component are similar to the disk motions in
component 1, the large values of $\sigma^{c2}$ and \vtsig$^{c2}$ are
indicative of an outflow, and it is classified as such. This outflow
may arise from the torus being misaligned with the host minor
axis. The bulk of the AGN radiation is pointed into the disk,
accelerating disk material (as in nearby Seyferts;
\citealt{fischer13a}). The lack of high bulk motions could mean the
outflow is in the plane of the sky, and the decrease in \vfifty$^{c2}$
at large radii may mimic NLR deceleration in lower-luminosity systems
\citep{fischer13a}.

There is also an increased velocity dispersion and shock-like line
ratios (high \nt/\ha, \sut/\ha, and \oo/\ha) along the galaxy minor
axis in \pgonesix, as seen in component 1. These latter effects have
been used as tracers of outflows in other resolved studies
\citep{lehnert96a,ho16a}. However, the dense velocity contours along
the minor axis of this system coupled with the $>$2~kpc PSF means that
the disk needs to be carefully modeled to account for beam smearing
\citep{davies11a}. In the absence of this detailed modeling, we do not
include these spaxels as part of the outflow.

Strikingly, in the five systems where both neutral and ionized
outflows are present, the neutral and ionized outflows are often
oriented opposite each other. This ionized/neutral asymmetry is not
obvious in other Type 2 quasars \citep{rupke13a}. Spatially resolved
observations show that \nad\ absorption correlates with optical dust
obscuration \citep{rupke13a,rupke15a}. An anti-correlation in the
outflow between line-emitting gas and dusty gas could reflect
asymmetric obscuration in the outflow. On one side, the obscuration
along the line of sight is small and the ionized outflow is
prominent. On the other, the obscuration is higher and the neutral
outflow dominates. One possible cause is the underlying gas disk,
though the obscuration correlated with the wind is thought to be in
the outflow rather than the disk \citep{rupke15a}.

The line ratio maps in Fig.~\ref{fig:emlrat_maps} (calculated on a
spaxel-by-spaxel basis using fluxes summed over velocity components)
are not used to determine whether or not an outflow is present, and a
quantitative comparison of how the outflows relate to line excitation
is reserved for future work. However, a qualitative understanding of
the excitation of the ionized gas provides an overview of possible
radiation physics in these outflows. The maps in
Figure~\ref{fig:eml_maps} and in the top panels of Figure
\ref{fig:emlrat_maps} often visually reveal correlations between
outflows and higher line ratios in many systems. However, this
correlation is not one-to-one; for instance, only the redshifted side
of the outflow in \pgoneseven\ has high excitation. And in some cases
there is no obvious correlation (\mrk).

Line ratio diagrams, seen in the bottom panels of
Figure~\ref{fig:emlrat_maps}, typically show patterns suggestive of
stellar and AGN photoionization in a sequence mixing pure stellar
photoionization and pure AGN photoionization. This mixing sequence,
which moves from the lower left to the upper middle of these diagrams,
is due to the fractional contribution of stellar and AGN
photoionization varying from one part of the galaxy to another
\citep[e.g.,][]{davies16b}. Stellar photoionization dominates in the
lower left, and AGN photoionization in the upper middle, of these
diagrams. There is also evidence of a stellar photoionization sequence
in metallicity in at least two hosts (\zw\ and \pgonefour); this
sequence is visible as a curving locus of points from the middle left
to the middle bottom \citep[e.g.,][]{kewley01a}. Shock ionization, a
common feature of galactic winds \citep{veilleux02b,sharp10a,rich11a},
is intermixed with AGN and stellar ionization in other systems
(\fohfive, \fohseven, \fonethreetwo, and \pgonesix), and is dominant
in at least one system (\fohfive). Shock ionization dominates the
right-hand side of these diagnostics \citep{dopita95b,rich11a}, and
mixing with other ionization mechanisms is revealed as loci of points
connecting regions of stellar, AGN, and shock ionization. Finally,
evidence of AGN photoionization variations due to changing ionization
parameter \citep{scharwachter11a,davies16a} are evident in at least
one quasar (\fonethreethree). These changes are visible as loci of
points moving from the upper middle toward the lower right.

\capstartfalse
\begin{deluxetable*}{cccrrrrr}
  \tablecolumns{8}
  \tablecaption{Outflow Velocity Statistics\label{tab:vels}}
  \tablewidth{\textwidth}

  \tablehead{ \colhead{} & \colhead{} &
    \multicolumn{2}{c}{FWHM$^\mathrm{c2}$} &
    \multicolumn{2}{c}{$\vfifty^\mathrm{c2}$} &
    \multicolumn{2}{c}{$\vtsig^\mathrm{c2}$} \\
    \cmidrule(lr){3-4} \cmidrule(lr){5-6} \cmidrule(lr){7-8}
    \colhead{Galaxy} & \colhead{Phase} & \colhead{avg} & \colhead{max} &
    \colhead{avg} & \colhead{max} &\colhead{avg} & \colhead{max} \\
    \colhead{(1)} & \colhead{(2)} & \colhead{(3)} & \colhead{(4)} &
    \colhead{(5)} & \colhead{(6)} & \colhead{(7)} & \colhead{(8)} }

  \startdata
         I Zw 1  &        neutral  &     202  &     654  &    -120  &    -326  &    -297  &    -635 \\
    F05189-2524  &        neutral  &     624  &    1728  &    -560  &   -1031  &   -1165  &   -2569 \\
        \nodata  &        ionized  &     592  &    1591  &    -423  &    -989  &    -926  &   -1598 \\
    F07599+6508  &        neutral  &     275  &     682  &     -96  &    -315  &    -337  &    -793 \\
        \nodata  &        ionized  &     310  &     641  &    -120  &    -839  &    -384  &   -1327 \\
        Mrk 231  &        neutral  &     395  &    1122  &    -416  &   -1112  &    -801  &   -1784 \\
        \nodata  &        ionized  &     768  &    1397  &    -672  &   -1036  &   -1324  &   -2003 \\
    F13218+0552  &        neutral  &     189  &     386  &     -47  &    -185  &    -213  &    -524 \\
        \nodata  &        ionized  &     641  &    1725  &    -169  &    -578  &    -714  &   -1897 \\
    F13342+3932  &        neutral  &     322  &    1027  &    -104  &    -432  &    -387  &   -1262 \\
        \nodata  &        ionized  &     386  &     882  &    -144  &    -418  &    -473  &   -1167 \\
     PG1411+442  &        neutral  &     909  &    2148  &    -486  &   -1274  &   -1044  &   -1780 \\
     PG1613+658  &        ionized  &     409  &     609  &    -109  &    -578  &    -457  &    -704 \\
    PG1700+518   &        ionized  &     895  &    2179  &    -570  &    -854  &   -1331  &   -2627 \\
    F21219-1757  &        neutral  &     239  &     646  &    -230  &    -487  &    -441  &    -700
  \enddata

  \tablecomments{\scriptsize Column 2: Gas phase. Column 3$-$4: Mean
    and maximum (over all spaxels) of FWHM of velocity component 2
    (c2), in /kms. Column 5$-$6: Mean and maximum values (over all
    spaxels) of c2 central velocity, in /kms. For the neutral phase,
    all components are combined in each spaxel into an optical depth
    weighted cumulative velocity distribution. The averages are taken
    over spaxels with $\vfifty<0$~\kms. For the ionized phase, only
    the c2 is used; with the exceptions of \pgonesix\ and \pgoneseven,
    the averages are taken over spaxels with
    $\vfifty^\mathrm{c2}<0$~\kms. For \pgonesix\ and \pgoneseven,
    spaxels with $\vfifty^\mathrm{c2}>0$~\kms\ are included, but the
    velocities are multiplied by $-$1 so that the red and blue outflow
    components can be combined into a single average. Column 7$-$8:
    Mean and maximum of $\vtsig \equiv \vfifty - 2\sigma$. $\sigma$ is
    the average velocity above and below \vfifty\ that encompasses
    34\%\ of the cumulative velocity distribution: the area between
    $\vfifty \pm \sigma$ contains 68\% of the distribution. As for
    \vfifty, the redshifted ionized gas spaxels in \pgoneseven\ have
    their distribution flipped in sign and combined with the
    blueshifted components.}

\end{deluxetable*}
\capstarttrue

\setlength{\tabcolsep}{3pt}
\capstartfalse
\begin{deluxetable*}{ccccccccccc}
  \tabletypesize{\scriptsize}
  \tablecaption{Mass, Momentum, and Energy\label{tab:mpe}}
  \tablewidth{\textwidth}

  \tablehead{\colhead{Galaxy} & \colhead{phase} &
    \colhead{$R_\mathrm{obs}$} & \colhead{$r_\mathrm{wind}$} &
    \colhead{log[$f_{H\alpha}$/} & \colhead{log($M$/} &
    \colhead{log[$(dM/dt)$/} & \colhead{log[$p$/} &
    \colhead{log[$(c~dp/dt)$/} & \colhead{log($E$/} & \colhead{log[$(dE/dt)$/}\\
    \colhead{~} & \colhead{~} & \colhead{(kpc)} & \colhead{(kpc)} &
    \colhead{(erg s$^{-1}$ cm$^{-2}$)]} & \colhead{\msun)} &
    \colhead{(\smpy)]} & \colhead{(dyne~s)]} &
    \colhead{\lsun]} & \colhead{erg)} & \colhead{(erg s$^{-1}$)]} \\
    \colhead{(1)} & \colhead{(2)} & \colhead{(3)} & \colhead{(4)} &
    \colhead{(5)} & \colhead{(6)} & \colhead{(7)} & \colhead{(8)} &
    \colhead{(9)} & \colhead{(10)} & \colhead{(11)}}

  \startdata
         I Zw 1  &  neutral  &    5.1  &    5.5  &\nodata  &   8.77$_{-0.03}^{+0.06}$  &   1.38$_{-0.06}^{+0.10}$  &  49.41$_{-0.06}^{+0.10}$  &  11.61$_{-0.09}^{+0.15}$  &  56.79$_{-0.09}^{+0.12}$  &  42.14$_{-0.12}^{+0.18}$ \\
    F05189-2524  &  neutral  &    2.7  &    3.0  &\nodata  &   8.58$_{-0.02}^{+0.03}$  &   1.98$_{-0.03}^{+0.05}$  &  49.75$_{-0.03}^{+0.05}$  &  12.66$_{-0.03}^{+0.06}$  &  57.52$_{-0.03}^{+0.06}$  &  43.58$_{-0.04}^{+0.07}$ \\
        \nodata  &  ionized  &    2.8  &    3.0  & -13.21  &   7.36$_{-0.14}^{+0.06}$  &   0.40$_{-0.14}^{+0.07}$  &  48.16$_{-0.14}^{+0.07}$  &  10.78$_{-0.14}^{+0.07}$  &  55.55$_{-0.14}^{+0.07}$  &  41.32$_{-0.13}^{+0.07}$ \\
        \nodata  &molecular  &\nodata  &   $<1$  &\nodata  &   8.28$_{-0.38}^{+0.38}$  &   2.43$_{-0.29}^{+0.03}$  &                  \nodata  &  12.72$_{-0.25}^{+0.08}$  &                  \nodata  &  43.20$_{-0.25}^{+0.10}$ \\
        \nodata  &    total  &\nodata  &\nodata  &\nodata  &   8.77$_{-0.09}^{+0.16}$  &   2.57$_{-0.19}^{+0.03}$  &  49.76$_{-0.03}^{+0.05}$  &  13.00$_{-0.11}^{+0.05}$  &  57.53$_{-0.03}^{+0.06}$  &  43.73$_{-0.07}^{+0.06}$ \\
    F07599+6508  &  neutral  &    7.5  &    8.1  &\nodata  &   9.27$_{-0.09}^{+0.09}$  &   1.27$_{-0.12}^{+0.39}$  &  49.46$_{-0.12}^{+0.39}$  &  11.20$_{-0.12}^{+0.19}$  &  57.24$_{-0.24}^{+0.19}$  &  41.81$_{-0.17}^{+0.44}$ \\
        \nodata  &  ionized  &    7.0  &    8.1  & -15.23  &   7.50$_{-0.15}^{+0.06}$  &  -0.64$_{-0.19}^{+0.06}$  &  47.56$_{-0.19}^{+0.06}$  &   9.45$_{-0.33}^{+0.10}$  &  54.65$_{-0.33}^{+0.10}$  &  39.95$_{-0.32}^{+0.12}$ \\
        \nodata  &    total  &\nodata  &\nodata  &\nodata  &   9.27$_{-0.09}^{+0.09}$  &   1.27$_{-0.12}^{+0.38}$  &  49.47$_{-0.12}^{+0.38}$  &  11.21$_{-0.12}^{+0.19}$  &  57.24$_{-0.24}^{+0.19}$  &  41.82$_{-0.17}^{+0.44}$ \\
        Mrk 231  &  neutral  &    3.4  &    3.6  &\nodata  &   8.84$_{-0.02}^{+0.02}$  &   2.00$_{-0.02}^{+0.03}$  &  49.85$_{-0.02}^{+0.03}$  &  12.54$_{-0.03}^{+0.05}$  &  57.46$_{-0.03}^{+0.05}$  &  43.35$_{-0.04}^{+0.06}$ \\
        \nodata  &  ionized  &    2.9  &    1.3  & -14.71  &   5.88$_{-0.04}^{+0.05}$  &  -0.30$_{-0.05}^{+0.06}$  &  47.10$_{-0.05}^{+0.06}$  &  10.37$_{-0.05}^{+0.06}$  &  54.78$_{-0.05}^{+0.06}$  &  41.18$_{-0.05}^{+0.06}$ \\
        \nodata  &molecular  &\nodata  &   $<1$  &\nodata  &   9.04$_{-0.03}^{+0.07}$  &   3.04$_{-0.06}^{+0.02}$  &                  \nodata  &  13.34$_{-0.18}^{+0.06}$  &                  \nodata  &  43.92$_{-0.16}^{+0.11}$ \\
        \nodata  &    total  &\nodata  &\nodata  &\nodata  &   9.25$_{-0.02}^{+0.05}$  &   3.08$_{-0.06}^{+0.01}$  &  49.85$_{-0.02}^{+0.03}$  &  13.40$_{-0.15}^{+0.05}$  &  57.47$_{-0.03}^{+0.05}$  &  44.02$_{-0.12}^{+0.09}$ \\
    F13218+0552  &  neutral  &   11.3  &   12.0  &\nodata  &   8.68$_{-0.39}^{+0.26}$  &   0.93$_{-1.09}^{+0.35}$  &  49.30$_{-1.09}^{+0.35}$  &  11.08$_{-1.52}^{+0.39}$  &  56.60$_{-1.01}^{+0.38}$  &  41.48$_{-1.39}^{+0.41}$ \\
        \nodata  &  ionized  &   11.5  &   12.0  & -13.76  &   9.21$_{-0.11}^{+0.04}$  &   1.03$_{-0.09}^{+0.04}$  &  49.40$_{-0.09}^{+0.04}$  &  10.80$_{-0.07}^{+0.04}$  &  56.18$_{-0.07}^{+0.04}$  &  40.87$_{-0.06}^{+0.05}$ \\
        \nodata  &    total  &\nodata  &\nodata  &\nodata  &   9.32$_{-0.11}^{+0.08}$  &   1.29$_{-0.24}^{+0.19}$  &  49.65$_{-0.24}^{+0.19}$  &  11.27$_{-0.44}^{+0.29}$  &  56.74$_{-0.46}^{+0.30}$  &  41.58$_{-0.64}^{+0.36}$ \\
    F13342+3932  &  neutral  &    9.2  &   10.7  &\nodata  &   8.50$_{-0.07}^{+0.17}$  &   0.71$_{-0.15}^{+0.22}$  &  49.03$_{-0.15}^{+0.22}$  &  10.91$_{-0.22}^{+0.24}$  &  56.80$_{-0.32}^{+0.24}$  &  41.83$_{-0.38}^{+0.27}$ \\
        \nodata  &  ionized  &    7.9  &   10.7  & -13.15  &   9.17$_{-0.04}^{+0.03}$  &   1.37$_{-0.04}^{+0.03}$  &  49.69$_{-0.04}^{+0.03}$  &  11.38$_{-0.04}^{+0.03}$  &  56.71$_{-0.04}^{+0.03}$  &  41.54$_{-0.05}^{+0.03}$ \\
        \nodata  &    total  &\nodata  &\nodata  &\nodata  &   9.25$_{-0.04}^{+0.04}$  &   1.46$_{-0.04}^{+0.05}$  &  49.78$_{-0.04}^{+0.05}$  &  11.51$_{-0.06}^{+0.08}$  &  57.06$_{-0.15}^{+0.15}$  &  42.01$_{-0.21}^{+0.19}$ \\
     PG1411+442  &  neutral  &    2.8  &    3.2  &\nodata  &   8.42$_{-0.12}^{+0.11}$  &   1.68$_{-0.11}^{+0.15}$  &  49.47$_{-0.11}^{+0.15}$  &  12.57$_{-0.13}^{+0.20}$  &  57.41$_{-0.13}^{+0.19}$  &  43.65$_{-0.17}^{+0.24}$ \\
     PG1613+658  &  ionized  &    7.5  &    8.0  & -14.11  &   7.73$_{-0.30}^{+0.15}$  &   0.04$_{-0.36}^{+0.16}$  &  48.23$_{-0.36}^{+0.16}$  &   9.98$_{-0.38}^{+0.17}$  &  55.18$_{-0.38}^{+0.17}$  &  40.11$_{-0.32}^{+0.15}$ \\
    PG1700+518   &  ionized  &    9.7  &   10.0  & -12.89  &   9.58$_{-0.09}^{+0.14}$  &   2.36$_{-0.10}^{+0.14}$  &  50.65$_{-0.10}^{+0.14}$  &  12.87$_{-0.10}^{+0.14}$  &  58.16$_{-0.10}^{+0.14}$  &  43.53$_{-0.11}^{+0.15}$ \\
    F21219-1757  &  neutral  &    5.5  &    5.9  &\nodata  &   8.79$_{-0.06}^{+0.09}$  &   1.56$_{-0.12}^{+0.12}$  &  49.62$_{-0.12}^{+0.12}$  &  11.87$_{-0.19}^{+0.16}$  &  57.04$_{-0.19}^{+0.16}$  &  42.41$_{-0.25}^{+0.20}$
  \enddata

  \tablecomments{Column 2: Gas phase. Column 3--4: Maximum observed
    wind radius (in projection) and assumed wind radius, in kpc.
    Column 5: Logarithm of the H$\alpha$ flux in the wind, corrected
    for reddening using the Balmer decrement. For \fohseven, the
    extincted flux is listed because the spectrum contains only one
    Balmer line. Column 6--11: Logarithms of the wind mass, momentum,
    energy, and their outflow rates, computed using a time-averaged
    thin shell model that depends inversely on the assumed shell
    radius (\S\,\ref{sec:ofprop};
    \citealt{rupke05b,shih10a,rupke13a}). The ionized gas values
    depend on electron density as $n_e^{-1}$. For quasars with \sutl\
    measurements, $n_e$ is calculated on a spaxel-by-spaxel basis; for
    others $n_e = 100$~cm$^{-2}$ is assumed
    (\S\,\ref{sec:emlfits}). The neutral gas values depends on
    ionization fraction, Na abundance, metallicity, and Na dust
    depletion (\S\,\ref{sec:ofprop} and \citealt{rupke05a}). Molecular
    gas values are from \citet{gonzalezalfonso17a}.}

\end{deluxetable*}
\capstarttrue
\setlength{\tabcolsep}{6pt}

\capstartfalse
\begin{deluxetable*}{ccrrrrrr}
  \tablecaption{Percent of Mass, Momentum, and Energy in Each
    Phase\label{tab:mperats_in}} \tablewidth{\textwidth}

  \tablehead{\colhead{Galaxy} & \colhead{Phase} & \colhead{$M$} &
    \colhead{$dM/dt$} & \colhead{$p$} &
    \colhead{$dp/dt$} & \colhead{$E$} & \colhead{$dE/dt$}\\
    \colhead{(1)} & \colhead{(2)} & \colhead{(3)} & \colhead{(4)} &
    \colhead{(5)} & \colhead{(6)} & \colhead{(7)} & \colhead{(8)} }

  \startdata
    F05189-2524  &   neutral  &   64.0  &   26.1  &\nodata  &   46.6  &\nodata  &   70.2 \\
        \nodata  &   ionized  &    3.9  &    0.7  &\nodata  &    0.6  &\nodata  &    0.4 \\
        \nodata  & molecular  &   32.1  &   73.2  &\nodata  &   52.8  &\nodata  &   29.5 \\
    F07599+6508  &   neutral  &   $<$98.3  &   $<$98.8  &   $<$98.8  &   $<$98.3  &   $<$99.7  &   $<$98.6 \\
        \nodata  &   ionized  &    $>$1.7  &    $>$1.2  &    $>$1.2  &    $>$1.7  &    $>$0.3  &    $>$1.4 \\
        Mrk 231  &   neutral  &   38.6  &    8.4  &\nodata  &   13.8  &\nodata  &   21.3 \\
        \nodata  &   ionized  &    0.0  &    0.0  &\nodata  &    0.1  &\nodata  &    0.1 \\
        \nodata  & molecular  &   61.3  &   91.6  &\nodata  &   86.1  &\nodata  &   78.5 \\
    F13218+0552  &   neutral  &   22.9  &   44.4  &   44.4  &   65.4  &   72.6  &   80.2 \\
        \nodata  &   ionized  &   77.1  &   55.6  &   55.6  &   34.6  &   27.4  &   19.8 \\
    F13342+3932  &   neutral  &   17.7  &   18.0  &   18.0  &   25.4  &   55.3  &   65.7 \\
        \nodata  &   ionized  &   82.3  &   82.0  &   82.0  &   74.6  &   44.7  &   34.3
  \enddata

  \tablecomments{Column 2: Gas Phase. Column 3$-$8: Percent of the
    mass, momentum, energy, and their outflow rates in each
    phase. Molecular gas properties are from
    \citet{gonzalezalfonso17a}. For \fohseven, the ionized gas
    measurements are not extinction corrected, so the neutral values
    are upper limits and the ionized values are lower limits to the
    true values.}

 \end{deluxetable*}
 \capstarttrue
 
\subsection{Global Outflow Properties} \label{sec:ofprop}

 \capstartfalse
\begin{deluxetable*}{ccrccccc}
  \tablecaption{Large-Scale Seyfert Galaxy Outflows \label{tab:seyferts}}
  \tablewidth{\textwidth}

  \tablehead{\colhead{Galaxy} & \colhead{Type} & \colhead{$dM/dt$} &
    \colhead{log[$(dE/dt)/$} & \colhead{log($M_\mathrm{BH}/$} &
    \colhead{log[$L_\mathrm{AGN}/$} &
    \colhead{$\sigma$} & \colhead{References} \\
    \colhead{~} & \colhead{~} & \colhead{(\smpy)} & \colhead{(erg
      s$^{-1}$)]} & \colhead{\msun)} & \colhead{(erg
      s$^{-1}$)]} & \colhead{(\kms)} & \colhead{~} \\
    \colhead{(1)} & \colhead{(2)} & \colhead{(3)} & \colhead{(4)} &
    \colhead{(5)} & \colhead{(6)} & \colhead{(7)} & \colhead{(8)}}

  \startdata
  NGC 1068 & 2   &   76$\;\quad$ & 42.87 & 6.934$_{-0.015}^{+0.015}$ & 44.95$\pm$0.50 & 176$\pm$9  & b, C, D, ff, 3, 7 \\
  NGC 1266 & 2   & 110$\;\quad$  & 41.04 & 6.738$_{-0.441}^{+0.441}$ & 43.53$\pm$0.30 &  94$\pm$5  & a, A, dd, 1 \\
  Mrk 79   & 1   &   10$\;\quad$   & 40.68 & 7.612$_{-0.565}^{+0.445}$ & 44.65$\pm$0.18 & 129$\pm$10 & d, F, gg, 2, 5 \\
  NGC 2992 & 2   & 168$\;\quad$  & 42.54 & 7.703$_{-0.443}^{+0.442}$ & 44.48$\pm$0.50 & 160$\pm$17 & d, D, dd, 5, 8 \\
  NGC 3783 & 1.5 &   3.5$\;\;$   & 40.99 & 7.367$_{-0.492}^{+0.442}$ & 43.70$\pm$0.25 & 110$\pm$19 & c, D, gg, 2, 5 \\
  NGC 4151 & 1.5 &   3$\;\quad$  & 41.63 & 7.553$_{-0.456}^{+0.441}$ & 43.28$\pm$0.28 &  96$\pm$10 & b, B, aa, 2, 5 \\
  NGC 4253 & 1   &  0.78  & \nodata & 6.822$_{-0.444}^{+0.443}$ & 43.66$\pm$0.21 &  93$\pm$32 & c, G, bb, 2, 5 \\
  NGC 4388 & 2   &   0.02        & 37.30 & 6.929$_{-0.010}^{+0.010}$ & 44.68$\pm$0.52 &  99$\pm$9  & b, I, ee, 3, 7 \\
  Circinus & 1   &   8.0$\;\;$   & 40.00 & 6.230$_{-0.084}^{+0.092}$ & 43.30$\pm$0.30 & 149$\pm$18 & c, J, cc, 4 \\
  NGC 5548 & 1.5 &   9.5$\;\;$   & 41.47 & 7.718$_{-0.440}^{+0.440}$ & 44.29$\pm$0.35 & 193$\pm$11 & b, H, bb, 2, 5 \\
  NGC 6814 & 1.5 &  10.5$\;\;$   & 41.05 & 7.163$_{-0.493}^{+0.442}$ & 43.24$\pm$0.34 & 106$\pm$17 & c, D, bb, 2, 5 \\
  NGC 7469 & 1   &   5.6$\;\;$  & 40.92 & 6.979$_{-0.460}^{+0.441}$ & 44.62$\pm$0.20 & 130$\pm$6  & b, D, gg, 2, 5 \\
  NGC 7582 & 1   &   3.4$\;\;$        & \nodata & 7.740$_{-0.097}^{+0.111}$ & 43.30$\pm$0.30 & 148$\pm$19 & b, E, hh, 6
  \enddata

  \tablecomments{Column 2: Seyfert type. Column 3--4: Mass and energy
    outflow rates. Error bars are not consistently available for these
    measurements; we assume 0.3~dex. Most estimates are based on
    ionized gas measurements. The exceptions are NGC 1068 (molecular
    and ionized gas measurements have been summed); NGC 1266 (only a
    molecular gas measurement is available); and Circinus (the
    molecular gas value is much greater than the ionized gas
    value). In most cases the ionized values are calculated
    geometrically and depend on the product of the electron density
    $n_e$ and filling factor $f$, which is typically assumed to be
    $n_ef = 5$~cm$^{-2}$. For consistency we have normalized geometric
    estimates to this value, except in the case of NGC~5548, for which
    f is calculated but not published. Some literature ionized gas
    estimates erroneously do not correct upward for mean particle
    mass; in these cases we multiply the published rates by
    1.4. Column 5: Logarithm of the black hole mass. Errors from
    reverberation mapping (references aa, bb, and gg) have the
    uncertainty in the virial coefficient $f$ added in quadrature
    ($\delta f = 0.44$; \citealt{woo10a}). Column 6: Logarithm of the
    bolometric AGN luminosity, as determined from the infrared
    luminosity, 5100~\AA\ luminosity, or \oth\ luminosity, in order of
    preference. In one case, NGC 7582, the only available measurement
    is from X-rays. Errors for $L_\mathrm{AGN}$ determined from
    5100~\AA\ or \oth\ combine the observational error and the scatter
    in the bolometric correction. Errors for $L_\mathrm{AGN}$
    determined from the infrared or X-rays are not listed; the errors
    are assumed to be 0.3~dex. Column 7: Velocity dispersion, from
    HyperLeda \citep{makarov14a} for all sources but NGC~4253
    \citep{woo10a} and NGC~1266 \citep{cappellari13a}. Column 8:
    References for data in columns 2--6.}

  \tablerefs{{\bf Seyfert type:}
    (a) \citealt{davis12a};
    (b) \citealt{osterbrock93a};
    (c) \citealt{veroncetty06a};
    (d) NED.
    {\bf Mass outflow rate:}
    (A) \citealt{alatalo15a};
    (B) \citealt{crenshaw15a};
    (C) \citealt{garciaburillo14a};
    (D) \citealt{mullersanchez11a};
    (E) \citealt{riffel09a};
    (F) \citealt{riffel13a};
    (G) \citealt{schonell14a};
    (H) \citealt{schonell17a};
    (I) \citealt{veilleux99c}; 
    (J) \citealt{zschaechner16a}.
    {\bf Black hole mass:}
    (aa) \citealt{bentz06a}; 
    (bb) \citealt{bentz10a};
    (cc) \citealt{greenhill03a};
    (dd) \citealt{gultekin09a};
    (ee) \citealt{kuo11a};
    (ff) \citealt{lodato03a};
    (gg) \citealt{peterson04a};
    (hh) \citealt{wold06a}.
    {\bf Bolometric AGN luminosity:}
    (1) \citealt{alatalo15a};
    (2) \citealt{bentz13a};
    (3) \citealt{liu09a};
    (4) \citealt{moorwood96b};
    (5) \citealt{runnoe12a};
    (6) \citealt{rivers15a};
    (7) \citealt{schmitt03a};
    (8) \citealt{ward80a}.}

\end{deluxetable*}
\capstarttrue

The key physical quantities that measure the potential impact of an
outflow on its host are velocities, size (maximum projected radius
$R$), mass ($M$), momentum ($p$), energy ($E$), and the outflow rates
of $M$, $p$, and $E$ in each gas phase. In Table~\ref{tab:vels} we
list mean and maximum values of FWHM$^{c2}$, \vfifty$^{c2}$, and
\vtsig$^{c2}$. The averages are taken over spaxels in the outflowing
component(s). The other quantities are listed in Table~\ref{tab:mpe}.

Mass, momentum, and energy and their outflow rates are calculated
using a time-averaged thin shell model (the single radius free wind,
or SRFW, model of \citealt{rupke13a}) in which these quantities depend
inversely on the assumed shell radius
\citep{rupke05b,shih10a,rupke13a}. The thin shell model is a fiducial
model for measuring such quantities. For spatially-resolved data, it
has the advantage of being able to associate each point in the plane
of the sky with a unique three-dimensional radius, assuming that only
the near side of the wind is observed. Real winds are almost certainly
more complex. However, thick shell models \citep[e.g.,][]{rupke05b}
require knowledge of the wind's density and/or radial distribution,
which in turn requires more detailed spatial information than is
available in the current data.

Some well-resolved winds are able to be fit with biconical or bipolar
bubble models \citep[e.g.,][]{veilleux94a,rupke13a,fischer13a}. Such
models tend to better account for wind projection and often lead to
higher values of mass, momentum, and energy \citep{rupke13a}. There
are indications of bipolar flows in at least two objects in our sample
(including the \pgonesix\ and \pgoneseven), and there are further
indications of minor axis collimation in others
(\S\,\ref{sec:of_presence}). However, the canonical sign of bipolar
shells, line splitting, is noticeably absent. Detailed model fits of
bipolar bubbles or cones would thus need to fully explore the
parameter space of possible geometries to address parameter
degeneracies. For simplicity and consistency with previous studies and
within the sample, the thin shell is used exclusively here. Fits to
bipolar models are reserved for future work, possibly using data of
higher spatial resolution.

The measurements of $M$, $p$, and $E$ may be lower limits to the true
values due to the fact that any spatially-unresolved outflow has been
removed from the data. Thus, outflows at scales $<$0.5~kpc in the
nearest systems, approaching $2-3$~kpc in the most distant systems,
are not considered here.

The maximum observed radius of each wind in two-dimensional projection
($R_\mathrm{obs}$) is measured in each phase. The observed range of
$R_\mathrm{obs}$ is 3--12~kpc. The (single) three-dimensional radius
adopted for the SRFW model, $r_\mathrm{wind}$, is a free parameter in
the model. Adopting $r_\mathrm{wind} = R_\mathrm{obs}$ exactly results
in edge effects: spaxels with projected radii $R = R_\mathrm{obs}$
yield formally infinite velocities when they are deprojected. The
radius adopted is thus slightly larger (by $\sim$0.5 spaxel) than the
observed radius. In most cases
$r_\mathrm{wind}^\mathrm{neutral} = r_\mathrm{wind}^\mathrm{ionized}$
is adopted. One exception is \mrk; while there is blueshifted,
AGN-photoionized emission at large radii, the line widths of this
$R\sim3$~kpc emission are very narrow and lie atop the southern arc of
star formation in this galaxy. It is thus assumed that these spaxels
represent AGN-photoionized gas that is forming stars, rather than an
outflow.

\begin{figure*}
  \subfloat{\includegraphics[width=0.5\textwidth]{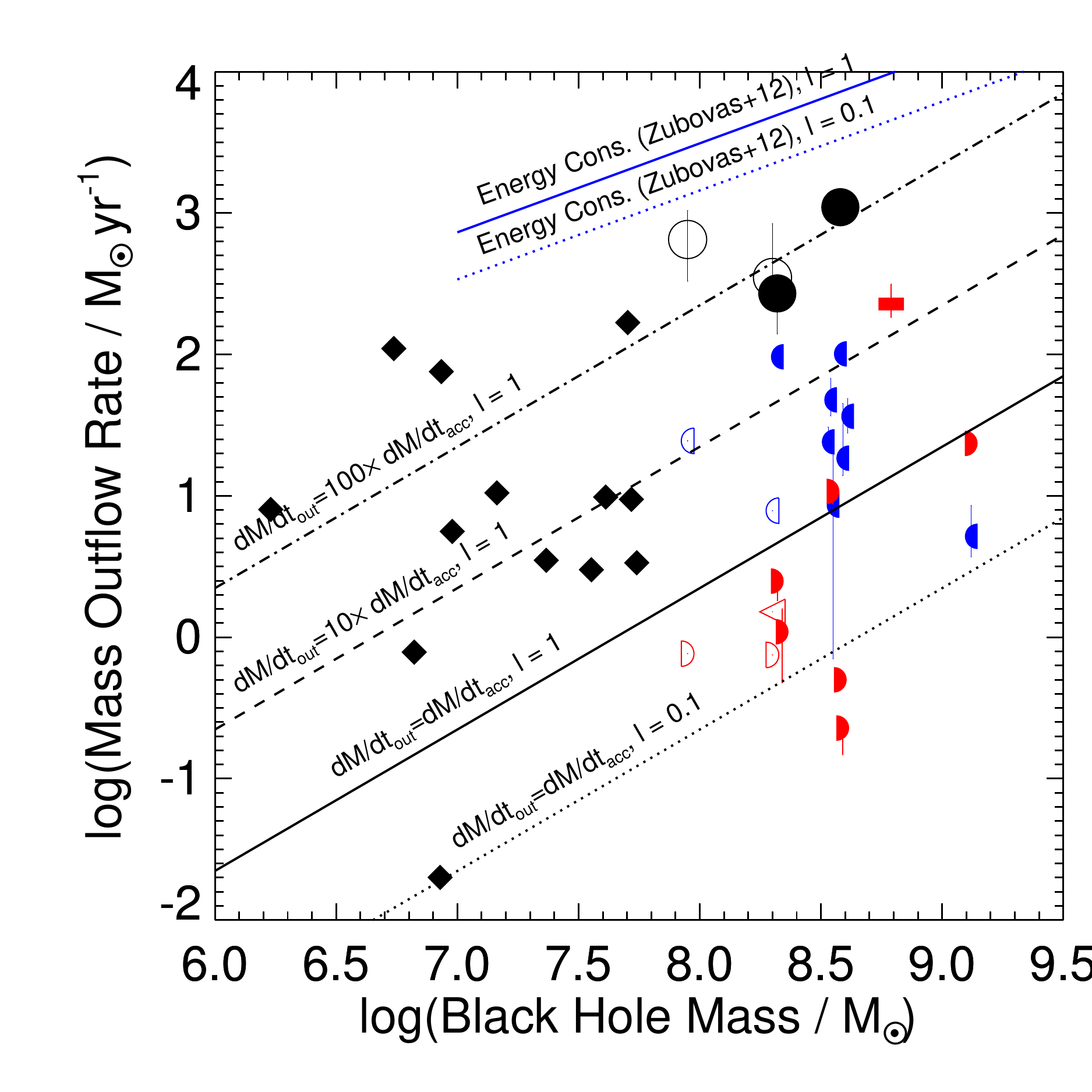}}
  \subfloat{\includegraphics[width=0.5\textwidth]{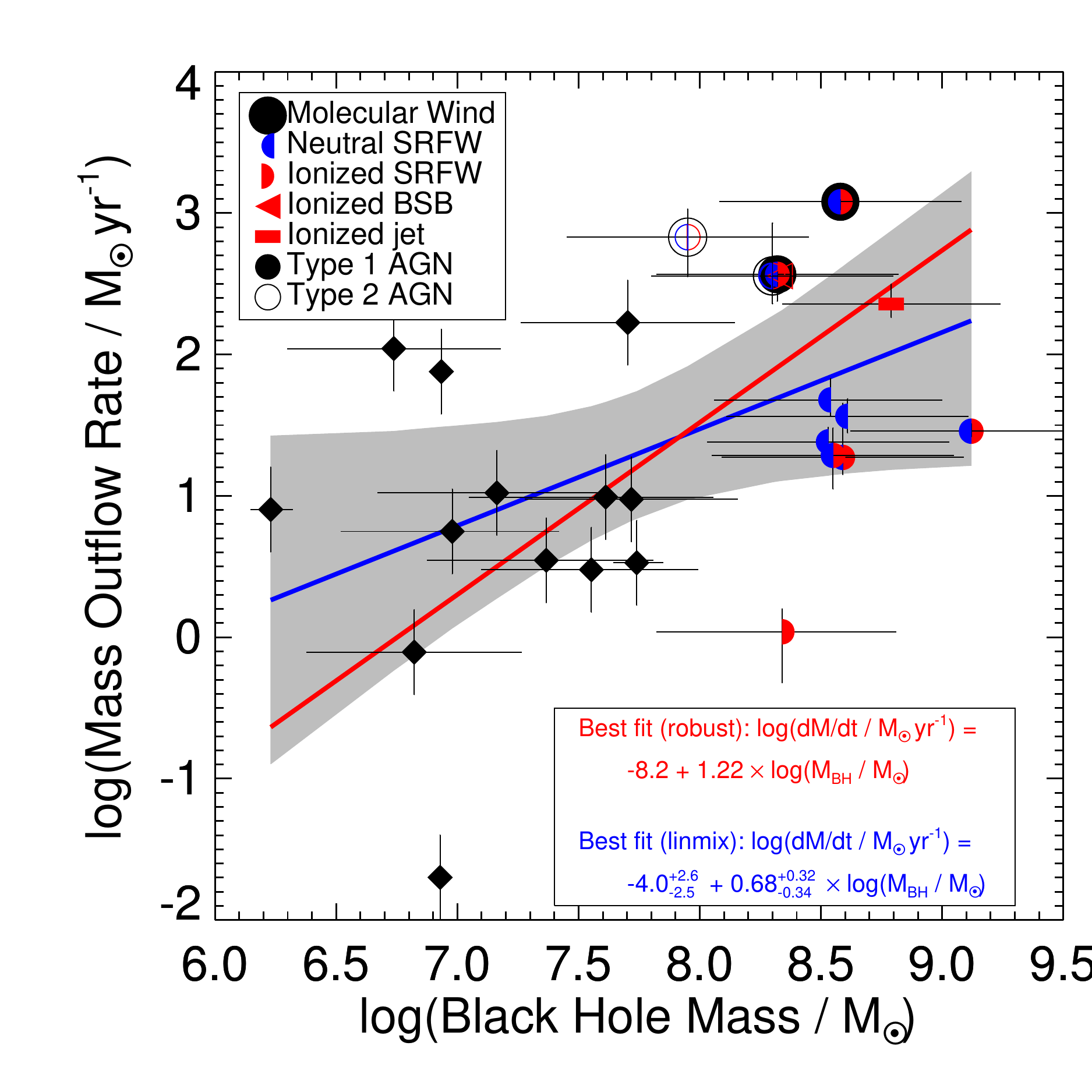}}
  \caption{Mass outflow rates as a function of black hole mass for
    this sample, two Type 2 quasars \citep{rupke13a}, and a sample of
    Seyfert galaxies with resolved outflows (shown as filled black
    diamonds; Table~\ref{tab:seyferts}). {\bf(Left)} Individual phases
    are shown. Ionized gas measurements in the Type 2 quasars and in
    \fohseven\ are lower limits because an extinction correction has
    not been performed. Ionized gas masses in the Type 2 quasars have
    been corrected to $n_e = 100$~cm$^{-2}$ to match the typical value
    measured in the current sample (\S\,\ref{sec:emlfits}). Most
    quasar measurements result from single-radius free wind (SRFW)
    models, though one galaxy also has masses from a bipolar
    superbubble (BSB) fit (\S\,\ref{sec:ofprop} and
    \citealt{rupke13a}). Models for energy-conserving winds driven by
    high-velocity blast waves (Eddington ratios $l = 0.1$ and 1) are
    overplotted as blue solid and dotted lines \citep{zubovas12a}, as
    well as lines of constant values of mass outflow rate divided by
    accretion rate (for $l = 0.1$ or 1 and 10\%\ efficiency of energy
    released by accretion) as black dotted, solid, dashed, and
    dotted-dashed lines. {\bf (Right)} Mass fluxes have been added up
    across measured phases. Robust and Bayesian fits (the latter
    including intrinsic scatter with LINMIX\_ERR; \citealt{kelly07a})
    to the correlation between outflow rate and $M_\mathrm{BH}$ are
    shown. The contours show 2$\sigma$ deviations from the median
    after applying the posterior distributions of slopes and
    intercepts to the measured dependent variables. The 95.5\%\ range
    of correlation coefficients and slopes ($-$0.01 to 0.88 and
    $-$0.02 to 1.37, respectively) shows a correlation at the
    2$\sigma$ level and a significant amount of intrinsic scatter
    (1.0~dex). Though the \citet{zubovas12a} models overpredict the
    mass outflow rates by 1--2 orders of magnitude, they do match the
    best-fit slope (0.63 in theory, 0.68 observed). }
  \label{fig:dmdt_v_mbh}
\end{figure*}

The ionized gas flux in each wind (except in \fohseven, for which
there is only one Balmer line in the present data) is corrected using
the extinction computed from the Balmer decrement and listed in
Table~\ref{tab:mpe}. As described in \S\,\ref{sec:emlfits}, Case B at
10$^4$~K and the \citet{cardelli89a} extinction curve are
assumed. This correction (to the flux in component 2, c2) is done on a
spaxel-by-spaxel basis where possible and uses the extinction computed
from the combined flux in both components (c1$+$c2;
Figure~\ref{fig:ebv}). Spaxels with uncertain extinction values are
corrected using the median extinction value in the host.

The outflowing ionized gas mass in each spaxel depends inversely on
electron density $n_e$. The electron density measured in each spaxel
(calculated as described in \S\,\ref{sec:emlfits} and shown in
Figure~\ref{fig:elecden}) is used to calculate the gas mass. In the
few spaxels where the \sut\ fluxes are too uncertain, a value of
100~cm$^{-2}$ is used (near the measured sample median;
\S\,\ref{sec:emlfits}). This value is also assumed for the two systems
without $n_e$ measurements (\mrk\ and \pgoneseven).

The neutral gas masses are proportional to the measured column density
of hydrogen \citep{rupke05b}. The hydrogen column density in turn is
related to the measured Na column through the ionization fraction, the
Na abundance relative to Solar and the galaxy's metallicity, and the
Na dust depletion \citep{rupke05a}. Solar metallicity is assumed
\citep{rupke08a} and the values for ionization fraction, Na abundance,
and Na dust depletion are taken from \citet{rupke05a}.

Five quasar hosts have detections of both neutral and ionized outflows
(though one, \fohseven, does not have extinction-corrected outflow
properties). Two hosts, \fohfive\ and \mrk, also have molecular phase
outflow measurements of critical wind properties. The latest
measurements from modeling of multiple OH absorption lines observed
with {\it Herschel} are adopted \citep{gonzalezalfonso17a}. These
molecular outflows have predicted sizes under $<$1~kpc. Their
properties are listed alongside the neutral and ionized properties in
Tables~\ref{tab:mpe} and \ref{tab:mperats_in}.

The four hosts with extinction-corrected ionized outflow measurements
and measurements in multiple gas phases show that the ionized gas
phase dominates the mass of the wind in 2/4 cases but is negligible in
the other two. However, the neutral and molecular phases typically
contribute most to the energetics. This illustrates the importance of
multiphase observations of winds, even in AGN outflows that are often
presumed to be dominated by ionized gas.

\subsection{Outflow Properties vs.\ Black Hole
  Properties} \label{sec:bhprop}

Observational studies of outflow properties are useful for informing
sub-grid prescriptions in cosmological simulations of galaxy formation
and evolution if observed outflow properties correlate with host, AGN,
or black hole properties. Previous studies have focused on how
quasar-mode outflows depend on host or AGN properties
\citep[e.g.,][]{sturm11a,veilleux13a,spoon13a,rupke13a,harrison14a,cicone14a,mcelroy15a,stone16a,sun17a,fiore17a}. Since
this sample of nearby luminous AGN has substantial ancillary data, and
thus useful constraints on the central black hole (Table
\ref{tab:sample}), this paper focuses on the dependence of outflow
properties on black hole properties.

To probe this dependence across a significant dynamic range, however,
a larger range of black hole mass is
required. Table~\ref{tab:seyferts} lists the outflow and host galaxy
properties in 13 nearby Seyfert galaxies with spatially-resolved
outflows. The outflow properties are mostly based on ionized gas
measurements of the NLR, except for those in NGC 1068, NGC~1266, and
Circinus. The measurements of NGC~1266 and Circinus are of the
molecular gas, which dominate the outflow mass and energy budgets. The
measurements of NGC 1068 combine ionized and molecular gas. These NLR
outflows have scales of 0.1 to a few kpc.

Figure~\ref{fig:dmdt_v_mbh} shows total mass outflow rate (summed over
all gas phases) versus black hole mass. Error bars on the IFS data
reflect statistical uncertainties in line fits to individual spaxels
(as determined from Monte Carlo simulations in the case of absorption
lines and least-square errors in the case of emission lines), which
have been propagated and summed in quadrature across all outflowing
spaxels. However, unquantified systematic errors (due to, e.g.,
uncertain 3D spatial distribution) could be larger than the
measurement uncertainties.

Because of the significant intrinsic scatter in this plot that is not
accounted for by the measurement errors, we estimate the strength and
properties of a possible linear relationship using a Bayesian
regression that fits the intrinsic scatter and uses a Gaussian mixture
model for the intrinsic distribution of independent variables
(LINMIX\_ERR; \citealt{kelly07a}). The default mixture of three
Gaussians and the Gibbs sampler are used. The resulting 95.5\%\
(2$\sigma$) ranges for the correlation coefficient and slope are
$-$0.01 to 0.88 and $-$0.02--1.37, respectively, with medians in the
posterior distributions of 0.50 and 0.68. The median intrinsic scatter
is 0.98, meaning there are unaccounted-for variables that contribute
significantly or that the error is underestimated. The best-fit linear
regression line (using the medians and 1$\sigma$ deviations from the
median of intercept and slope) is then
\begin{multline}
\mathrm{log}[(dM/dt)/(\smpy)] = \\ -(4.0_{-2.5}^{+2.6}) +
(0.68_{-0.34}^{+0.32})\times \mathrm{log}(M_\mathrm{BH} / \msun).
\end{multline}
A robust fit that does not account for errors and bisects the fit of X
on Y and Y on X gives a steeper slope of 1.2.

\begin{figure}
  \includegraphics[width=0.9\columnwidth,center]{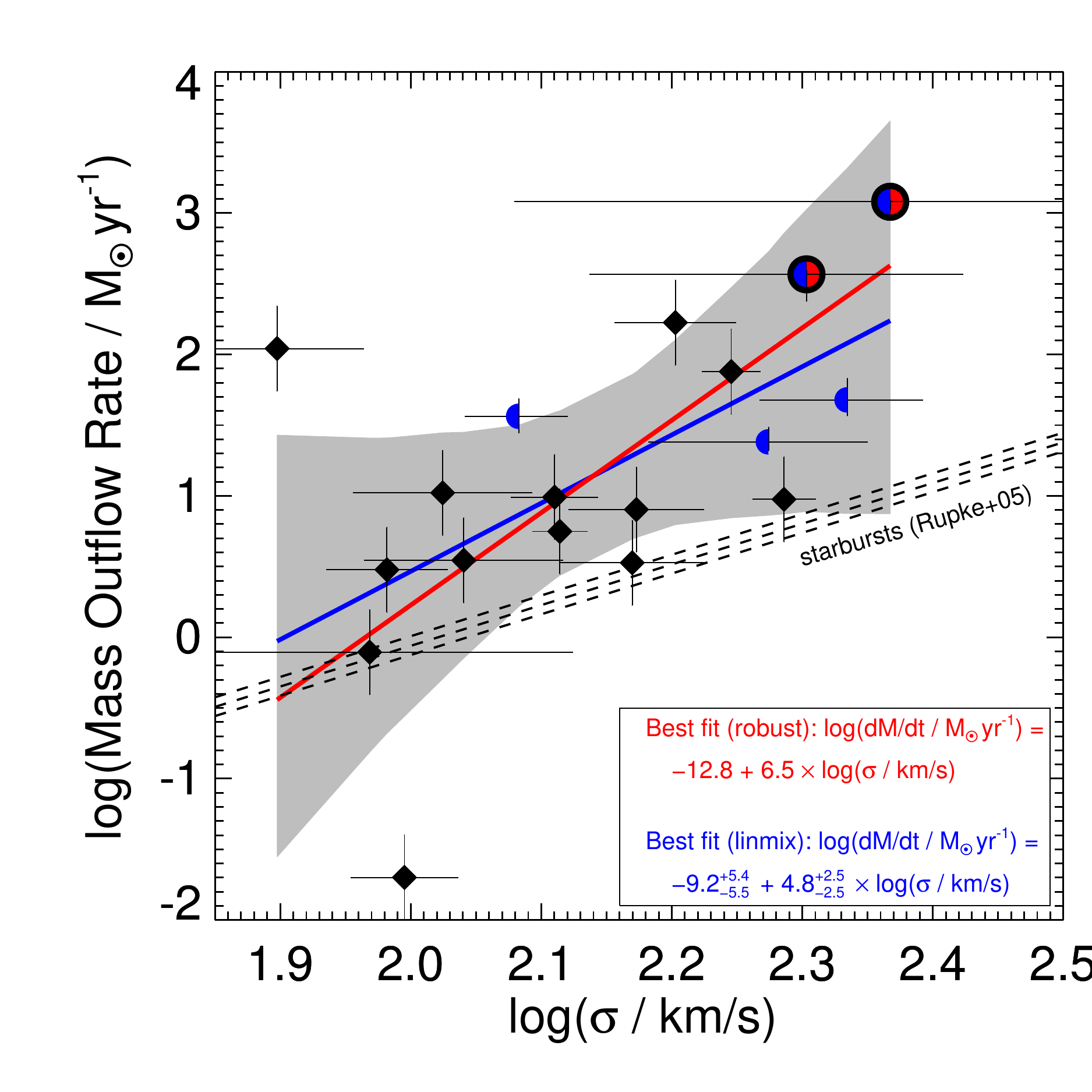}
  \caption{Mass outflow rates as a function of stellar velocity
    dispersion for quasars and Seyferts. See Figure
    \ref{fig:dmdt_v_mbh} and \S\,\ref{sec:bhprop} for more
    information. The dashed lines show the relationship between these
    two quantities for starburst winds (slope 2.9; \citealt{rupke05b})
    for three different values of $v_\mathrm{rot}/2\sigma$ (0.42,
    0.50, and 0.58). The correlation is not significant at the
    2$\sigma$ level, but the best-fit slope is consistent with an
    extrapolation from the slope of $dM/dt$ vs. $M_\mathrm{BH}$ and
    the $M_\mathrm{BH}$--$\sigma$ relation. It is apparent that
    AGN-driven outflows in massive black holes are driving mass out at
    greater rates than in pure starbursts.}
  \label{fig:dmdt_v_sigma}
\end{figure}

\begin{figure*}
  \subfloat{\includegraphics[width=0.5\textwidth]{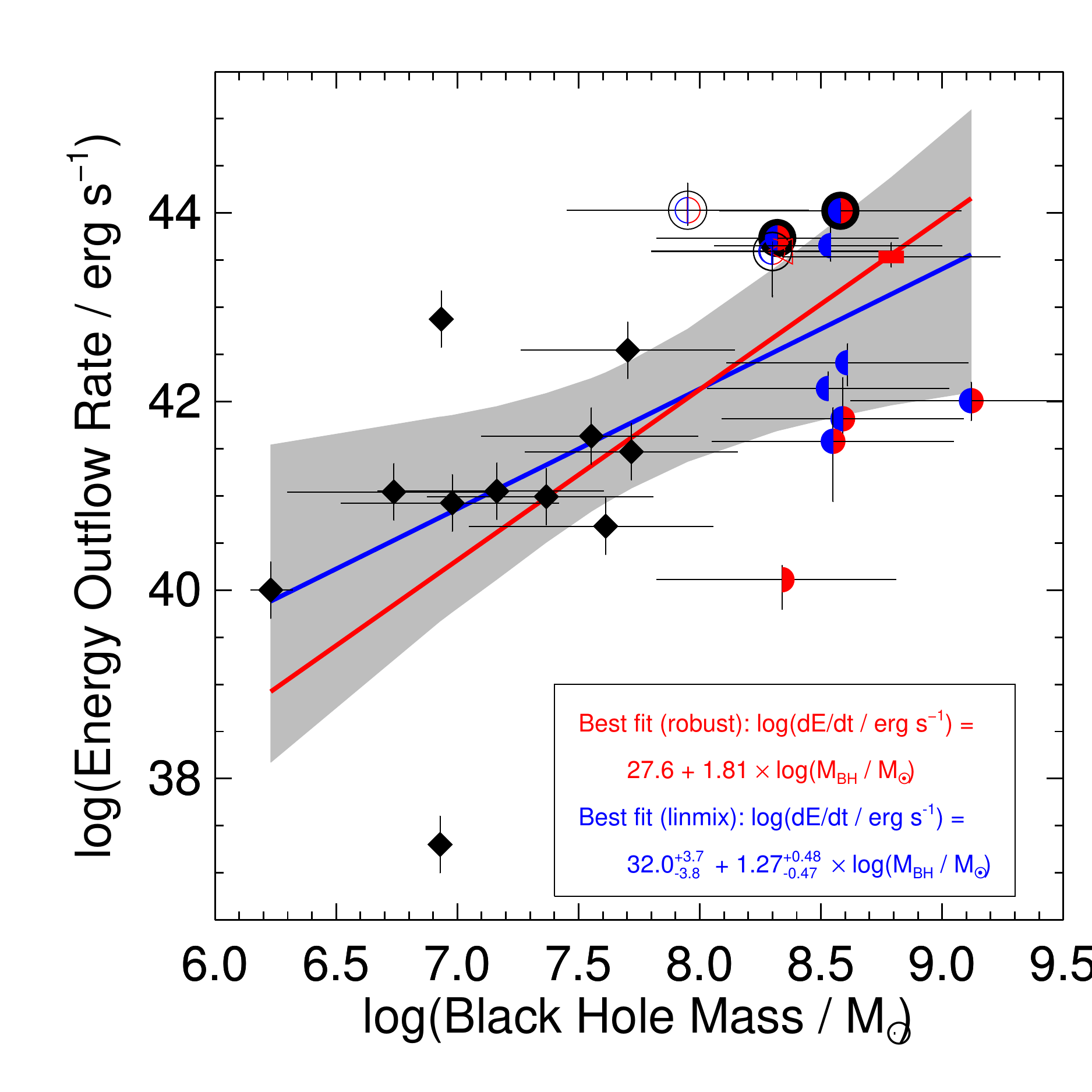}}
  \subfloat{\includegraphics[width=0.5\textwidth]{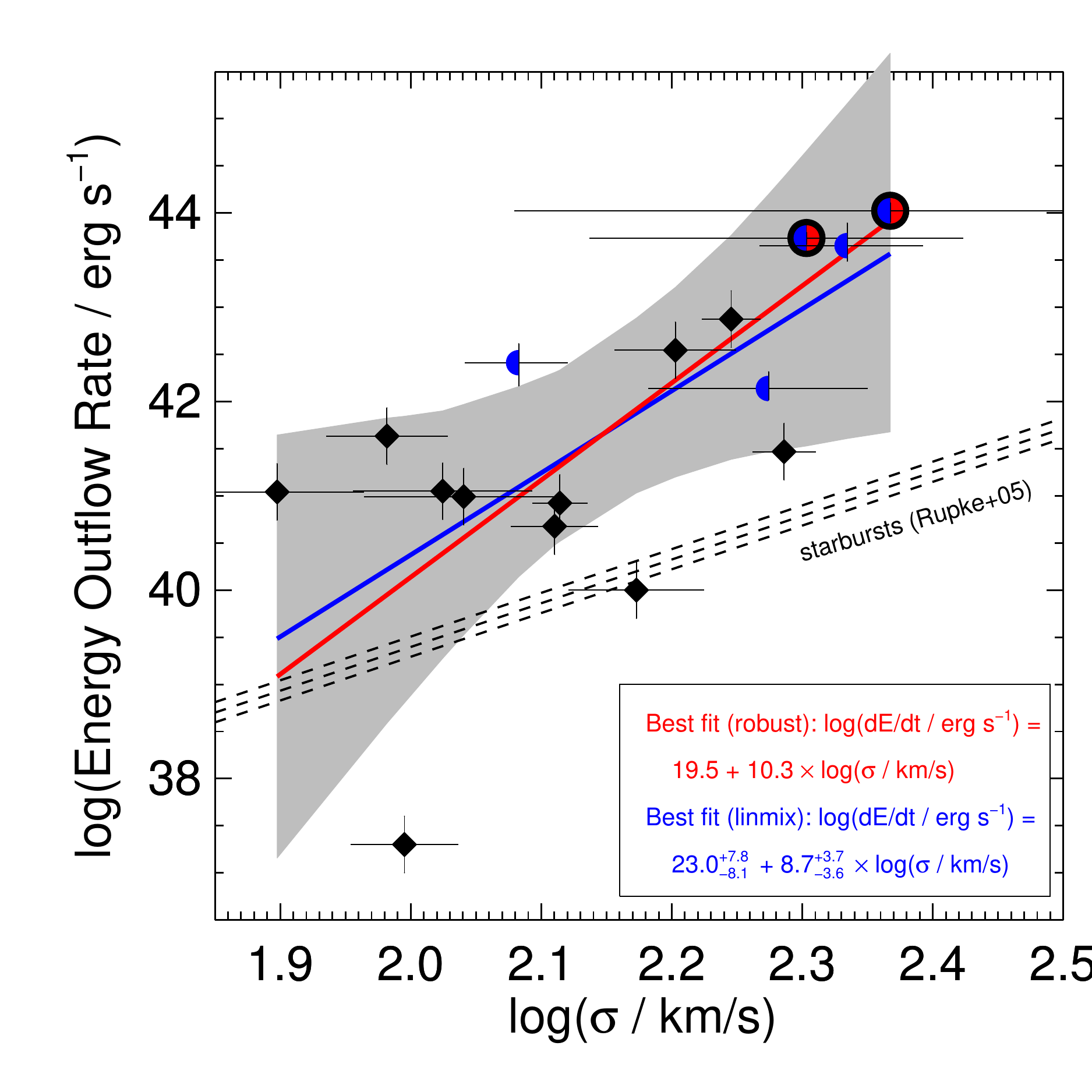}}
  \caption{Energy outflow rates as a function of $M_\mathrm{BH}$ and
    $\sigma$. See Figures~\ref{fig:dmdt_v_mbh}--\ref{fig:dmdt_v_sigma}
    and \S\,\ref{sec:bhprop} for more details. The correlations are
    significant at greater than 2$\sigma$, and the intrinsic scatter
    is 1.4~dex.}
  \label{fig:dedt_v}
\end{figure*}

This statistical test points to a positive correlation between black
hole mass and total mass outflow rate in this sample at 95.5\%\
confidence, though the slope of this correlation is poorly constrained
(95.5\%\ range 0.0--1.4). It also suggests that either unaccounted-for
systematic errors or other variables play a significant role in their
relationship. For instance, while the relationship among the different
gas phases is not fully understood spatially, it may be that adding
the phases together means that some phases are counted twice. The
ionization correction to the \nad\ measurement tries to account for
all gas phases, but it is entirely possible that the gas clouds traced
by \nad\ are not spatially coincident with those detected by ionized
or molecular gas tracers. In fact, they are often spatially disjoint
(Fig.~\ref{fig:ofgeometry}). In this case adding together the masses
inferred from different phases is appropriate, since they trace
different parts of the wind.

Selection effects may also play a role. Since most or all galaxies
host nuclear black holes, they will appear on this plot
somewhere. Studies of galaxy-scale outflows in complete, unbiased
samples of galaxies will show whether the outflows in this sample are
simply the tip of the iceberg (i.e., there is a broad distribution of
mass outflow rates extending to zero) or that the mass outflow rates
measured here are representative of all AGN with high Eddington ratio.

Because black hole mass correlates with velocity dispersion, mass
outflow rate should correlate with stellar velocity
dispersion. Stellar velocity dispersion is known for only a subset of
the current sample (Table~\ref{tab:sample}). Figure
\ref{fig:dmdt_v_sigma} shows that there may be a correlation between
these two quantities, but the correlation is not significant at the
2$\sigma$ level. The 95.5\%\ range of possible correlation
coefficients and slopes is $-$0.08 to 0.97 and $-$0.7 to 10.6. The
best-fit line is
\begin{multline}
\mathrm{log}[(dM/dt)/(\smpy)] = \\ -(9.2_{-5.5}^{+5.4}) +
(4.8_{-2.5}^{+2.5})\times \mathrm{log}[\sigma/(\kms)].
\end{multline}
Based on the correlation between log($dM/dt$) and log($M_\mathrm{BH}$)
and the $M_\mathrm{BH}$--$\sigma$ relation \citep{gultekin09a}, the
slope should be 2.9$\pm$1.5, which is consistent with the fitted slope
within the (large) errors.

Energy outflow rates are also published for most of the Seyferts in
Table~\ref{tab:seyferts}. The correlation between $dE/dt$ and
$M_\mathrm{BH}$ or $\sigma$ (Figure~\ref{fig:dedt_v}) is more
significant than the correlations with mass outflow rate. The
posterior distributions of correlation coefficients and slopes for
$dE/dt$ vs. $M_\mathrm{BH}$ have a 95.5\%\ range of 0.15--0.94 and
0.31--2.28, respectively. For $dE/dt$ vs. $\sigma$, these ranges are
0.08--0.98 and 1.1--17.3. The best fit lines are
\begin{multline}
\mathrm{log}[(dE/dt)/(\mathrm{erg}~\mathrm{s}^{-1})] = \\ -(32.0_{-3.8}^{+3.7}) + (1.27_{-0.47}^{+0.48})\times
\mathrm{log}(M_\mathrm{BH} / \msun),
\end{multline} and\
\begin{multline}
\mathrm{log}[(dE/dt)/(\mathrm{erg}~\mathrm{s}^{-1})] = \\ -(23.0_{-8.1}^{+7.8}) +
(8.7_{-3.6}^{+3.7})\times \mathrm{log}[\sigma/(\kms)],
\end{multline}
the best-fit correlation coefficients are 0.61 and 0.68, and the
intrinsic scatter is 1.4~dex.

\section{DISCUSSION} \label{sec:discussion}

The following section addresses the implications of the outflow
properties of the current sample and the observed correlations with
black hole properties for the wind power source, quasar wind models,
and quasar-mode feedback.

\subsection{Wind Power Source} \label{sec:powersource}

The quasars in this sample host both a starburst and a luminous AGN,
and it is important to determine which power source is driving the
observed large-scale outflows. Previous work has argued that the \mrk\
outflow is powered primarily by the AGN
\citep[e.g.,][]{fischer10a,feruglio10a,rupke11a,fiore15a,morganti16a}. In
\pgoneseven\ the culprit is clearly the jet spawned by black hole
accretion, while in \pgonesix\ the outflow is driven by the AGN within
the narrow-line region.

\begin{figure*}
  \subfloat{\includegraphics[width=0.5\textwidth]{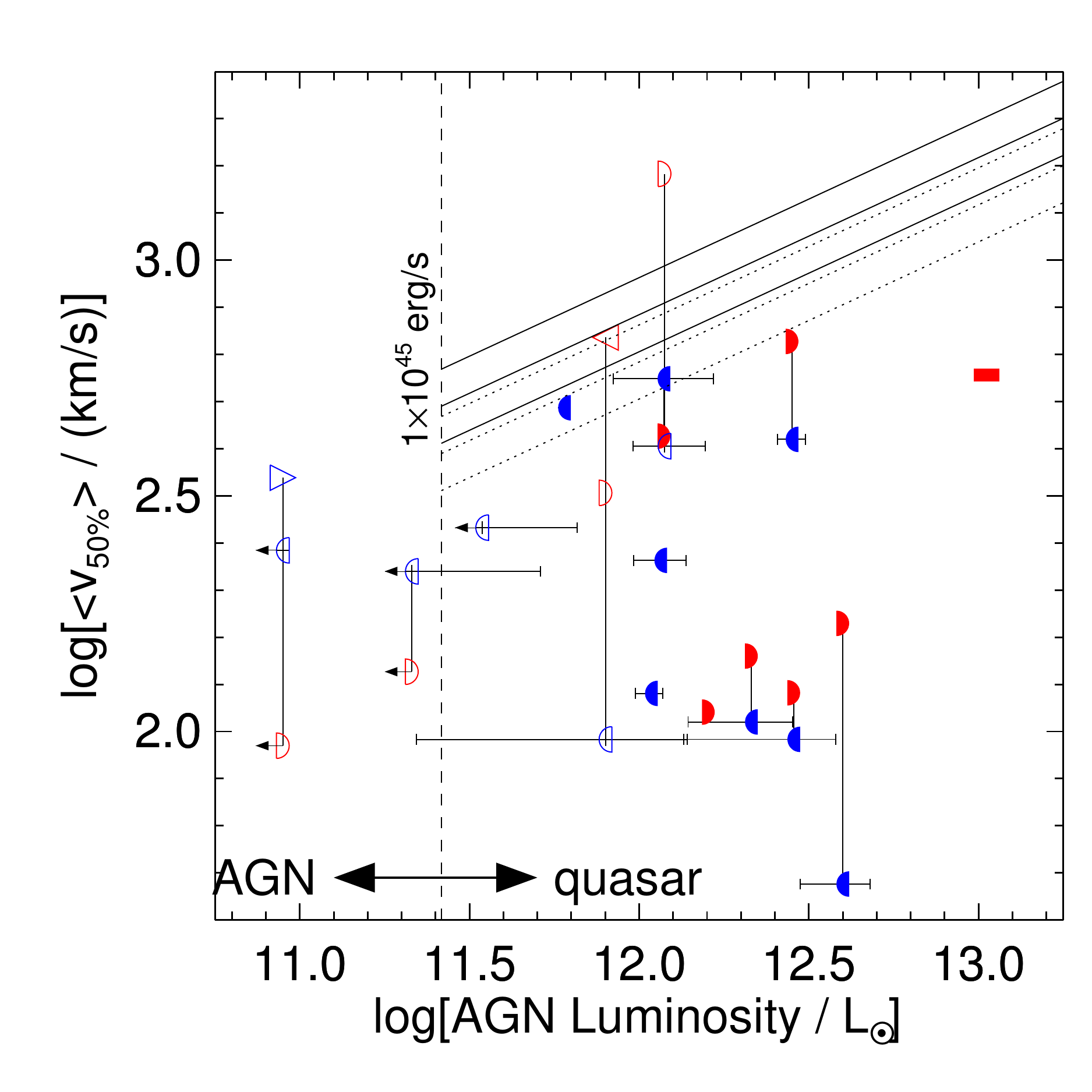}}
  \subfloat{\includegraphics[width=0.5\textwidth]{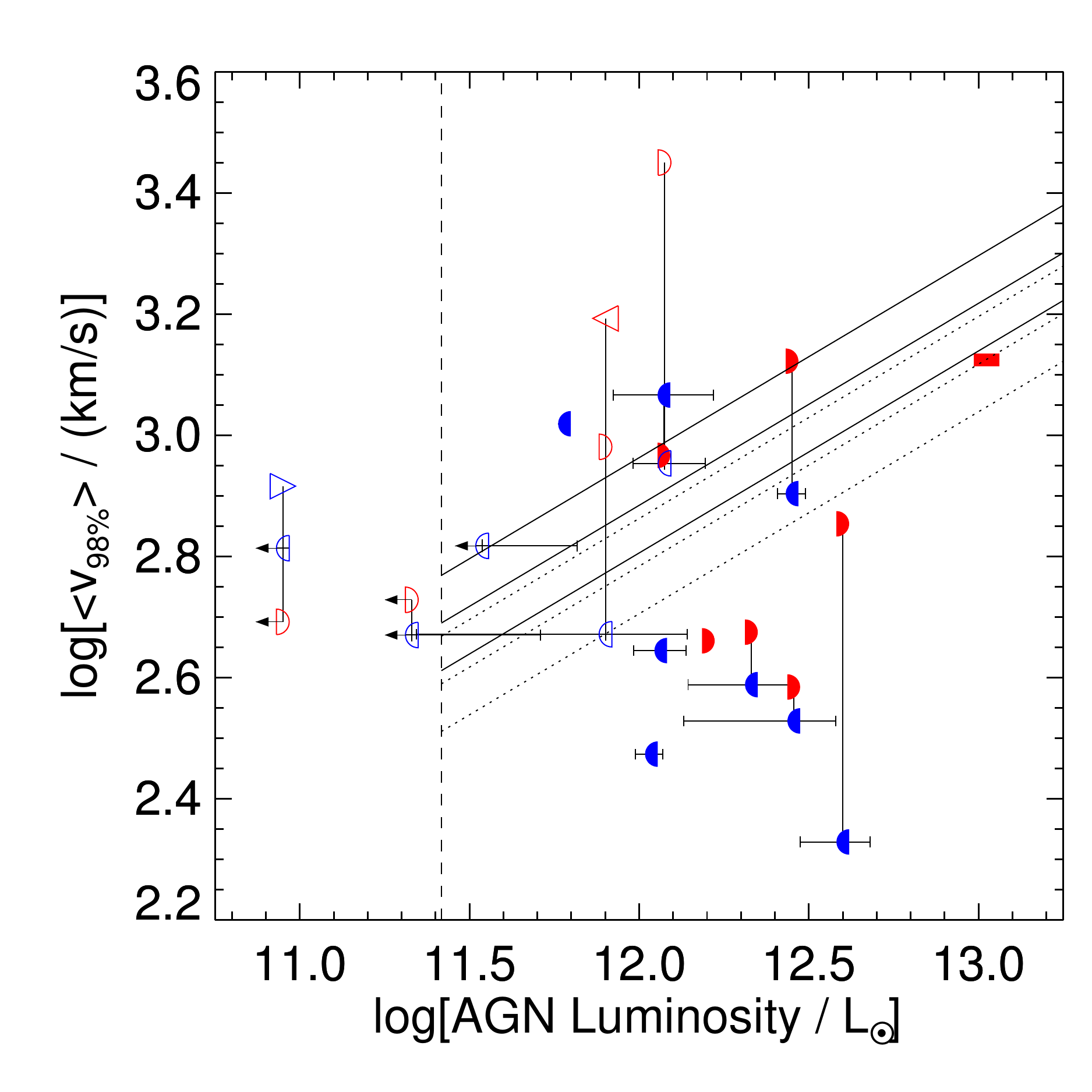}}
  \caption{Spatially-averaged velocities (\vfifty\ or \vtsig) vs.\ AGN
    luminosity from ionized and neutral IFS observations of nearby
    Type 1 and 2 quasars and starbursts (\citealt{rupke13a} and this
    work). See Figure~\ref{fig:dmdt_v_mbh} for an explanation of
    symbols. Measurements of the same galaxy are connected by vertical
    solid black lines. The typical quasar threshold is labeled with a
    vertical dashed line. The prediction of an energy-conserving model
    from \citet{zubovas12a} is overlaid. Solid lines show predictions
    for black hole masses of log($M_\mathrm{BH}$)$=$8.0, 8.5, and 9.0;
    a galaxy gas mass fraction $f_g = 0.1$ (typical for ULIRGs;
    \citealt{rupke08a}); and ratio of gas density to background
    density of $f_c = 0.16$ (as in \citealt{zubovas12a}; see this
    paper for more details on $f_g$ and $f_c$).  The black hole mass
    is converted to velocity dispersion using the
    $M_\mathrm{BH}$--$\sigma$ relation \citep{gultekin09a}. Dotted
    lines show $f_g = 0.2$. There is evidence that the highest
    velocities are only observed in quasars, but not all quasars have
    high-velocity winds. No correlation is evident. Energy-conserving
    models can reproduce the highest velocities in many systems, but
    overpredict average velocities and even the highest velocities in
    many radio-quiet quasars.}
  \label{fig:vel_v_lagn}
\end{figure*}

In other systems (particularly \fohfive\ and \pgonefour, and to a
lesser extent in \fonethreetwo\ and \fonethreethree) the velocities
($\langle\vtsig\rangle$) exceed those observed in local starbursts
\citep[e.g.,][]{rupke05b}. Previous studies have found that velocities
of kpc-scale outflows in AGN reach the highest velocities
($\sim$1000~\kms\ or higher) only at quasar-like luminosities
($\sim$10$^{45}$ erg s$^{-1}$), and that there is a correlation
between outflow velocity and AGN luminosity
\citep{sturm11a,spoon13a,veilleux13a,rupke13a,harrison14a,cicone14a,stone16a,gonzalezalfonso17a,sun17a,fiore17a}. Figure~\ref{fig:vel_v_lagn}
shows how velocity depends on AGN luminosity in IFS observations of
ionized and neutral gas velocities in nearby Type 1 and 2 quasars and
luminous starbursts (\citealt{rupke13a} and this study). While there
is still evidence that the highest outflow velocities only appear in
quasars, there are many quasars with velocities comparable to those in
starbursts. No correlation is evident. However, the dynamic range in
luminosity of the current sample is relatively small.

\begin{figure*}
  \subfloat{\includegraphics[width=0.5\textwidth]{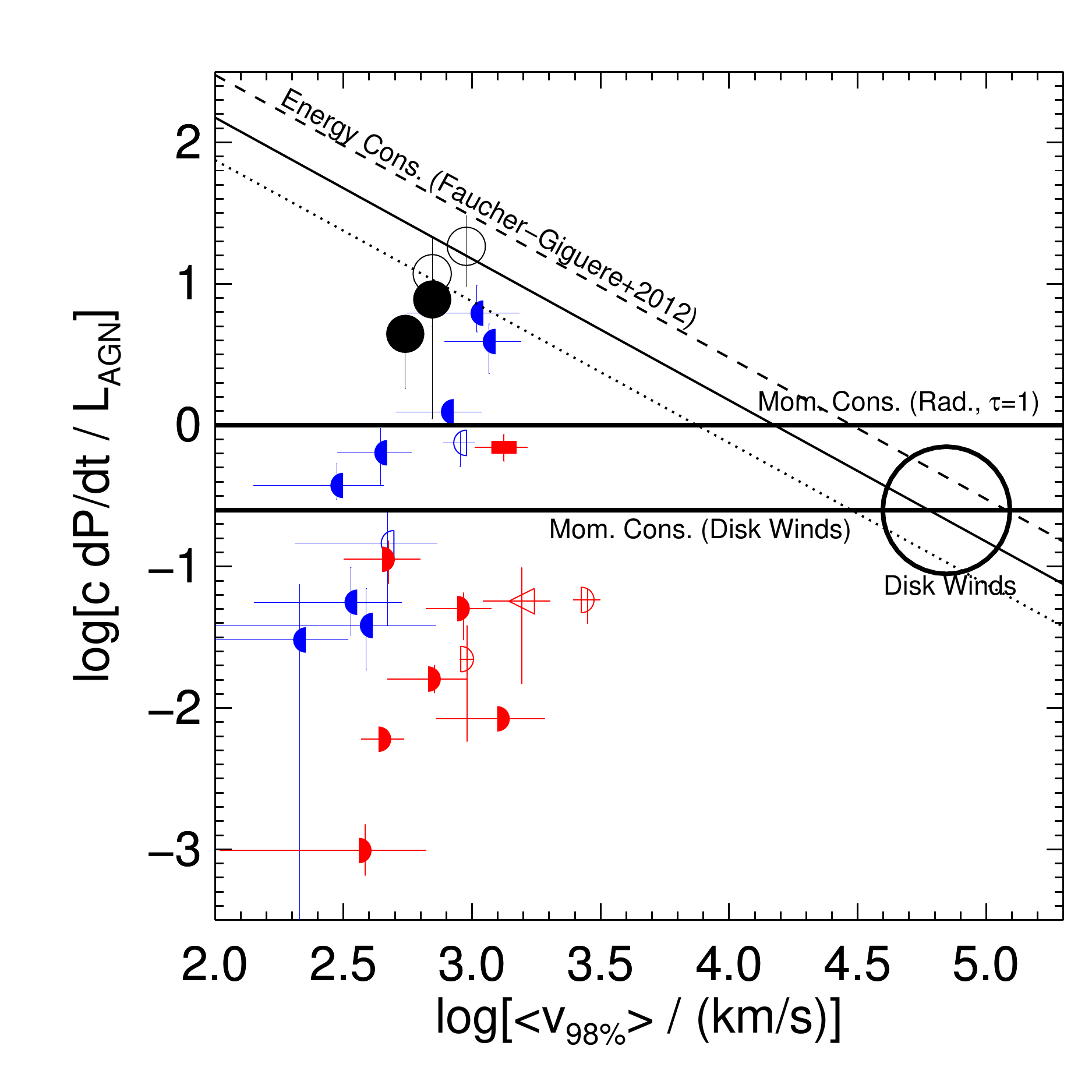}}
  \subfloat{\includegraphics[width=0.5\textwidth]{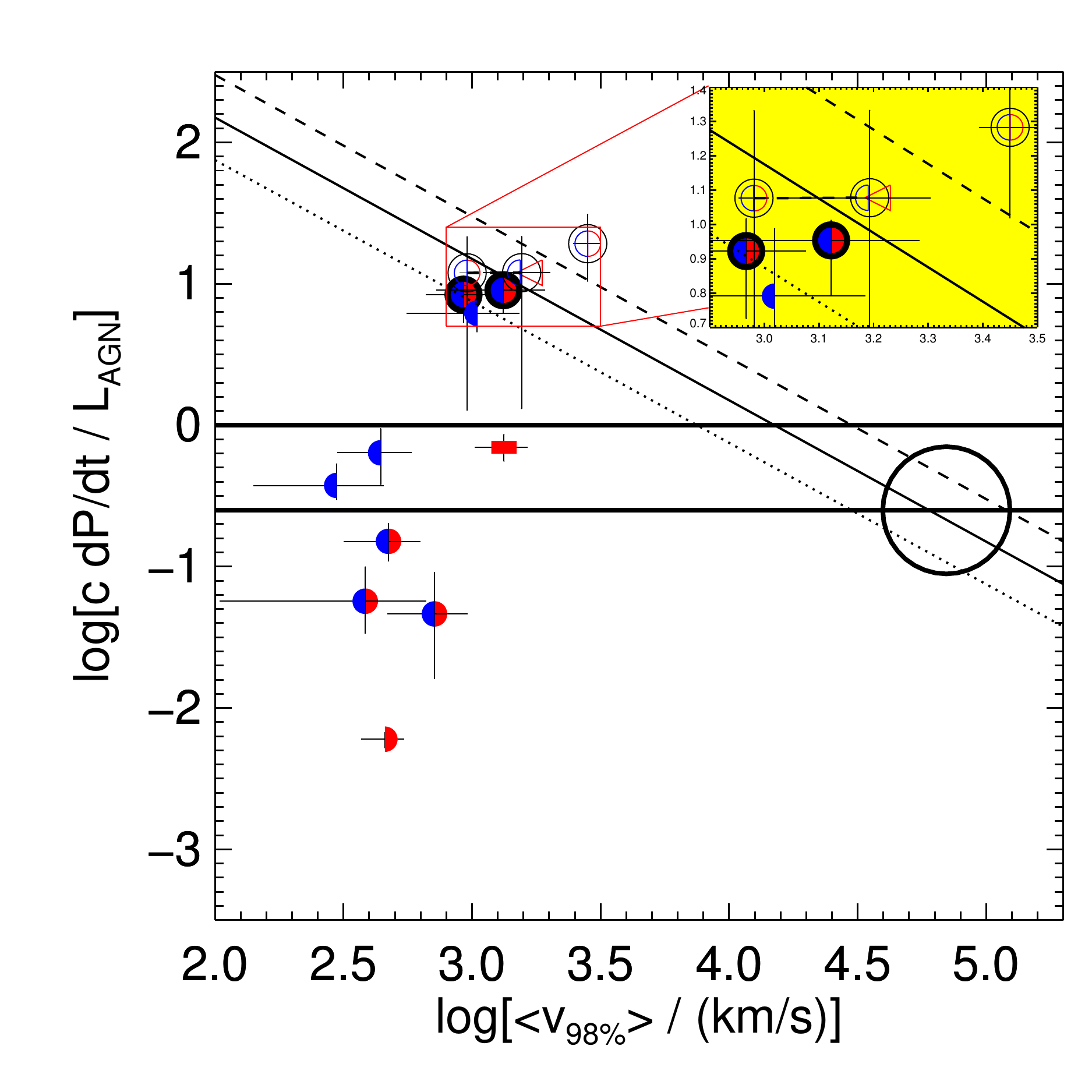}}
  \caption{Spatially-averaged velocities (\vtsig) vs.\ momentum
    outflow rate normalized to AGN luminosity. The same galaxies as in
    Figure~\ref{fig:vel_v_lagn} are displayed, but the starbursts have
    been removed. On the left, individual phases are shown. On the
    right, the gas momenta have been added up across measured phases,
    and the maximum $\langle \vtsig \rangle$ across phases is
    used. Ionized gas measurements in the Type 2 quasars and in
    \fohseven\ are lower limits because extinction correction has not
    been performed. Models for energy-conserving winds driven by blast
    waves from high-velocity accretion disk winds are overplotted
    (velocities of 0.1$c$, 0.2$c$, and 0.5$c$;
    \citealt{fauchergiguere12a}), as well as predictions based on pure
    momentum conservation from accretion disk winds (the circle shows
    the locus of representative accretion disk winds;
    \citealt[e.g.,][]{tombesi15a}) or from radiation pressure
    (assuming one photon scattering, though more photon scattering can
    arise from high infrared opacities). While some Type 1 and 2
    systems are consistent with energy-conserving models, half of the
    quasars have lower momentum boosts by 1 to several orders of
    magnitude. This suggests either a loss of wind momentum and energy
    or a different driving mechanism for the quasar wind.}
  \label{fig:vel_v_dpdt}
\end{figure*}

\capstartfalse
\setlength{\tabcolsep}{3pt}
\begin{deluxetable*}{ccrrrrrrr}
  \tablecaption{Power Sources and Feedback \label{tab:mperats_host}}
  \tabletypesize{\footnotesize}
  \tablewidth{\textwidth}

  \tablehead{ \colhead{Galaxy} & \colhead{phase} &
    \colhead{log[$(dM/dt)/$} & \colhead{log[$(dM/dt)/$} &
    \colhead{log[$(c~dp/dt)/$} & \colhead{log[$(dE/dt)/$} &
    \colhead{log[$(dM/dt)/$} &
    \colhead{log[$(c~dp/dt)/$} & \colhead{log[$(dE/dt)/$} \\
    \colhead{~} & \colhead{~} & \colhead{$SFR$]} &
    \colhead{$(dM/dt)_\mathrm{SB}$]} &
    \colhead{($3.5\times L_\mathrm{SB}$)]} &
    \colhead{$(dE/dt)_\mathrm{SB}$]} &
    \colhead{$(dM/dt)_\mathrm{acc}$]} & \colhead{$L_\mathrm{AGN}$]} &
    \colhead{$L_\mathrm{AGN}$]} \\
    \colhead{(1)} & \colhead{(2)} & \colhead{(3)} & \colhead{(4)} &
    \colhead{(5)} & \colhead{(6)} & \colhead{(7)} & \colhead{(8)} &
    \colhead{(9)} }

  \startdata
         I Zw 1  &neutral  &    0.40  &    1.38  &    0.12  &   -0.28  &    1.51  &   -0.43  &   -3.48 \\
    F05189-2524  &neutral  &    0.12  &    1.10  &    0.44  &    0.28  &    2.08  &    0.59  &   -2.08 \\
        \nodata  &ionized  &   -1.47  &   -0.49  &   -1.45  &   -1.99  &    0.49  &   -1.30  &   -4.34 \\
        \nodata  &molecular  &    0.57  &    1.55  &    0.50  &   -0.10  &    2.53  &    0.65  &   -2.45 \\
        \nodata  &  total  &    0.70  &    1.68  &    0.77  &    0.43  &    2.66  &    0.92  &   -1.92 \\
    F07599+6508  &neutral  &   -0.89  &    0.09  &   -1.32  &   -1.78  &    0.98  &   -1.25  &   -4.23 \\
        \nodata  &ionized  &   -2.79  &   -1.82  &   -3.07  &   -3.64  &   -0.93  &   -3.01  &   -6.09 \\
        \nodata  &  total  &   -0.88  &    0.10  &    0.50  &   -1.78  &    1.03  &   -1.25  &   -4.22 \\
        Mrk 231  &neutral  &   -0.24  &    0.74  &   -0.06  &   -0.33  &    1.72  &    0.09  &   -2.68 \\
        \nodata  &ionized  &   -2.54  &   -1.56  &   -2.23  &   -2.50  &   -0.58  &   -2.08  &   -4.85 \\
        \nodata  &molecular  &    0.80  &    1.78  &    0.73  &    0.24  &    2.76  &    0.89  &   -2.12 \\
        \nodata  &  total  &    0.84  &    1.82  &    0.80  &    0.34  &    2.80  &    0.95  &   -2.01 \\
    F13218+0552  &neutral  &   -1.15  &   -0.17  &   -1.37  &   -2.04  &    0.50  &   -1.52  &   -4.70 \\
        \nodata  &ionized  &   -1.05  &   -0.07  &   -1.64  &   -2.65  &    0.60  &   -1.80  &   -5.31 \\
        \nodata  &  total  &   -0.80  &    0.18  &    0.74  &   -1.94  &    0.91  &   -1.33  &   -4.61 \\
    F13342+3932  &neutral  &   -1.44  &   -0.46  &   -1.61  &   -1.77  &    0.55  &   -1.42  &   -4.09 \\
        \nodata  &ionized  &   -0.78  &    0.20  &   -1.14  &   -2.05  &    1.21  &   -0.95  &   -4.37 \\
        \nodata  &  total  &   -0.70  &    0.28  &    0.74  &   -1.58  &    1.32  &   -0.82  &   -3.91 \\
     PG1411+442  &neutral  & \nodata  & \nodata  & \nodata  & \nodata  &    2.07  &    0.79  &   -1.71 \\
     PG1613+658  &ionized  &   -1.46  &   -0.48  &   -2.11  &   -2.83  &    0.00  &   -2.22  &   -5.68 \\
    PG1700+518   &ionized  &    0.23  &    1.21  &   -0.10  &   -0.03  &    1.50  &   -0.16  &   -3.07 \\
    F21219-1757  &neutral  &   -0.12  &    0.86  &   -0.18  &   -0.71  &    1.67  &   -0.20  &   -3.24
  \enddata

  \tablecomments{Column 2: Gas phase. Column 3: Logarithm of mass
    outflow rate normalized to the star formation rate. Column 4:
    Logarithm of the mass outflow rate divided by the hot gas mass
    production rate from a continuous starburst
    \citep{leitherer99a}. Column 5: Logarithm of the momentum outflow
    rate divided by the momentum injection rate from a continuous
    starburst
    \citep{leitherer99a,veilleux05a,heckman15a,gonzalezalfonso17a}. Column
    6: Logarithm of the energy outflow rate divided by the mechanical
    energy production rate from a continuous starburst
    \citep{leitherer99a}. Column 7: Logarithm of the mass outflow rate
    divided by the accretion rate, calculated using a 10\%\ efficiency
    of radiative energy released by accretion. Column 8: Logarithm of
    the momentum outflow rate divided by the momentum input rate if
    the AGN bolometric luminosity accelerates the wind and each photon
    is intercepted once. Column 9: Logarithm of the energy outflow
    rate divided by the AGN luminosity.}

\end{deluxetable*}
\setlength{\tabcolsep}{6pt}
\capstarttrue

To further quantify how these outflows relate to their potential power
sources, Table \ref{tab:mperats_host} lists ratios of wind properties
to properties of the star-forming host or AGN.

To compare to the star-forming host, the mass entrainment efficiency
$\eta$ \citep{rupke05b} is computed as $dM/dt$ normalized to the star
formation rate, which comes in turn from the AGN fraction in
Table~\ref{tab:sample} and the total infrared luminosity
\citep{veilleux09a}. The mass, momentum, and energy outflow rates are
also compared to predicted rates from the starburst itself
\citep{leitherer99a,heckman15a,gonzalezalfonso17a}. (The AGN fraction
for \pgonefour\ is formally 100\%, so these quantities are not
calculated.) $\eta$ ranges from 0.03 to 7, with a median value of
0.8. This median is a factor of several higher than in the purely
neutral phase of starburst-driven winds in ULIRGs
\citep{rupke05b}. The median increase in wind mass from entrainment,
compared to the expected input from starbursts, is a factor of 7,
which is again higher than in ULIRG starbursts \citep{rupke05b}. 7/9
systems require more momentum than can be input by the starburst
(without multiple photon scatterings from the radiation field), and
one of the remaining two is AGN-driven (\pgonesix). 3/9 require more
energy than produced by the starburst, and 5/9 require more than 20\%\
of the starburst energy (meaning very little energy can be lost to
radiation or dissipative shocks). Two systems, \fohseven\ and
\fonethreetwo, have winds that could plausibly be explained by
starbursts. Of the four remaining systems that require $<10\%$ of the
starburst energy, one is clearly an AGN outflow (\pgonesix) and in
another the ionized outflow rates are lower limits.

This conclusion that the large-scale outflows in this sample are
largely powered by the AGN rather than star formation are strengthened
by comparing to starbursts of similar mass. Though the errors on the
slopes are large, the observed correlations between quasar$+$Seyfert
outflow rates and $\sigma$ may be steeper than the correlations
between log($dM/dt$) or log($dE/dt$) and logarithm of circular
velocity, log($v_\mathrm{c}$), determined for starbursts (slopes
2.9$\pm$0.7 and 4.64$\pm$0.8, respectively;
\citealt{rupke05b}). Figures~\ref{fig:dmdt_v_sigma} and
\ref{fig:dedt_v} show the relationship between log($dM/dt$) and
log($v_\mathrm{c}$) for starbursts, after applying the definition
$v_\mathrm{c}^2 \equiv 2\sigma^2 + v_\mathrm{rot}^2$. The result is
insensitive to the choice of $v_\mathrm{rot}/2\sigma$, which is 0.42
on average for ULIRGs \citep{dasyra06b} and 0.58 for the Seyferts in
Table~\ref{tab:seyferts} (based on HyperLeda data). The observed
best-fit steep correlations with $\sigma$ imply that while NLR
outflows in nearby Seyferts are not too different from their starburst
counterparts, galaxies hosting luminous black holes are ejecting mass
and energy at rates many times those of starburst hosts of similar
$\sigma$.

The conclusion is that most or all of the observed outflows are
AGN-driven; however, star formation could play a role in driving the
lowest-velocity and/or lowest-energy outflows.

\subsection{Comparison to Quasar Wind Models} \label{sec:models}

Models for the powering of quasar-driven, large-scale outflows can be
coarsely grouped into two categories: energy-conserving blast waves
powered by small-scale accretion disk winds
\citep[e.g.,][]{king11a,fauchergiguere12a,zubovas12a,wagner13a,hopkins16a,gaspari17a}
and momentum-conserving winds driven by multiple photon scatterings in
dusty gas with high optical depths in the infrared
\citep[e.g.,][]{thompson14a,ishibashi15a,costa17a}. Besides the high
observed velocities in quasar-mode winds, both types of models have
striven to explain large ``momentum boosts'': the momentum outflow
rate is factors of a few to a few tens larger than the outflow rate
expected from pure radiation pressure driving assuming one photon
scattering
\citep{sturm11a,veilleux13a,spoon13a,rupke13a,cicone14a,stone16a,gonzalezalfonso17a}. However,
these observational studies have contained few examples of Type 1
quasars.

To facilitate comparison to wind models and to assess feedback
(\S\,\ref{sec:feedback}), the outflow rate is compared to the
accretion rate, and the momentum and energy outflow rate are compared
to the AGN luminosity, in Table~\ref{tab:mperats_host}.

The momentum boosts in the neutral and ionized phases of nearby quasar
winds (Table~\ref{tab:mperats_host} and Figure \ref{fig:vel_v_dpdt})
show a wide dynamic range, from $(c~dp/dt)/L_\mathrm{AGN} = 0.001$ to
6. For comparison, molecular gas measurements for this sample are
taken to be total momentum outflow rates and maximum velocities from
\citet{gonzalezalfonso17a}. Momentum boosts tend to be higher, but
velocities lower, in the molecular phase of individual systems.  When
all phases are summed, half of the IFS quasar sample shows
$(c~dp/dt)/L_\mathrm{AGN} > 1$, with a range of 0.01--20.

Roughly half the sample lies 2--3 orders of magnitude below
predictions of the momentum boost from energy-conserving blast wave
models \citep{fauchergiguere12a}. This implies either that the blast
wave is not energy-conserving in these systems, that some other
mechanism drives the wind (such as radiation pressure), or that
another phase dominates the momentum budget. Measurements of one of
these systems, \zw, have not detected a molecular outflow, though the
neutral outflow velocities in this system are comparable to the disk
velocities and the outflow could be hidden in the core of a molecular
line profile \citep{veilleux13a,cicone14a}.

Comparison of the observed velocities at a given $L_\mathrm{AGN}$
(Figure~\ref{fig:vel_v_lagn}) with energy-conserving models
\citep{zubovas12a} show that the models can reproduce the
spatially-averaged highest velocities in most, though not all,
quasars. These models also match the best-fit slope in the $dM/dt$
vs. $M_\mathrm{BH}$ plane (Figure~\ref{fig:dmdt_v_mbh}): 0.63 (theory)
for 0.68 (observed). However, these models over-predict the average
velocity in most systems and the highest velocities in some Type 1
quasars. They also over-predict the observed mass outflow rates by 1--2
orders of magnitude (Figure~\ref{fig:dmdt_v_mbh}). However, the
observed mass outflow boost in AGN-driven winds compared to
starburst-only winds (0.01--10; \S\,\ref{sec:powersource}) is of the
same order of magnitude as predicted by some simulations
\citep{hopkins16a}.

\begin{figure}
  \includegraphics[width=0.9\columnwidth,center]{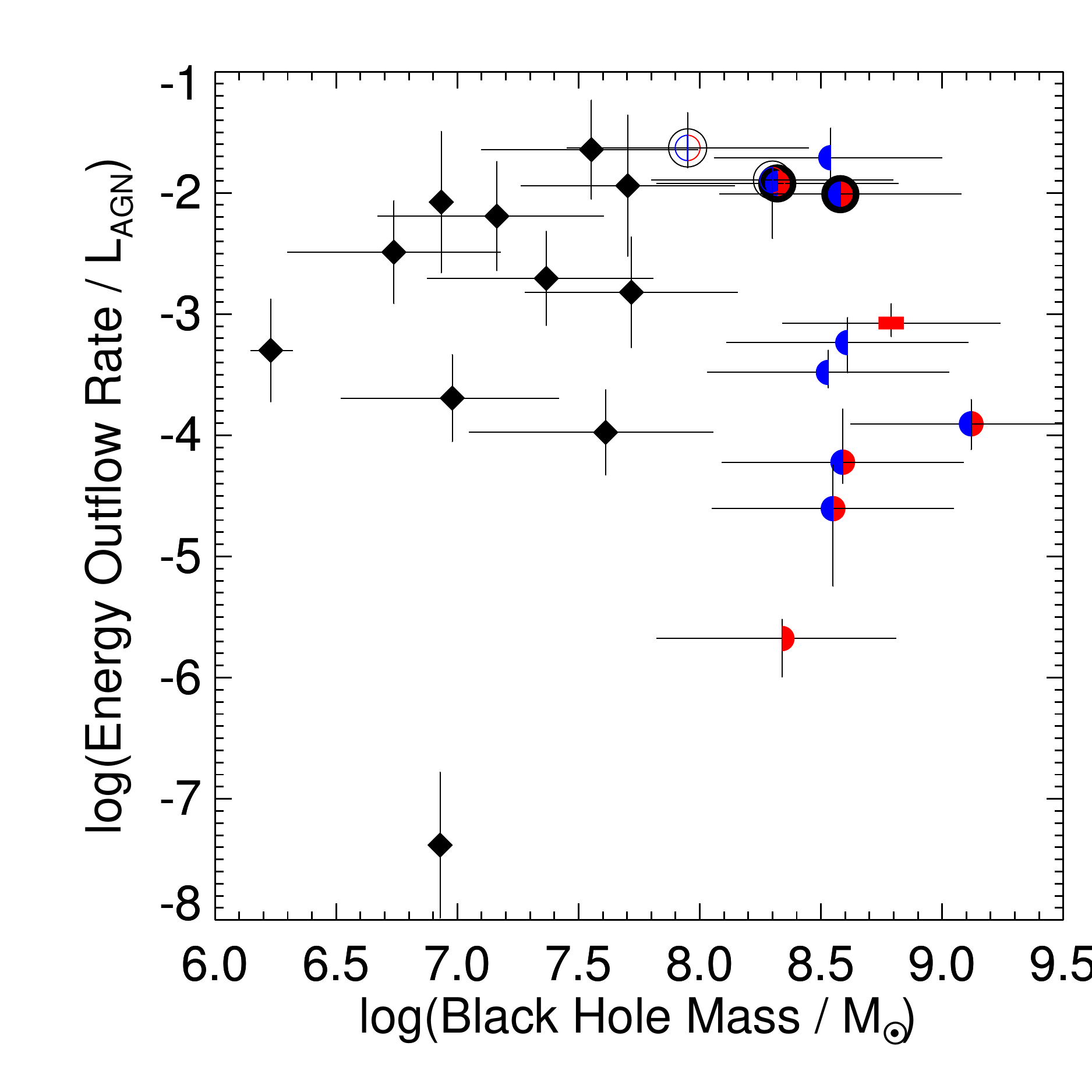}
  \caption{Energy outflow rate divided by (bolometric) AGN luminosity
    as a function of black hole mass. There is no dependence on
    $M_\mathrm{BH}$, and the values are consistent with (or lower
    than) values inferred from or assumed by current models and
    simulations
    \citep[e.g.,][]{hopkins10a,zubovas12a,hopkins16a}. Though the
    plots are not shown, there is similarly no dependence on
    $L_\mathrm{AGN}$ or Eddington ratio.}
  \label{fig:dedtlagn}
\end{figure}

The efficiencies with which the AGNs power these winds are high in the
sense that the mechanical energy required is typically a very small
fraction of the AGN luminosity. The range of $(dE/dt)/L_\mathrm{AGN}$
for both Seyferts and quasars is large, with most sources in the range
0.002\%\ to 2\%; such values are low enough for models to successfully
drive AGN outflows \citep[e.g.,][]{hopkins10a,zubovas12a} and are
similar to efficiencies assumed in numerical simulations
\citep[e.g.,][]{hopkins16a}. There is no correlation of this
efficiency measure with $M_\mathrm{BH}$ (Figure~\ref{fig:dedtlagn}),
$L_\mathrm{AGN}$, or Eddington ratio.

The conclusion is that it is difficult to distinguish among competing
models or constrain individual models. However, none of these models
make predictions about how outflow properties scale with black hole
mass -- such predictions would optimally leverage the results of the
current study.

\subsection{Kpc-Scale Quasar-Mode Winds as AGN
  Feedback} \label{sec:feedback}

A key remaining question is how these large-scale, quasar-driven
outflows will affect the star formation and AGN activity in their
hosts. This question is difficult to answer conclusively with the
current data.

It is clear that in an absolute sense, the most massive black holes
produce the most massive and energetic outflows. High-velocity,
high-momentum boost feedback, as observed in part of this sample, may
be the most effective at shaping the $M_\mathrm{BH}$--$\sigma$
relation at high black hole masses in simulations, as well
\citep{anglesalcazar17a}. In essence, these outflow rates illustrate
how well the black hole, through its current accretion rate, is moving
away gas that could at some future time feed the black hole. This
future potential depends on (1) the uncertain timescales for gas at
large radii to inflow to the accretion disk and then to the black hole
and (2) what fraction of gas at large radii will actually fall into
the black hole at some future time.


\begin{figure*}
  \subfloat{\includegraphics[width=0.5\textwidth]{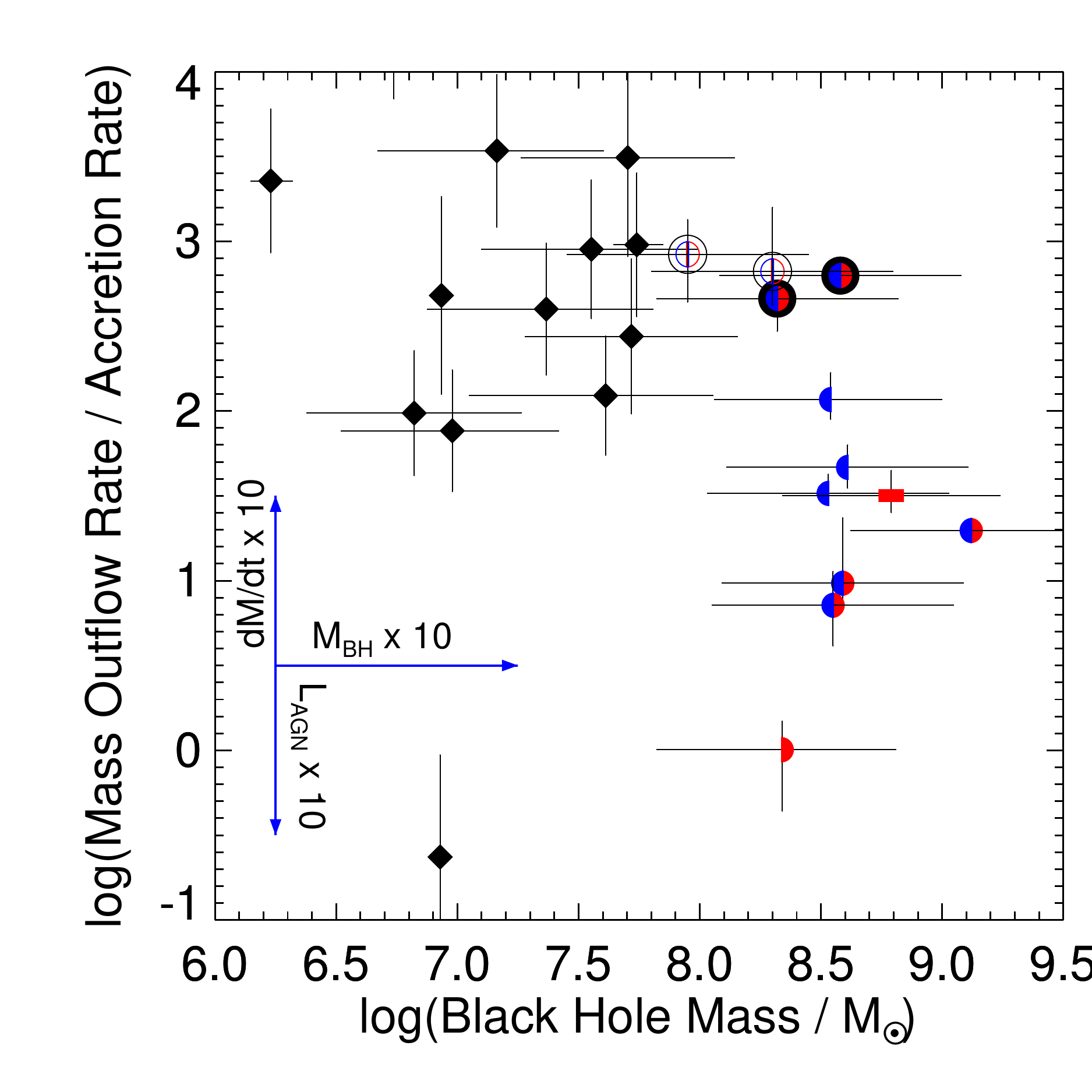}}
  \subfloat{\includegraphics[width=0.5\textwidth]{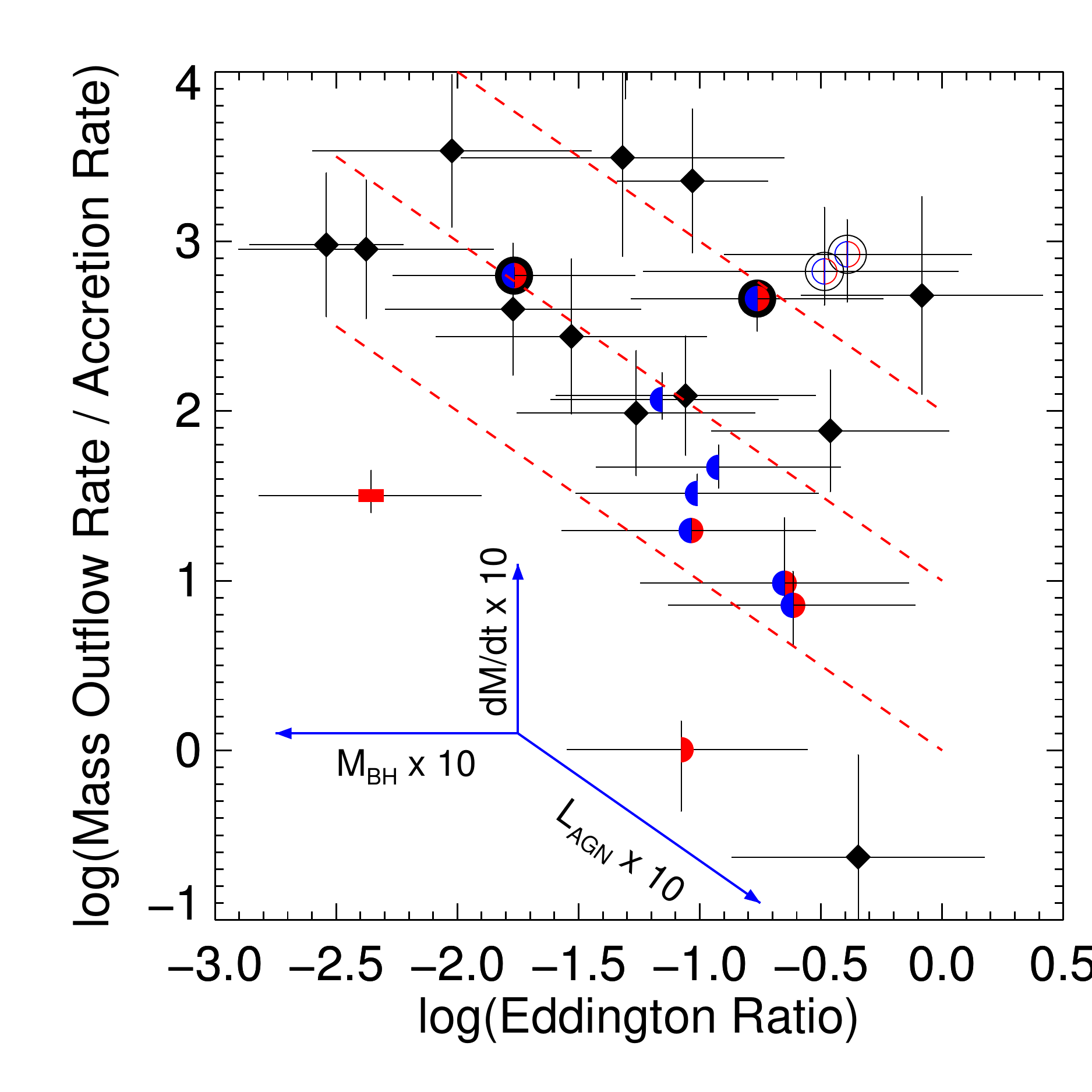}}
  \caption{Mass outflow rates, normalized to the accretion rates of
    individual AGNs (assuming 10\%\ efficiency of radiative energy
    released by accretion; Table \ref{tab:mperats_host}), as a
    function of black hole mass and Eddington ratio. Arrows show how
    an order of magnitude change in mass outflow rate, black hole
    mass, or AGN luminosity affects the position of points. The ratio
    of mass out (on large scales) to mass in (to the black hole) may
    decrease with increasing black hole mass and increasing Eddington
    ratio. Another possible interpretation is that the radiative
    efficiency increases with black hole mass, such that more massive
    black holes have lower accretion rates for a given luminosity.
    Finally, in the right panel, dashed lines of constant $dM/dt$ and
    $M_\mathrm{BH}$, but varying $L_\mathrm{AGN}$, show that accretion
    variations (reflected in changes in $L_\mathrm{AGN}$) on
    timescales much shorter than the outflow dynamical time could
    explain some of the scatter or apparent correlation.}
  \label{fig:dmdtrat}
\end{figure*}

A possible relative measure of feedback on the AGN itself is the mass
outflow rate divided by the present black hole accretion rate. In the
quasar$+$Seyfert sample, this ratio is mostly in the range 1--1000,
which is larger than at radii $<$100~pc in the simulations of
\citet{hopkins16a}. Figure~\ref{fig:dmdtrat} shows how
$\frac{(dM/dt)_\mathrm{out}}{(dM/dt)_\mathrm{acc}}$ depends on black
hole mass and Eddington ratio. There is not a significant correlation
with $M_\mathrm{BH}$. However, the Seyfert subsample on average has
higher $\frac{(dM/dt)_\mathrm{out}}{(dM/dt)_\mathrm{acc}}$ and lower
$M_\mathrm{BH}$ than the quasar subsample. The correlation with
Eddington ratio is almost 2$\sigma$ (using Bayesian fitting as in
\S\,\ref{sec:bhprop}), with a median slope of -1.0 and a 95.5\%\ range
of $-$2.4 to 0.1. However, the Bayesian test does not account for the
fact that the variables are not completely independent, since both
depend on $L_\mathrm{AGN}$ ($dM/dt_\mathrm{acc}\sim L_\mathrm{AGN}$
and Eddington ratio is proportional to $L_\mathrm{AGN}$).

We conclude that there is weak evidence of inverse correlations of
$\frac{(dM/dt)_\mathrm{out}}{(dM/dt)_\mathrm{acc}}$ with
$M_\mathrm{BH}$ and Eddington ratio. If such correlations exist, then
one consequence could be that relative feedback (gas out on large
scales over gas in at small scales) is higher in systems with lower
Eddington ratio and lower black hole mass. Alternatively, it could be
that the assumption of a constant radiative efficiency from accretion
(how well the black hole converts accretion energy into luminosity) is
incorrect. Instead, the efficiency may be higher in higher-mass black
holes \citep{davis11a}. Higher radiative efficiency means a lower
accretion rate $(dM/dt)_\mathrm{acc}$ produces the same luminosity
$L_\mathrm{AGN}$, which could remove any correlation of
$\frac{(dM/dt)_\mathrm{out}}{(dM/dt)_\mathrm{acc}}$ with
$M_\mathrm{BH}$. A higher efficiency could result from higher black
hole spin, since in standard thin-disk accretion \citep{shakura73a}
the efficiency depends on the spin \citep{reynolds13a}. However,
current measurements of black hole spin do not show higher spin in
higher mass black holes \citep{reynolds13a}.

The issue of timescale is also an important consideration. Quasar
outflows are powered by current and past AGN activity. The dynamical
timescales of these flows are of order 1--100 Myr, while AGN accretion
may have a smaller duty cycle of $\sim$0.1~Myr
\citep[e.g.,][]{schawinski15a,king15a}. Thus, the accretion rate (and
thus the AGN luminosity and Eddington ratio) may change significantly
on shorter timescales than the outflow. Some of the scatter (and the
correlation itself) in the right panel of Figure~\ref{fig:dmdtrat}
could result from short timescale variations of $L_\mathrm{AGN}$ if
$dM/dt_\mathrm{out}$ and $M_\mathrm{BH}$ indeed vary on longer
timescales.

As with the other correlations found in this work, selection biases
likely exist, and larger samples are needed. For instance, the lower
regions of Figure~\ref{fig:dmdtrat} could be occupied by galaxies with
lower mass outflow rates but similar accretion rates and black hole
masses (e.g., the radio-mode feedback regime in
radiatively-inefficient accretion modes), which could also fill in
other parts of Figures~\ref{fig:dmdt_v_mbh}--\ref{fig:dedt_v}. An open
question is whether the distributions in outflow properties, taken
over all galaxies with accreting black holes, are continuous or
significantly bimodal. For instance, in a hypothetical on-or-off
model, outflows are either present and look like the ones present
here, or they are absent and any outflow rates are exactly zero. The
result would be a bimodal distribution in outflow rates. In this
on-or-off model, the correlations presented here may accurately
describe active feedback modes.

\section{SUMMARY} \label{sec:summary}

Quasar-mode feedback takes the form of kpc-scale, massive,
high-velocity winds driven by radiation and energetic winds from the
hearts of radio-quiet quasars. These winds potentially play a crucial
role in feedback that regulates the growth of supermassive black holes
and star formation in galaxies. This study reports on the first
multiphase study of such winds in a sample of nearby Type 1
quasars. The subsample, which is selected from the QUEST sample of
optically- and IR-selected quasars, consists of 10 quasars with
$z < 0.3$. The quasar PSF is carefully removed using a robust spectral
method that includes iterative fitting to each spaxel of a scaled PSF,
a stellar host continuum, multicomponent emission lines, and
multicomponent interstellar absorption lines. This method recovers a
high spatial resolution PSF and stellar, emission, and absorption line
properties even at radii less than the PSF FWHM.

Outflows of ionized and/or neutral gas are revealed by the presence of
broad, blueshifted absorption lines and broad, blueshifted emission
lines. Kpc-scale outflows (maximum size 3--12 kpc, with the size
measurement frequently limited by the instrument FOV) are found in the
entire sample. By phase, 70\%\ have ionized outflows, 80\%\ have
neutral outflows, and 50\%\ have both. The outflows are oriented
preferentially along the galaxy minor axes, and their properties are
most consistent with being powered by the central AGN. The ionized
outflows typically show a mix of stellar photoionization with AGN
photoionization, but shocks also sometimes play a role. The quasar
with the most massive and energetic outflow (\pgoneseven) arises due
to a kpc-scale radio jet, though the system is not formally
radio-loud.

There is a wide range of typical outflow
velocities. Spatially-averaged maximum velocities
($\langle\vtsig\rangle$) are in the range $200-1300$~\kms, while peak
velocities (max. \vtsig) are $500-2600$~\kms. A correlation with AGN
luminosity may exist in samples with a larger dynamic range of
$L_\mathrm{AGN}$, but there is no such correlation in this sample.

Outflowing masses, momenta, and energies are computed using the
single-radius free wind model. These data are combined with the
results of molecular gas outflow measurements, previous IFS
measurements of a few Type 2 quasars, and mass outflow rates from
nearby Seyferts. Quasar mass outflow rates are in the range 1 to
$>$1000~\smpy. The quasar ``momentum boosts'' range from 0.01 to 20,
with half of the sample having $(c~dp/dt)/L_\mathrm{AGN}\ga1$. Energy
outflow rates are primarily in the range 0.002\%\ to 2\%\ of
$L_\mathrm{AGN}$. When compared with properties of the star forming
hosts, and with other starbursts of similar mass, the data are most
consistent with the outflows being driven by the central AGN rather
than star formation (except in the lowest-velocity and lowest-energy
cases).

Previous studies have focused on how kpc-scale outflow properties
depend on host galaxy or AGN properties. The large range of black hole
masses (three orders of magnitude) covered by the current subsample
and previous studies of Seyferts allows an accounting of how
large-scale outflow properties depend on black hole mass, and thereby
accretion rate and Eddington ratio. Significantly, the total mass and
energy outflow rates summed over phases correlate at 2$\sigma$ with
black hole mass according to $dM/dt \sim M_\mathrm{BH}^{0.7\pm0.3}$
and $dE/dt \sim M_\mathrm{BH}^{1.3\pm0.5}$. They correlate similarly
with $\sigma$ in a way that is consistent with the
$M_\mathrm{BH}$--$\sigma$ relation.

Despite these correlations, the present data are unable to constrain
or distinguish among competing quasar wind models, and cannot provide
strong conclusions about quasar-mode winds as AGN feedback. It is
clear that the most massive black holes do have the most massive and
energetic winds. A possible relative feedback measure, the mass
outflow rate normalized to the accretion rate, decreases with
increasing black hole mass and Eddington ratio (though the correlation
significance is low). However, these tentative correlations may be
affected by variations of radiative efficiency with black hole mass or
variations in the accretion rate on shorter timescales than the
outflow dynamical time.

This study exemplifies the importance of deep, 3D observations of
large-scale outflows in quasars for gaining a complete picture of
their properties. Only by studying the multiphase (warm ionized$+$cool
neutral with optical IFS, cold molecular with mm interferometry, and
other phases) structure of these winds do we gain a full accounting of
the phase space (velocity$+$spatial geometry) and mass, momentum, and
energy budget of these winds. Furthermore, it demonstrates the power
of high-fidelity IFS for probing the central regions of PSF-dominated
systems.

Future work can improve on this study in several ways. (1) More IFS
observations of quasars. It is clear that the error bars on the
correlations presented here are limited by statistics -- the combined
quasar$+$Seyfert sample in this paper numbers only 23. (2) Deep,
wide-field spectroscopy to determine the wind extents. Measurements of
the full extent of the winds discussed here are largely limited by the
GMOS FOV. (3) Adaptive optics IFS of the unresolved PSF. This study is
limited to the large-scale outflows in these systems, but the
unresolved PSF contains higher-velocity outflows that couple to these
large-scale outflows. IFS observations at higher spatial resolution
are necessary to connect these two phenomena.

\acknowledgments The authors thank Yuval Birnboim, Fran\c{c}oise
Combes, Annette Ferguson, Fred Hamann, Bernd Husemann, Dusan Keres,
and James Turner for helpful comments and discussion. They also thank
the referee for their careful reading of the manuscript and helpful
comments to improve it. D.S.N.R. thanks Lisa Kewley and the GEARS3D
group at Australian National University for its hospitality while this
work was completed. The authors thank the Gemini Director and User's
Committee for awarding the observing time under which some of this
data was obtained.

This work was based on observations obtained at the Gemini Observatory
(program IDs GN-2003B-C-5, GS-2005B-Q-50, GN-2007A-Q-12,
GS-2011B-Q-64, GN-2012A-Q-15, GN2013A-Q-51, and GN-2015A-DD-9), which
is operated by the Association of Universities for Research in
Astronomy, Inc., under a cooperative agreement with the NSF on behalf
of the Gemini partnership: the National Science Foundation (United
States), the National Research Council (Canada), CONICYT (Chile),
Minist\'{e}rio da Ci\^{e}ncia, Tecnologia e Inova\c{c}\~{a}o (Brazil)
and Ministerio de Ciencia, Tecnolog\'{i}a e Innovaci\'{o}n Productiva
(Argentina).

D.S.N.R. was supported in part by the J. Lester Crain Chair of Physics
at Rhodes College and by a Distinguished Visitor grant from the
Research School of Astronomy \&\ Astrophysics at Australian National
University. S.V. was supported in part by NSF grant AST1009583 and
NASA grant ADAP NNX16AF24G.

The \hst\ observations described here were obtained from the Hubble
Legacy Archive, which is a collaboration between the Space Telescope
Science Institute (STScI/NASA), the Space Telescope European
Coordinating Facility (ST-ECF/ESA) and the Canadian Astronomy Data
Centre (CADC/NRC/CSA).

Revision 114 of the IFUDR GMOS package, based on the Gemini IRAF
package, was provided by James Turner and Bryan Miller via the Gemini
Data Reduction User Forum.

We acknowledge the usage of the HyperLeda database
(http://leda.univ-lyon1.fr) and the AGN Black Hole Mass Database
(http://www.astro.gsu.edu/AGNmass/; \citealt{bentz15a}).

\bibliography{apj-jour,dsr-refs}

\clearpage

\setcounter{figure}{0}
\begin{figure*}
  \includegraphics[left]{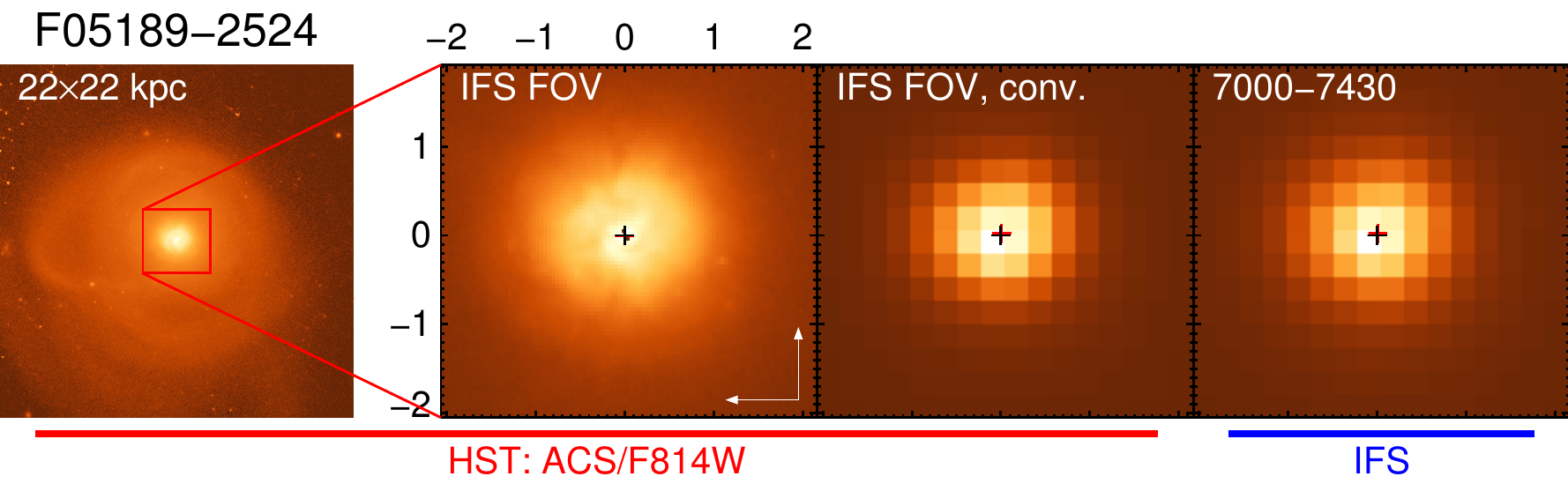}
  \caption{\it Continued.}
\end{figure*}
\setcounter{figure}{0}
\begin{figure*}
  \includegraphics[left]{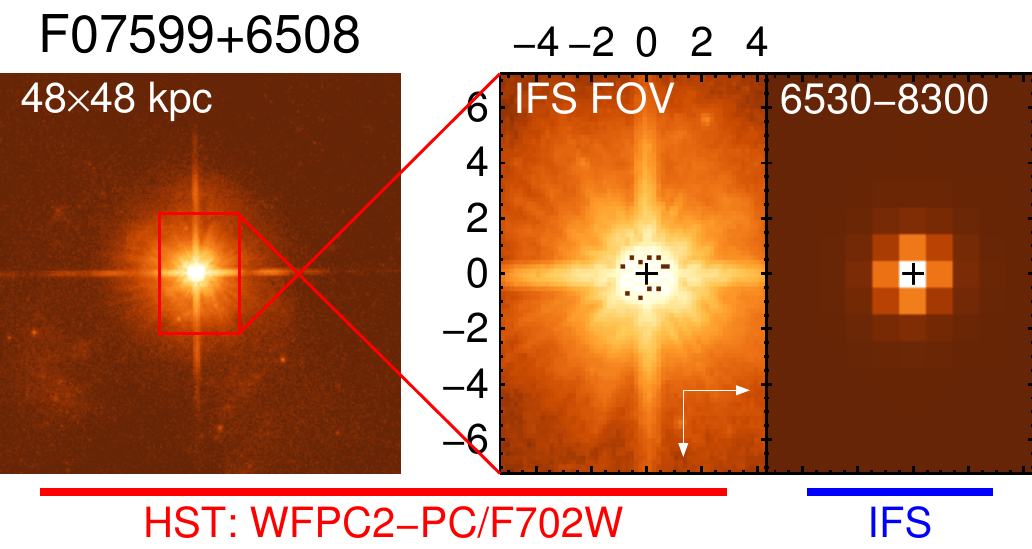}
  \caption{\it Continued.}
\end{figure*}
\setcounter{figure}{0}
\begin{figure*}
  \includegraphics[left]{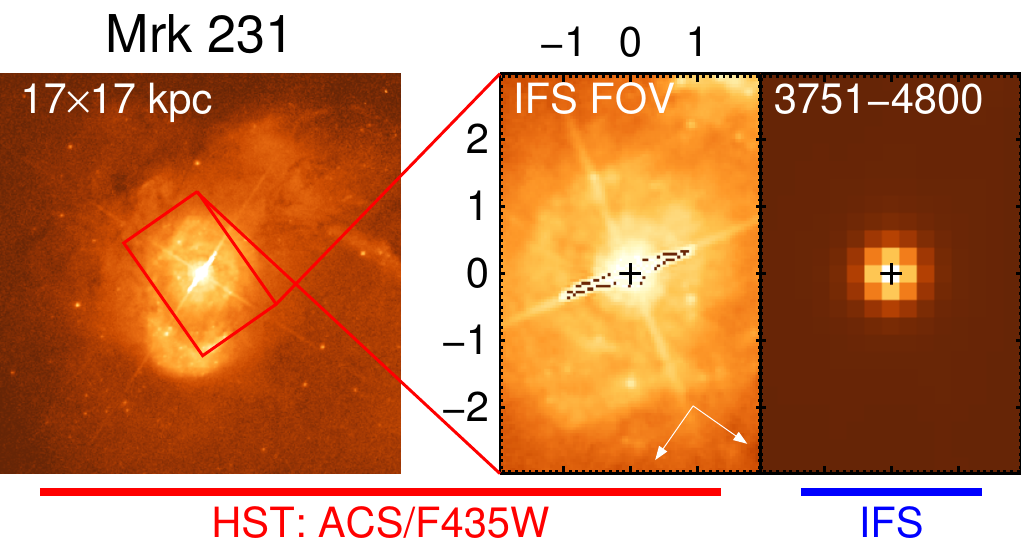}
  \caption{\it Continued.}
\end{figure*}
\setcounter{figure}{0}
\begin{figure*}
  \includegraphics[left]{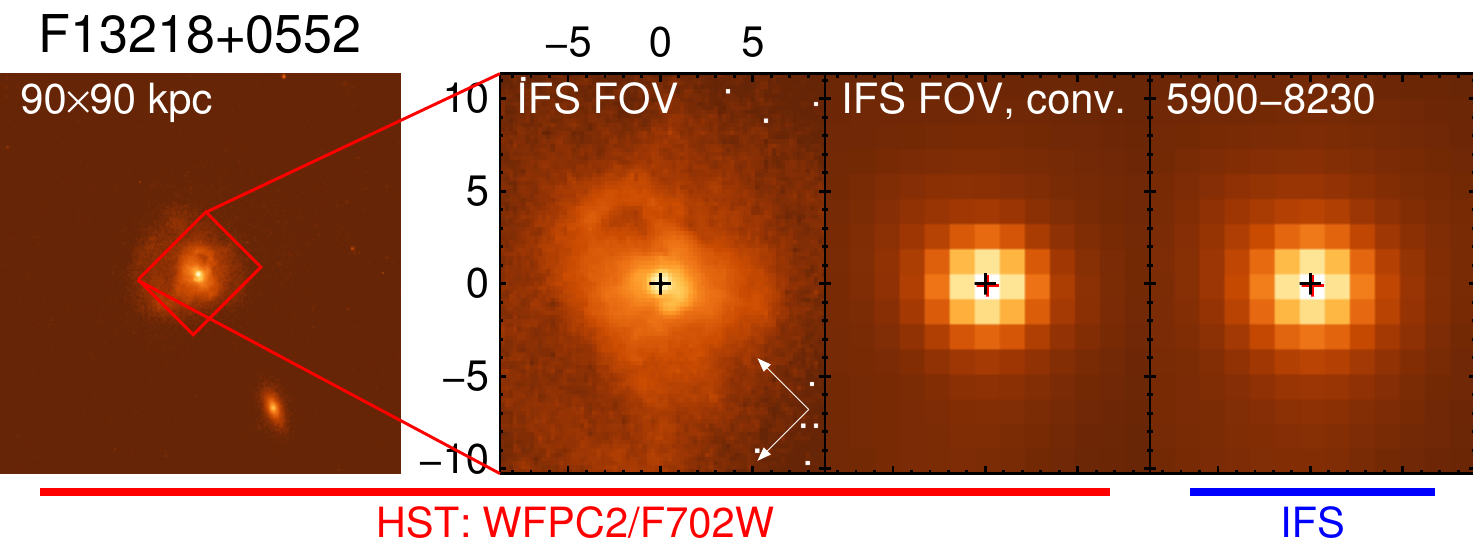}
  \caption{\it Continued.}
\end{figure*}
\setcounter{figure}{0}
\begin{figure*}
  \includegraphics[left]{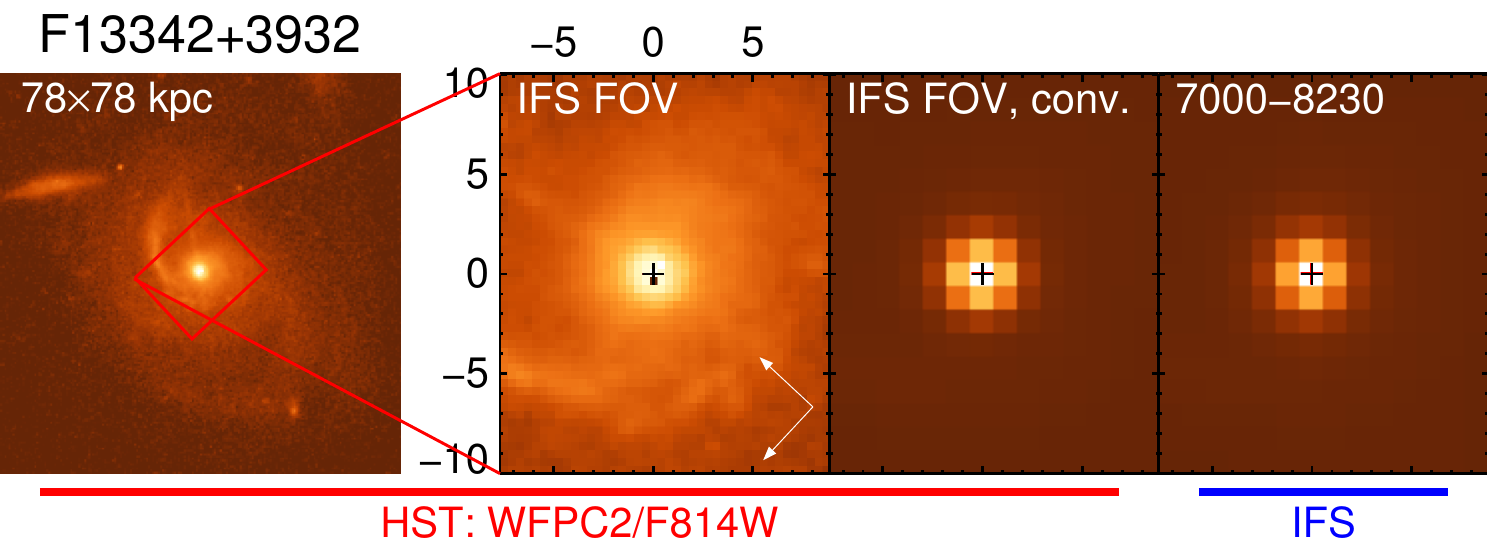}
  \caption{\it Continued.}
\end{figure*}
\setcounter{figure}{0}
\begin{figure*}
  \includegraphics[left]{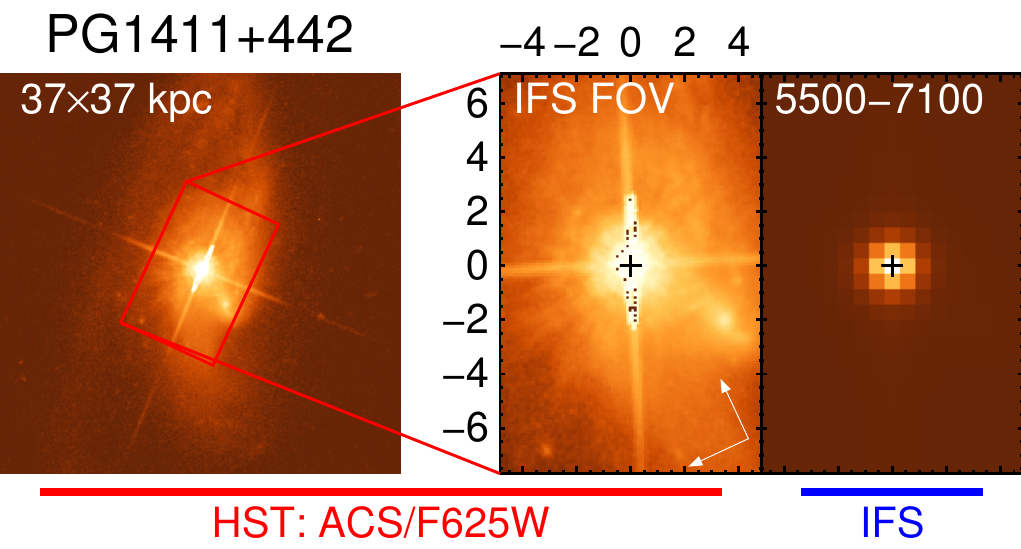}
  \caption{\it Continued.}
\end{figure*}
\setcounter{figure}{0}
\begin{figure*}
  \includegraphics[left]{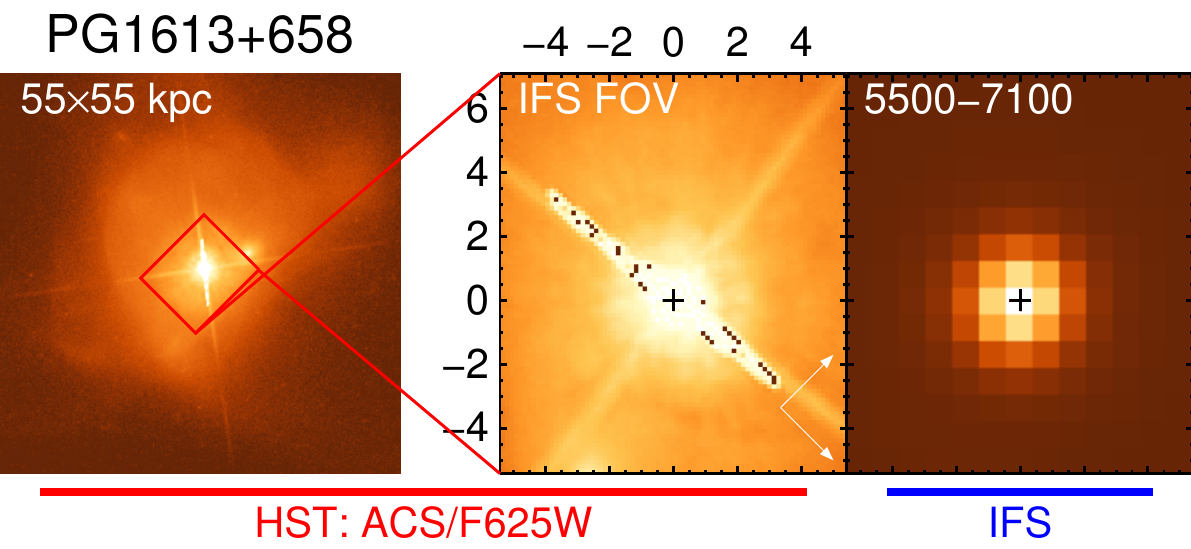}
  \caption{\it Continued.}
\end{figure*}
\setcounter{figure}{0}
\begin{figure*}
  \includegraphics[left]{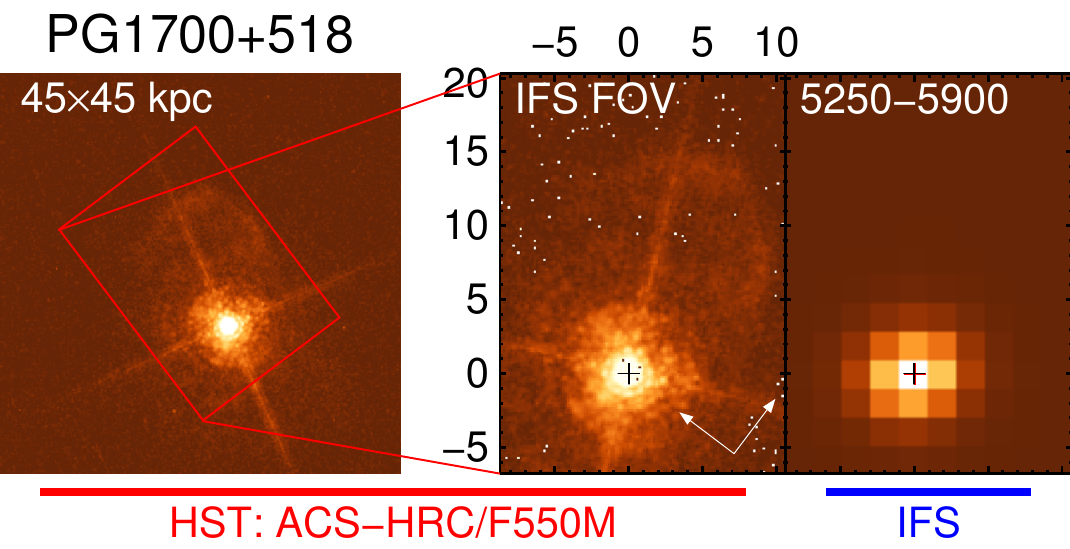}
  \caption{\it Continued.}
\end{figure*}
\setcounter{figure}{0}
\begin{figure*}
  \includegraphics[left]{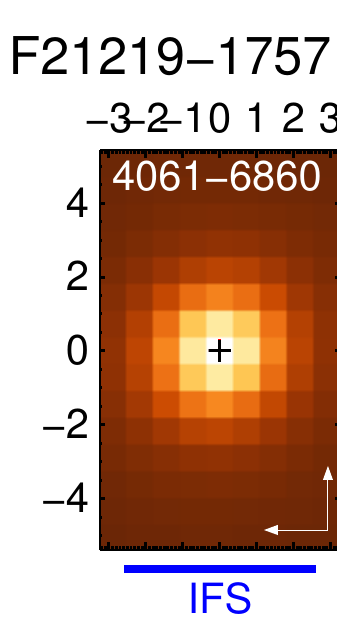}
  \caption{\it Continued.}
\end{figure*}

\clearpage

\setcounter{figure}{4}
\begin{figure*}
  \includegraphics[center]{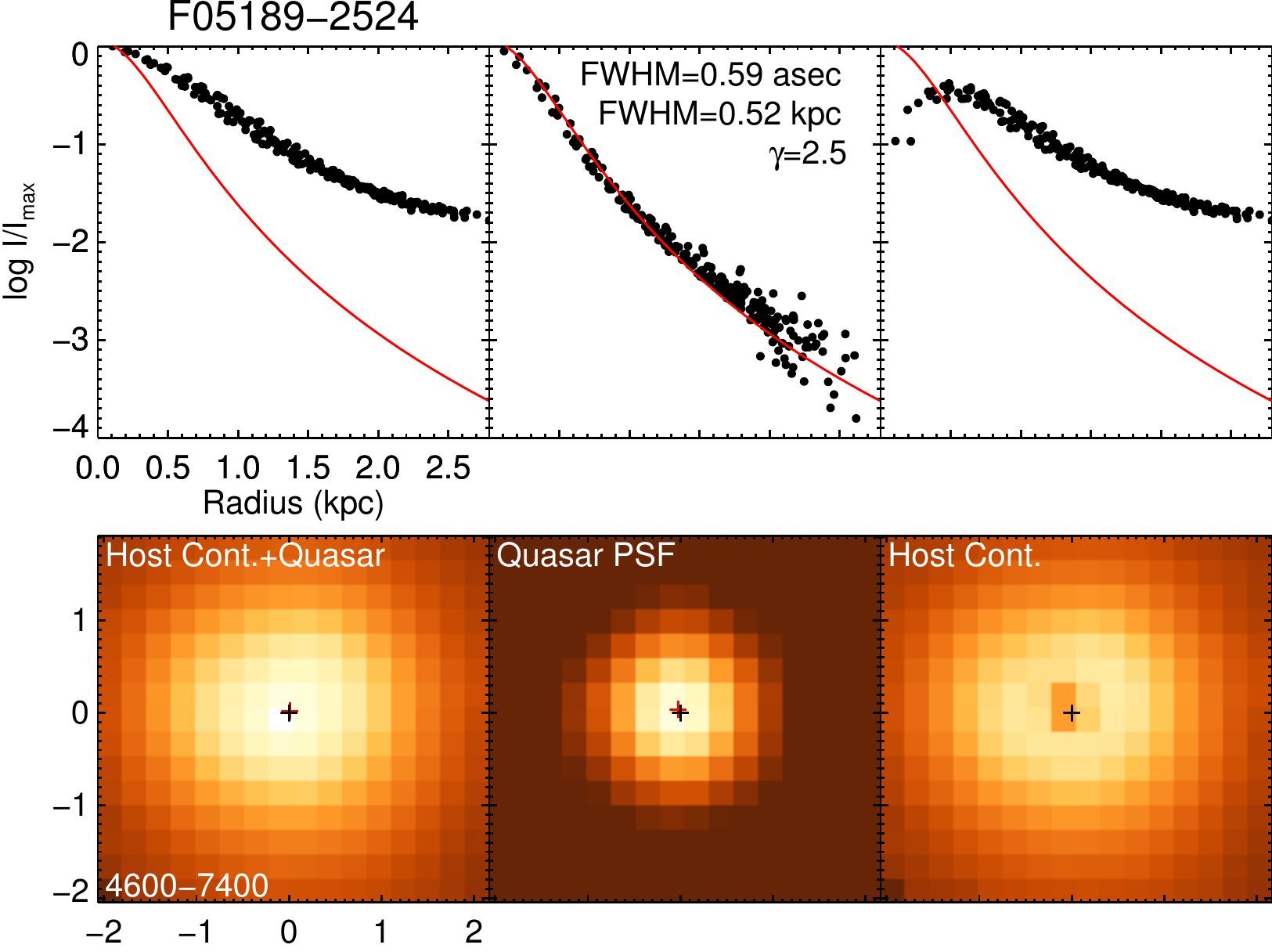}
  \caption{\it Continued.}
\end{figure*}
\setcounter{figure}{4}
\begin{figure*}
  \includegraphics[center]{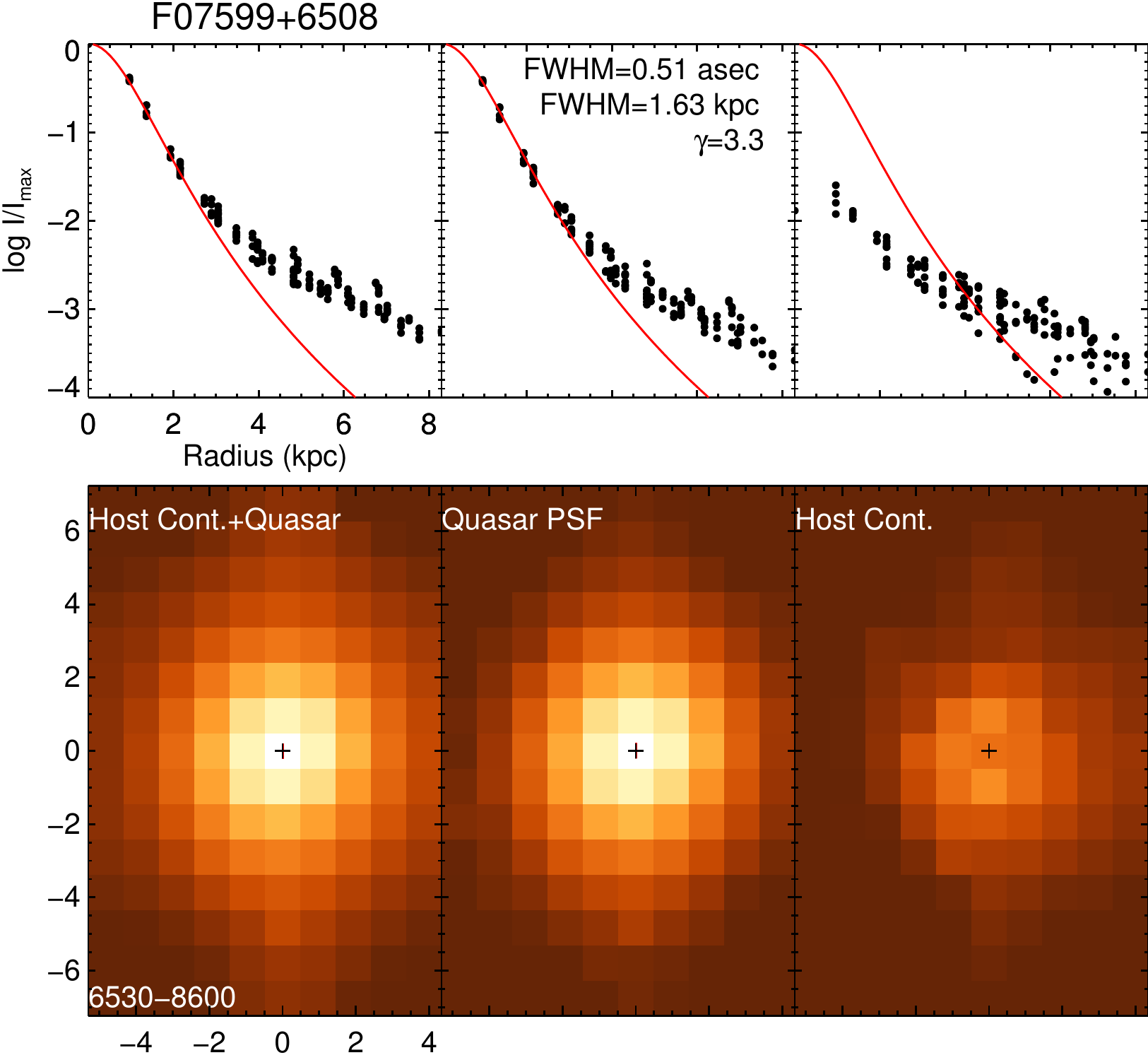}
  \caption{\it Continued.}
\end{figure*}
\setcounter{figure}{4}
\begin{figure*}
  \includegraphics[center]{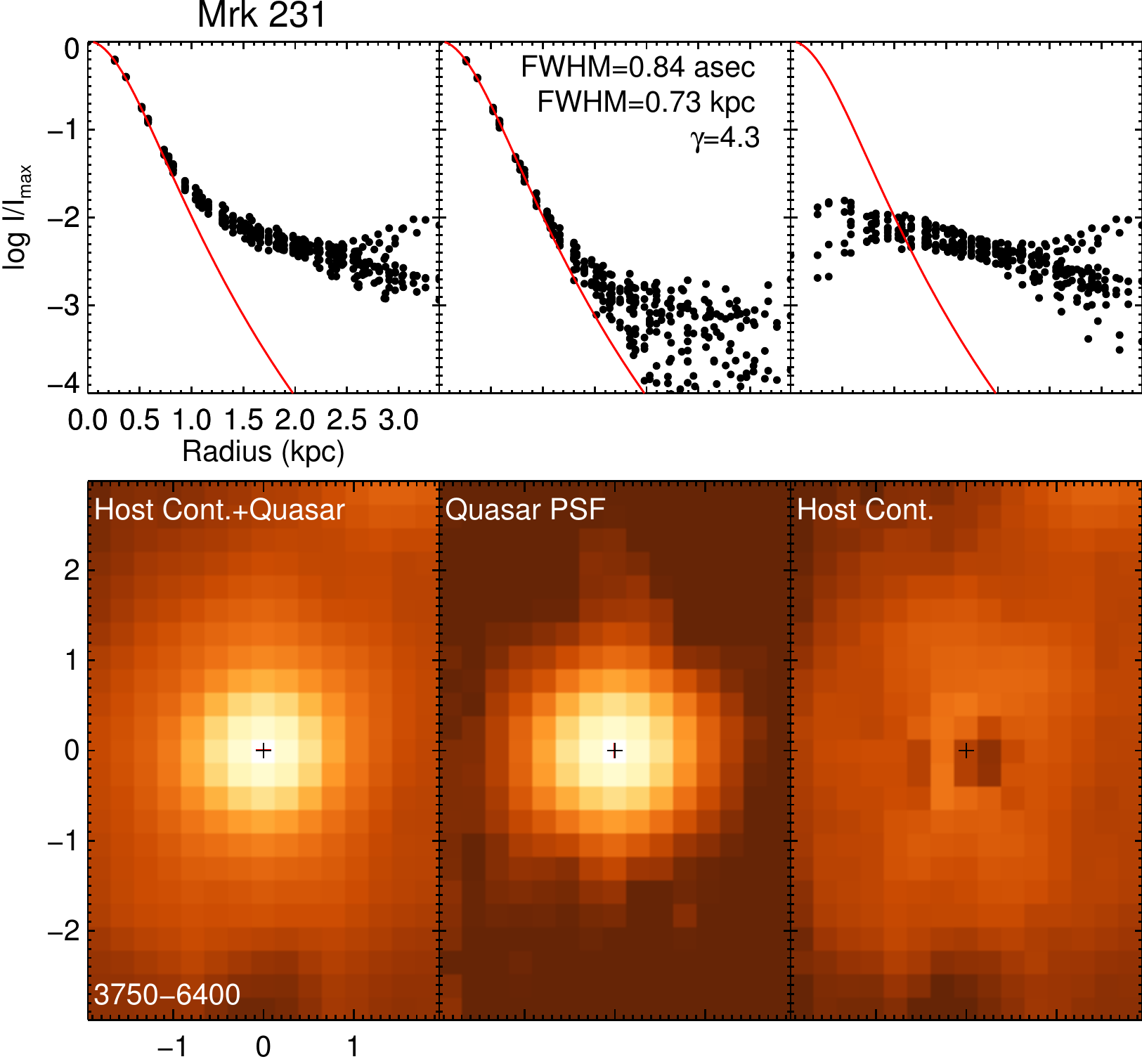}
  \caption{\it Continued.}
\end{figure*}
\setcounter{figure}{4}
\begin{figure*}
  \includegraphics[center]{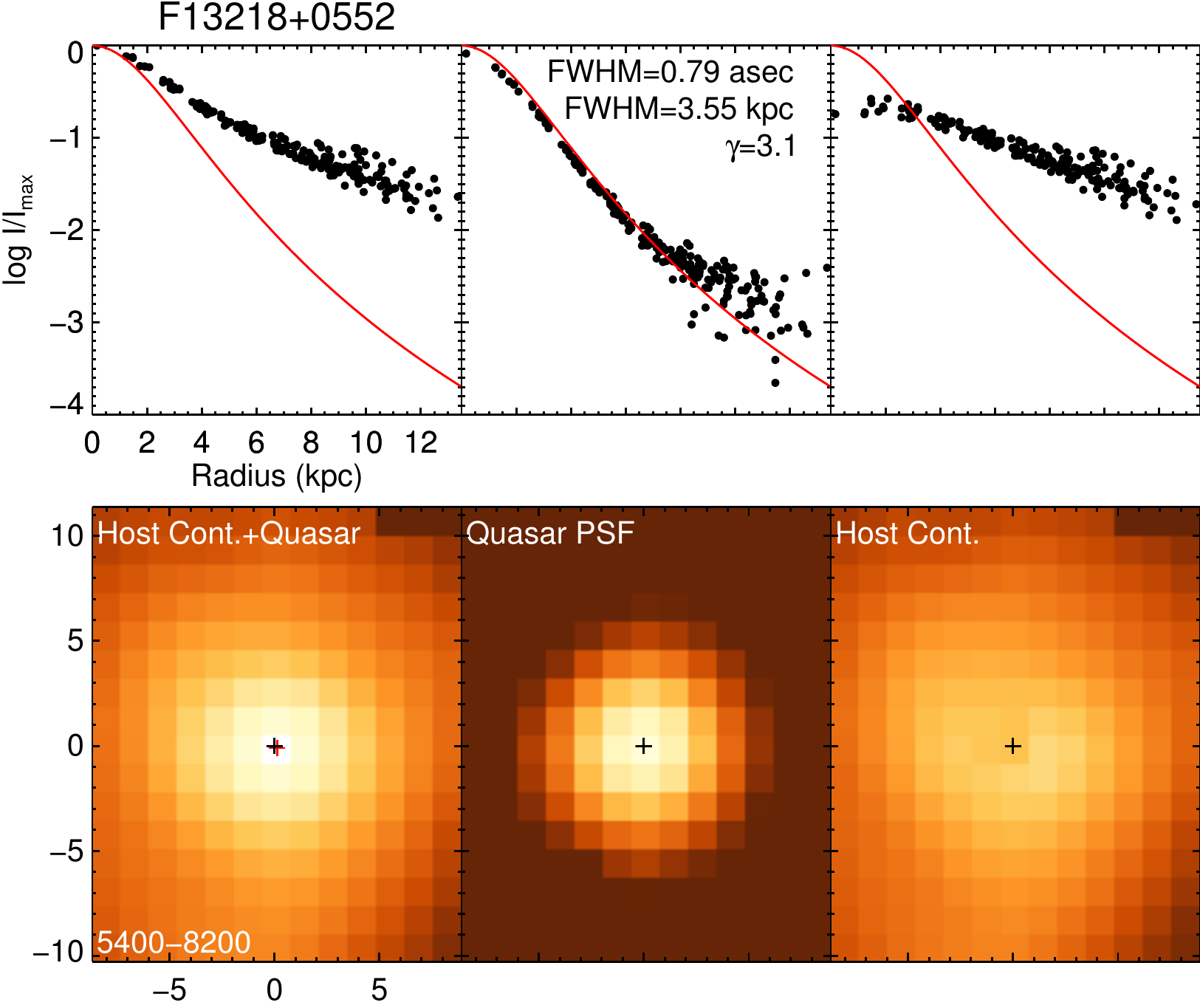}
  \caption{\it Continued.}
\end{figure*}
\setcounter{figure}{4}
\begin{figure*}
  \includegraphics[center]{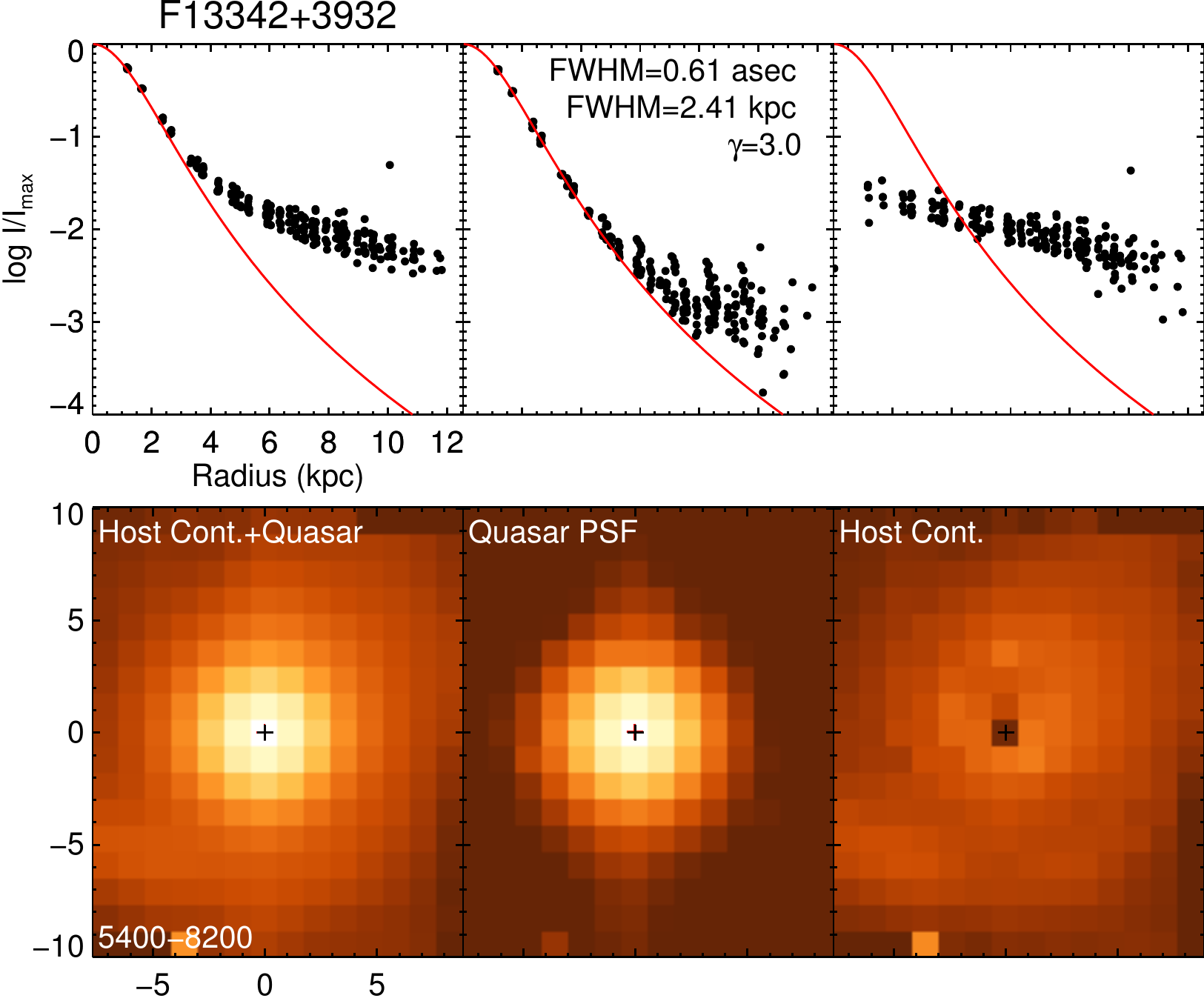}
  \caption{\it Continued.}
\end{figure*}
\setcounter{figure}{4}
\begin{figure*}
  \includegraphics[center]{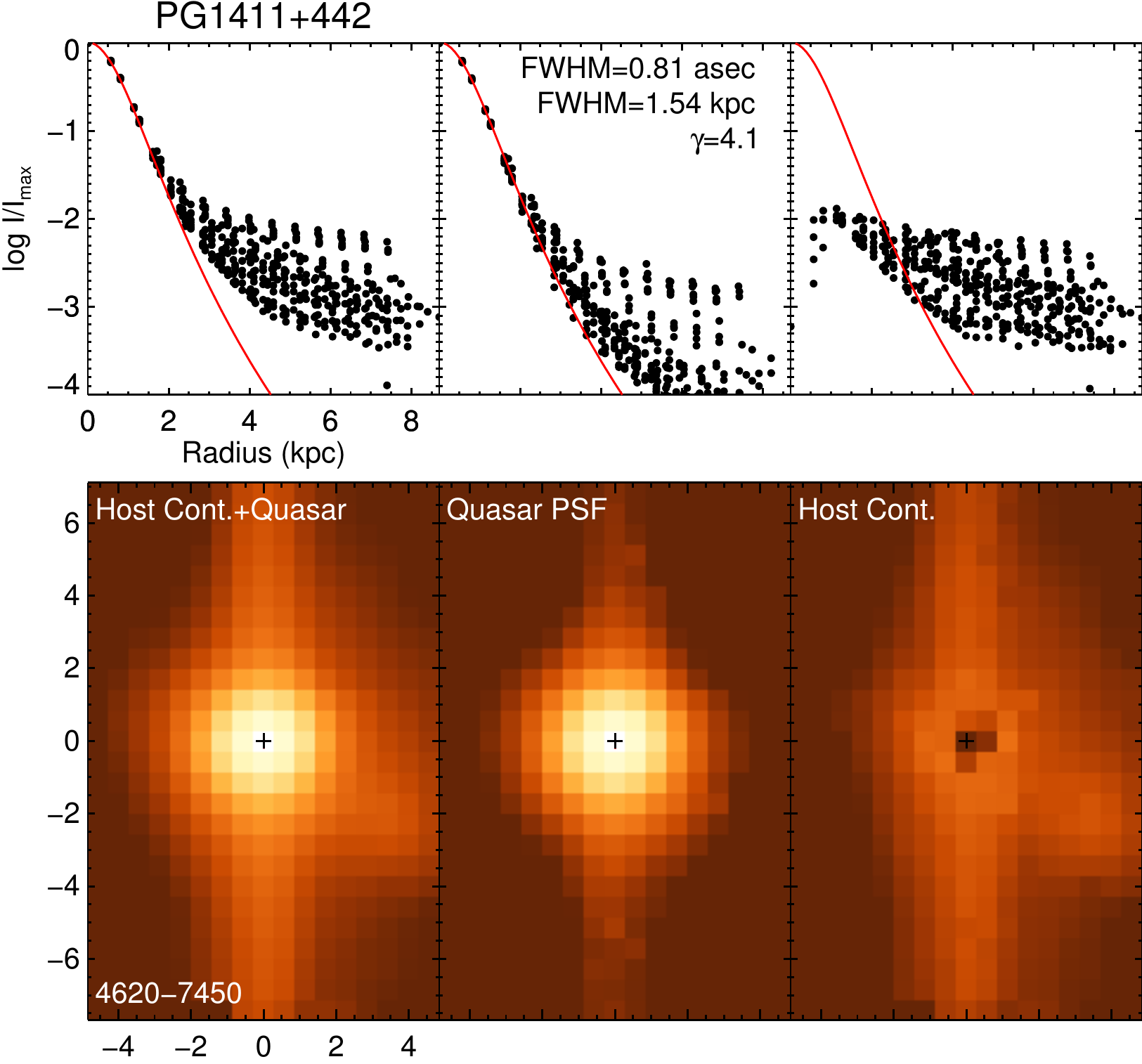}
  \caption{\it Continued.}
\end{figure*}
\setcounter{figure}{4}
\begin{figure*}
  \includegraphics[center]{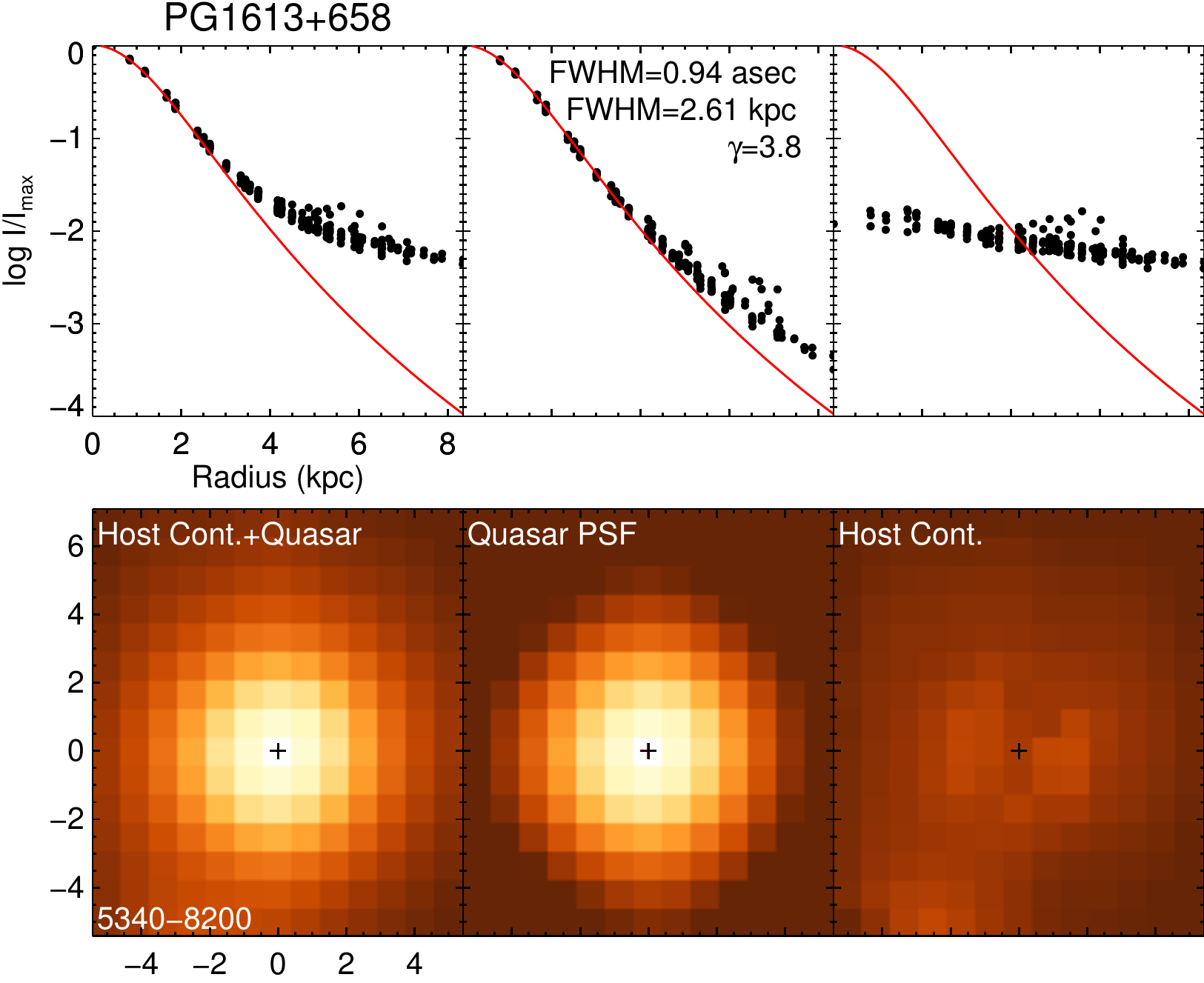}
  \caption{\it Continued.}
\end{figure*}
\setcounter{figure}{4}
\begin{figure*}
  \includegraphics[center]{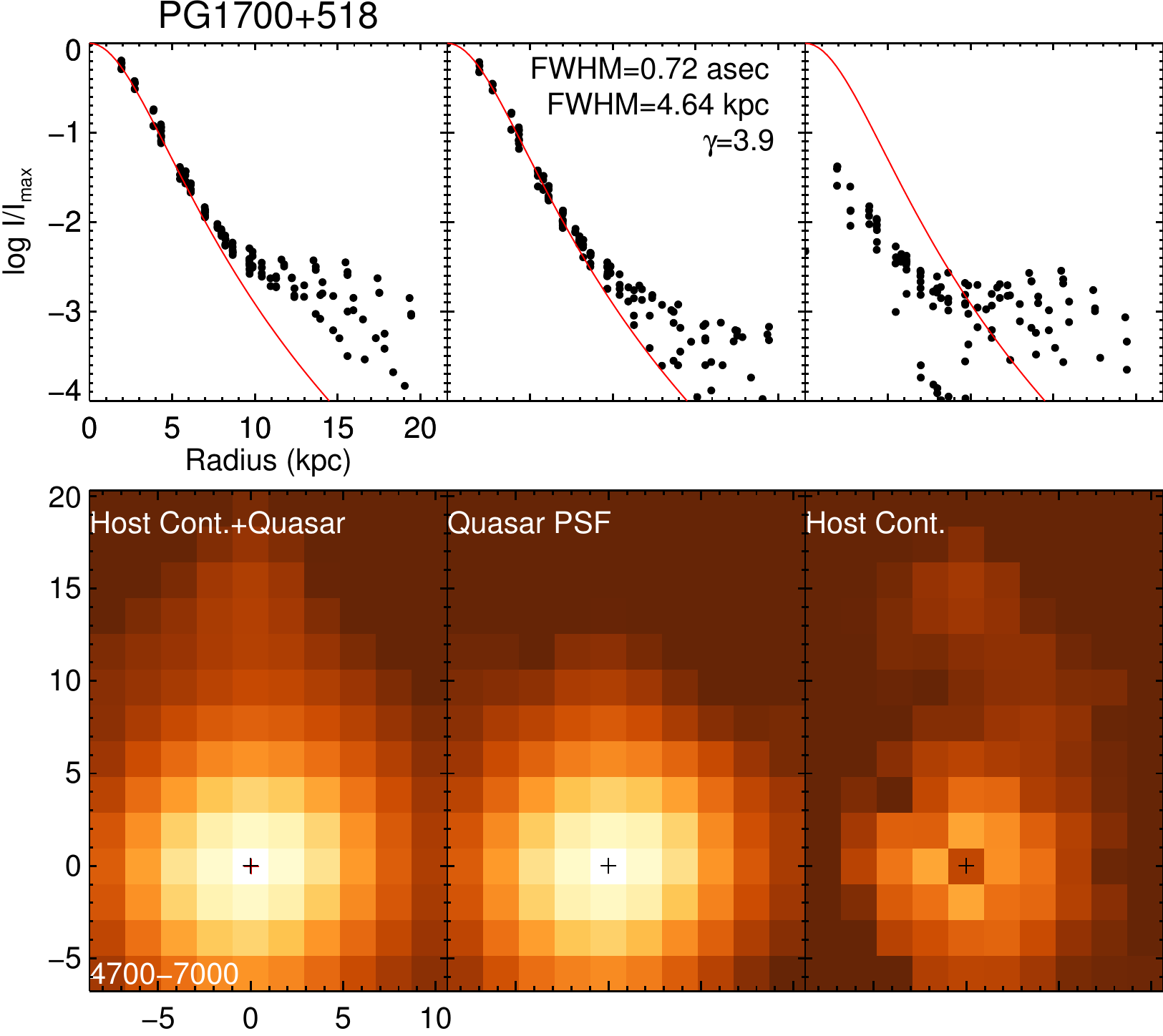}
  \caption{\it Continued.}
\end{figure*}
\setcounter{figure}{4}
\begin{figure*}
  \includegraphics[center]{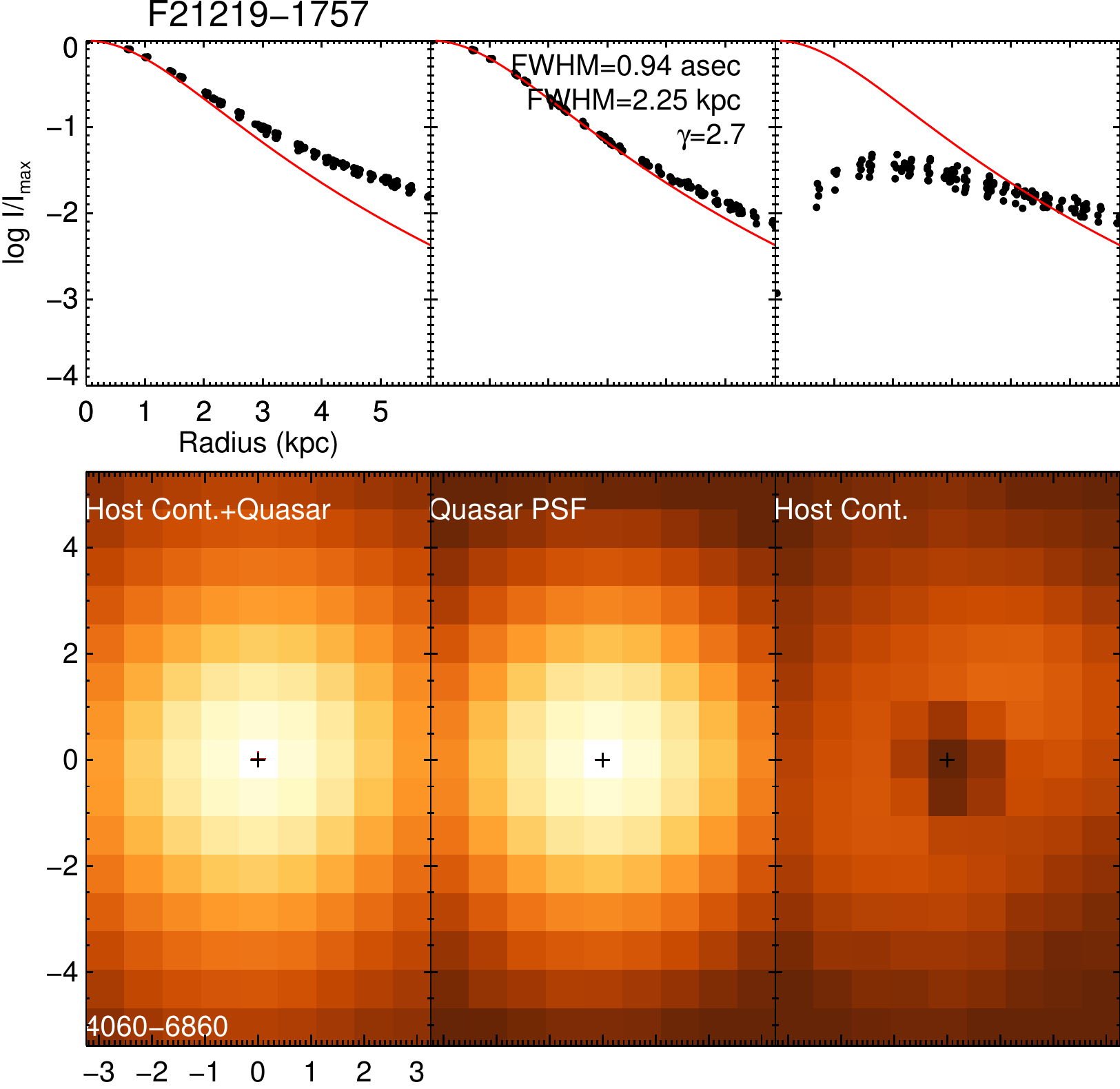}
  \caption{\it Continued.}
\end{figure*}

\setcounter{figure}{6}
\begin{figure*}
  \includegraphics[center]{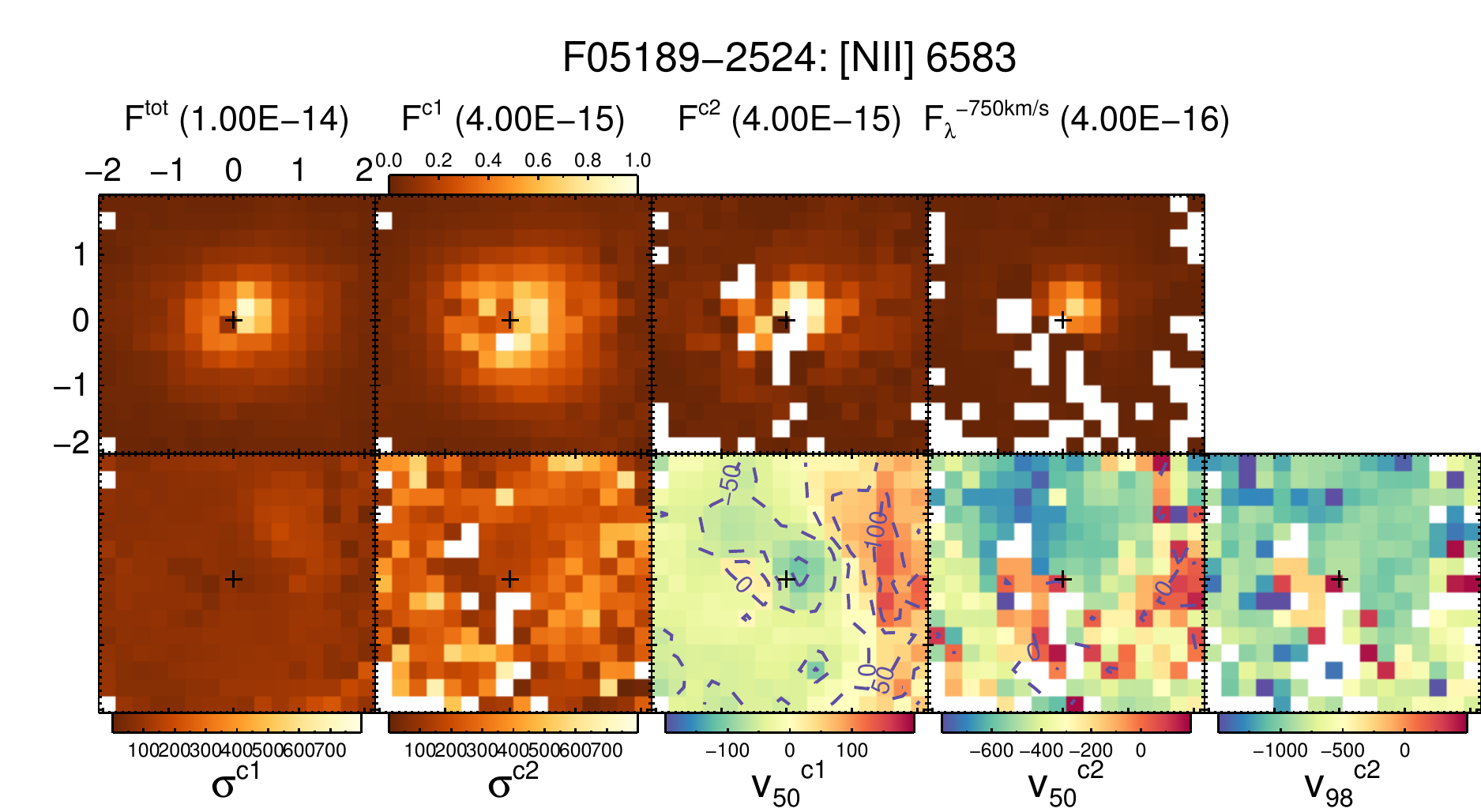}
  \caption{\it Continued.}
\end{figure*}
\setcounter{figure}{6}
\begin{figure*}
  \includegraphics[center]{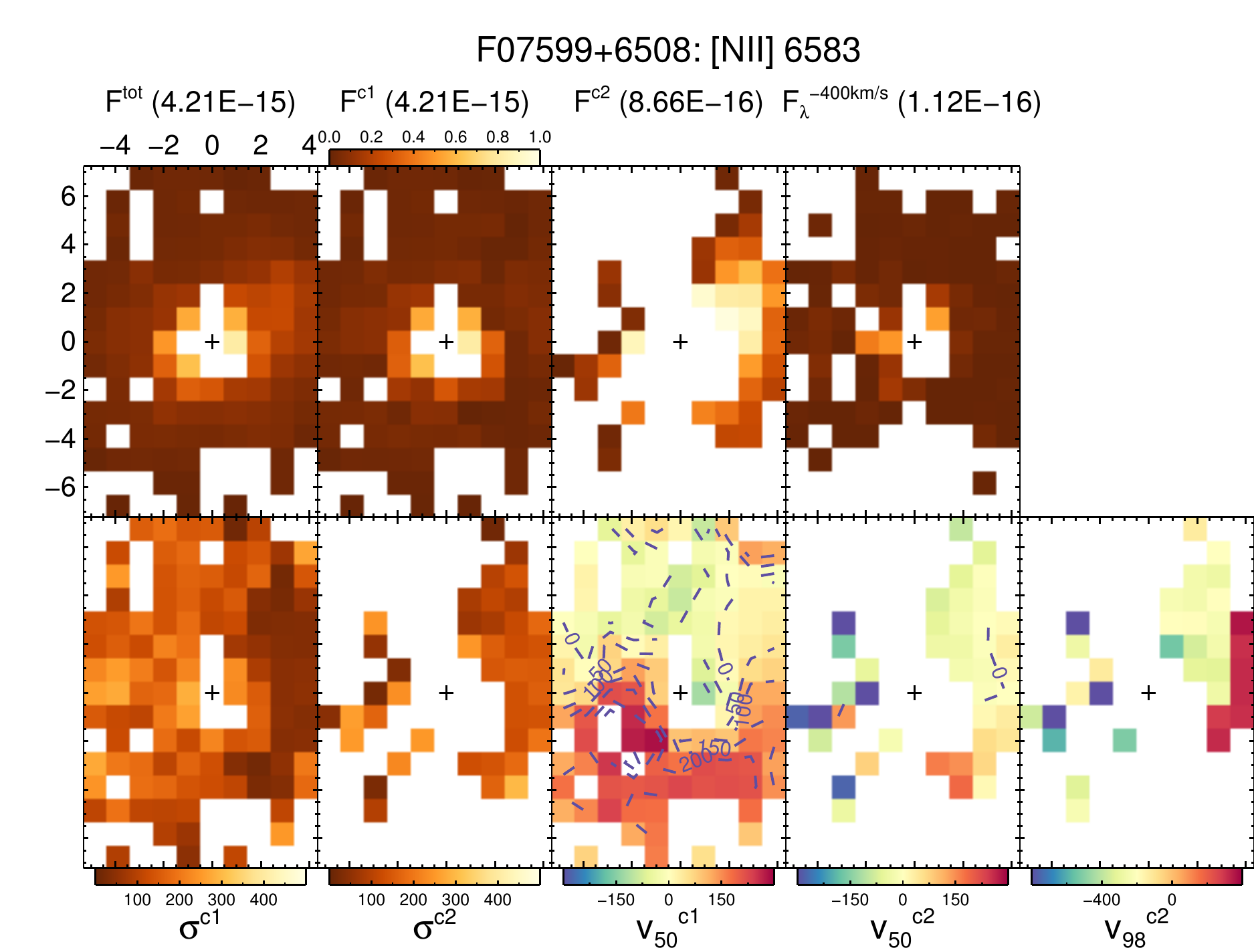}
  \caption{\it Continued.}
\end{figure*}
\setcounter{figure}{6}
\begin{figure*}
  \includegraphics[center]{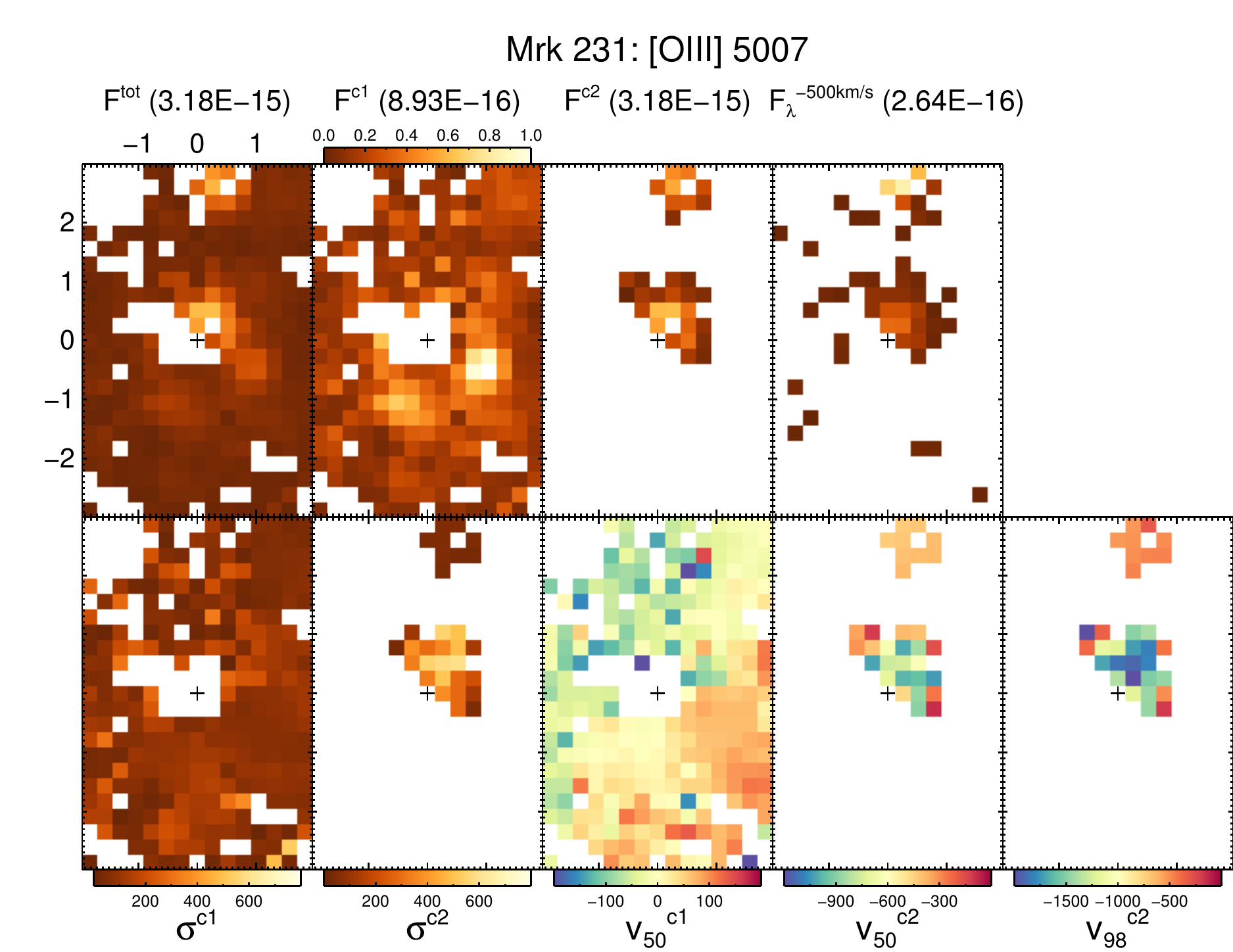}
  \caption{\it Continued.}
\end{figure*}
\setcounter{figure}{6}
\begin{figure*}
  \includegraphics[center]{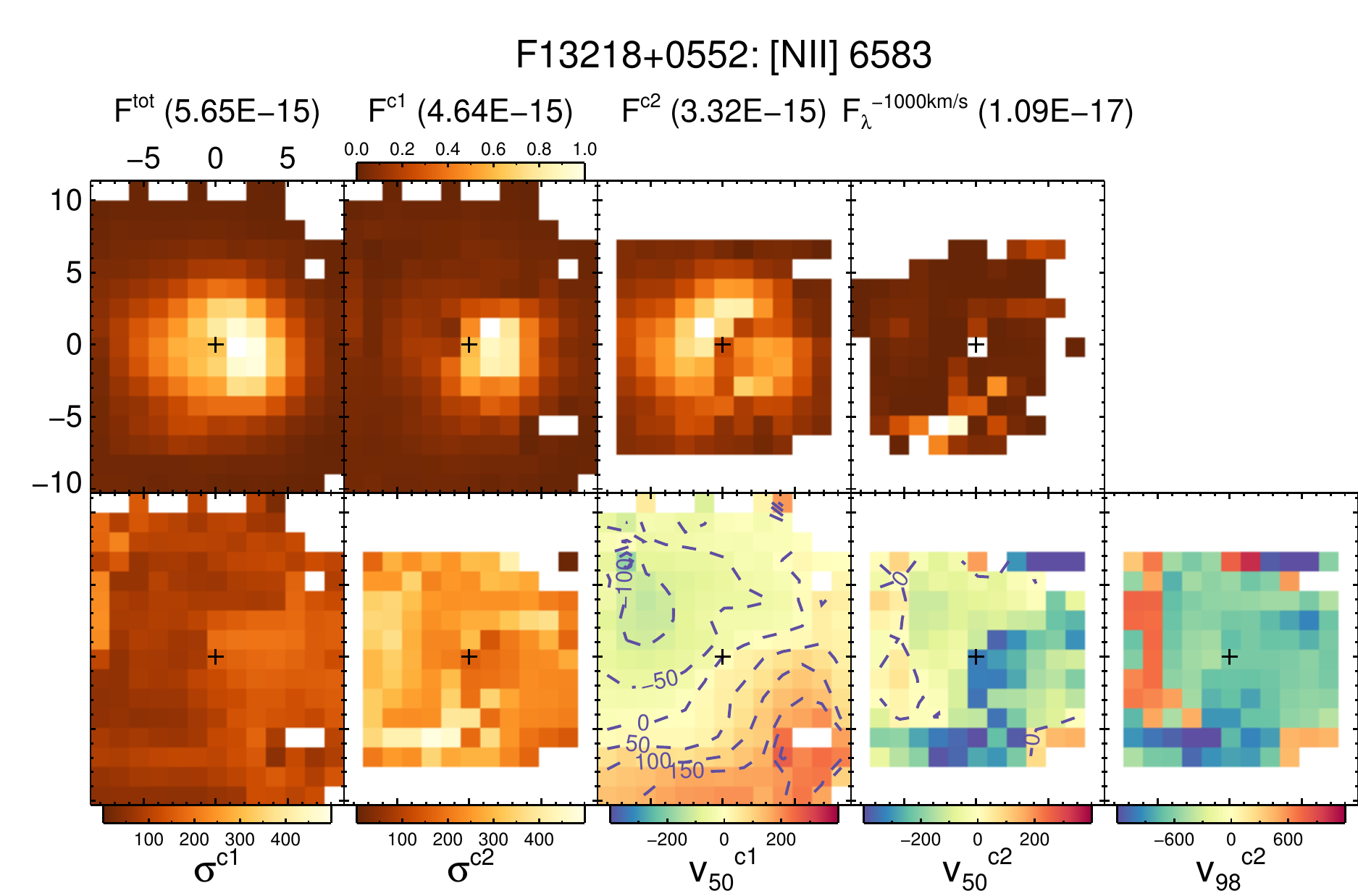}
  \caption{\it Continued.}
\end{figure*}
\setcounter{figure}{6}
\begin{figure*}
  \includegraphics[center]{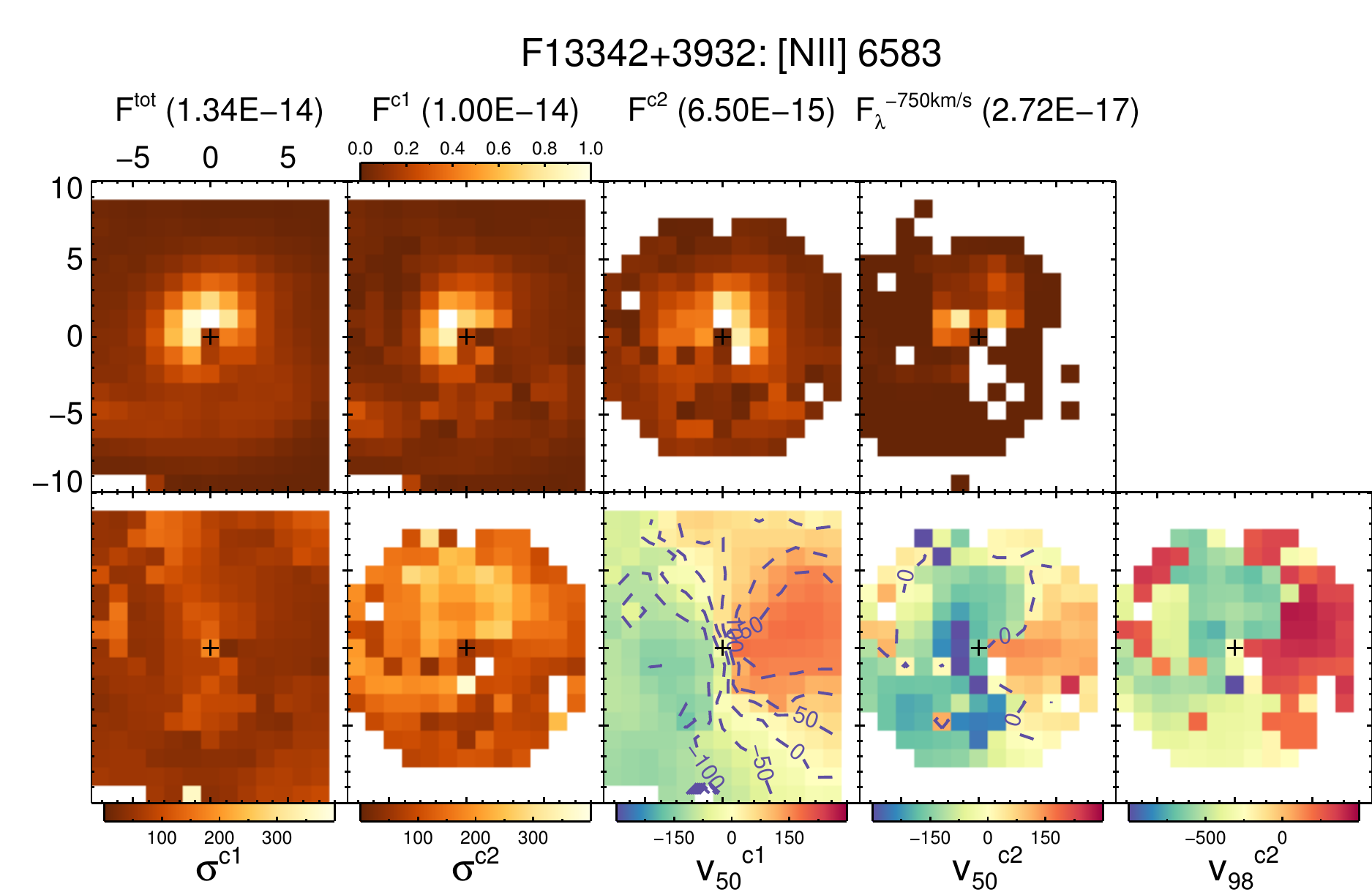}
  \caption{\it Continued.}
\end{figure*}
\setcounter{figure}{6}
\begin{figure*}
  \includegraphics[center]{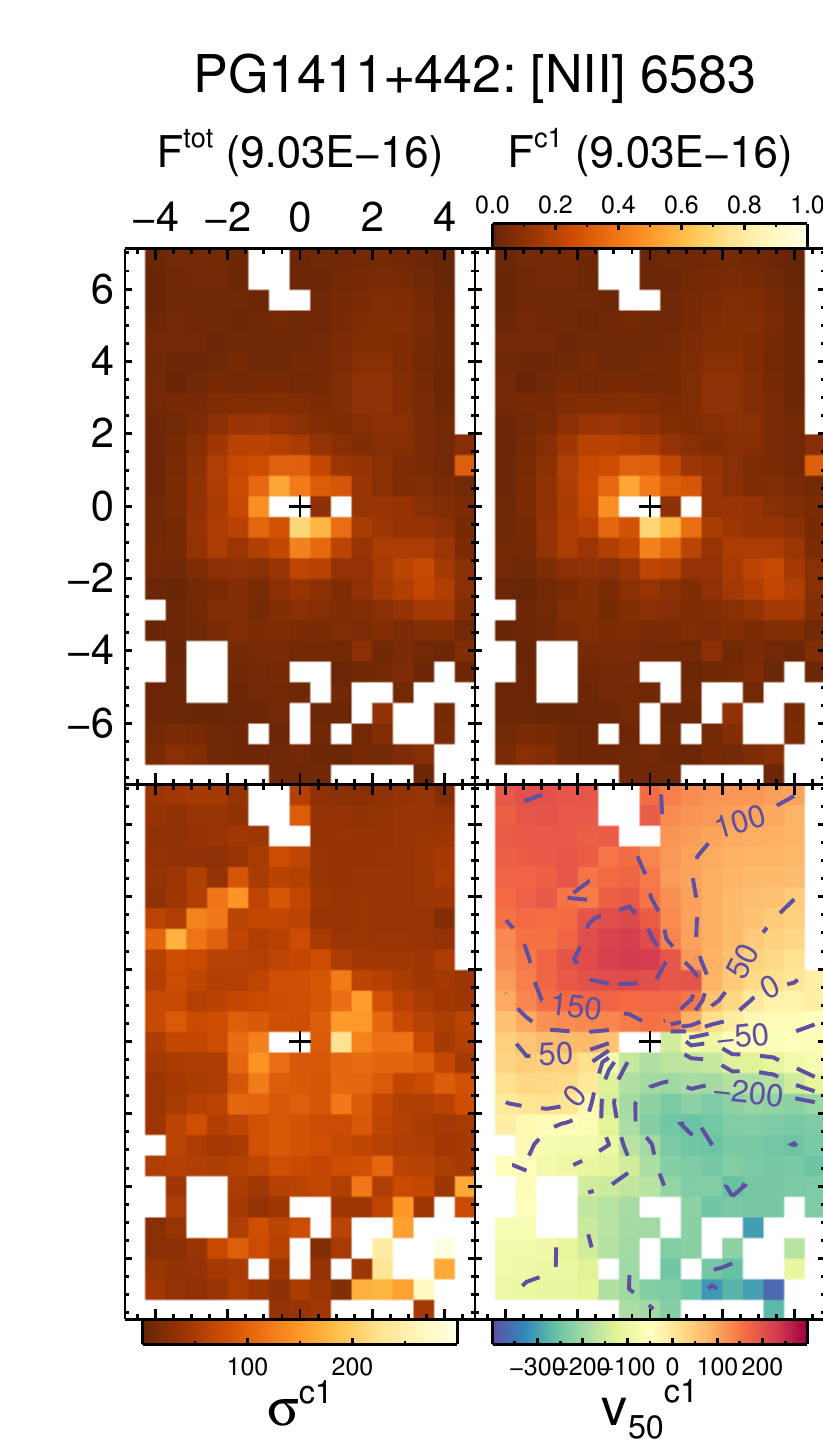}
  \caption{\it Continued.}
\end{figure*}
\setcounter{figure}{6}
\begin{figure*}
  \includegraphics[center]{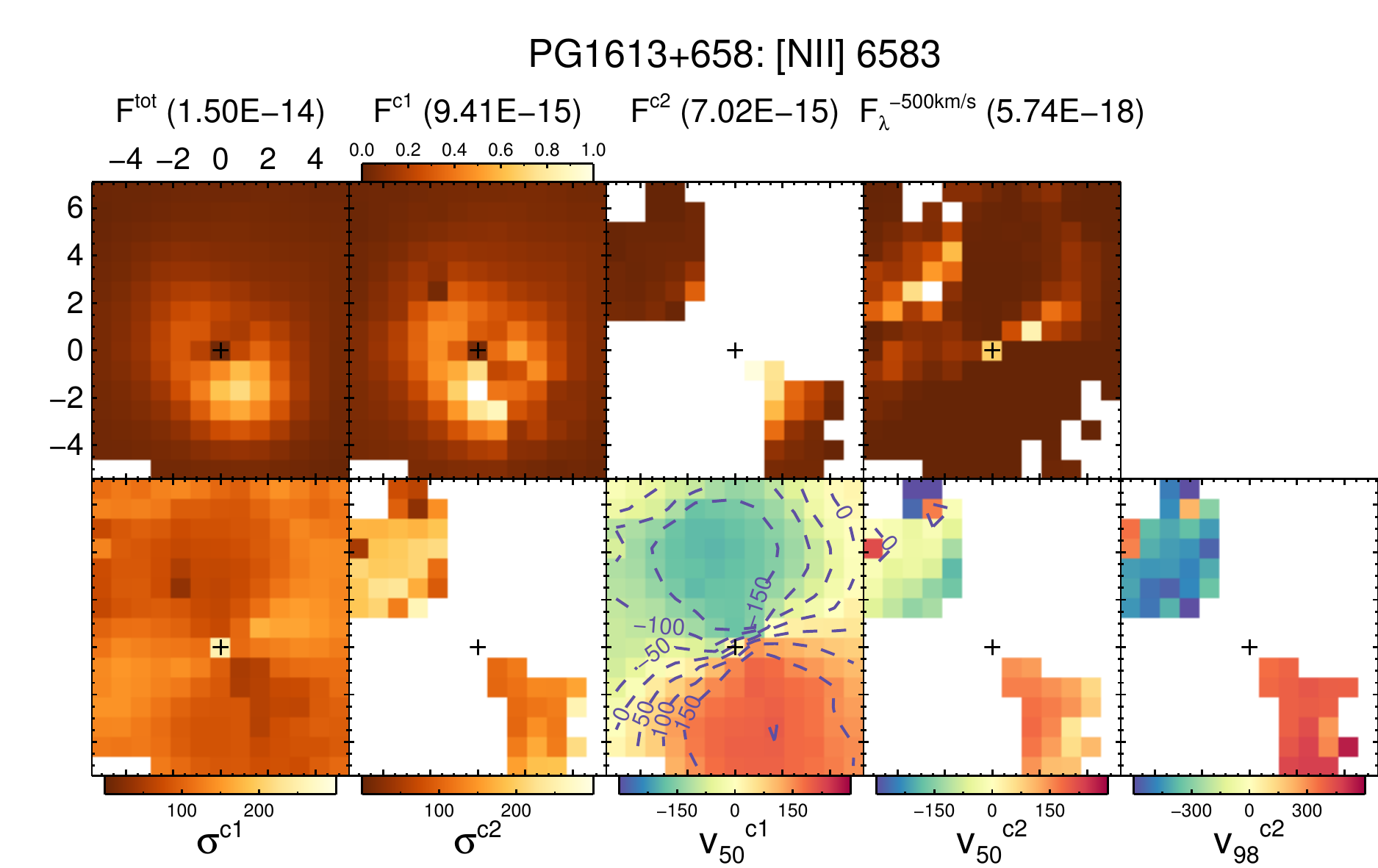}
  \caption{\it Continued.}
\end{figure*}
\setcounter{figure}{6}
\begin{figure*}
  \includegraphics[center]{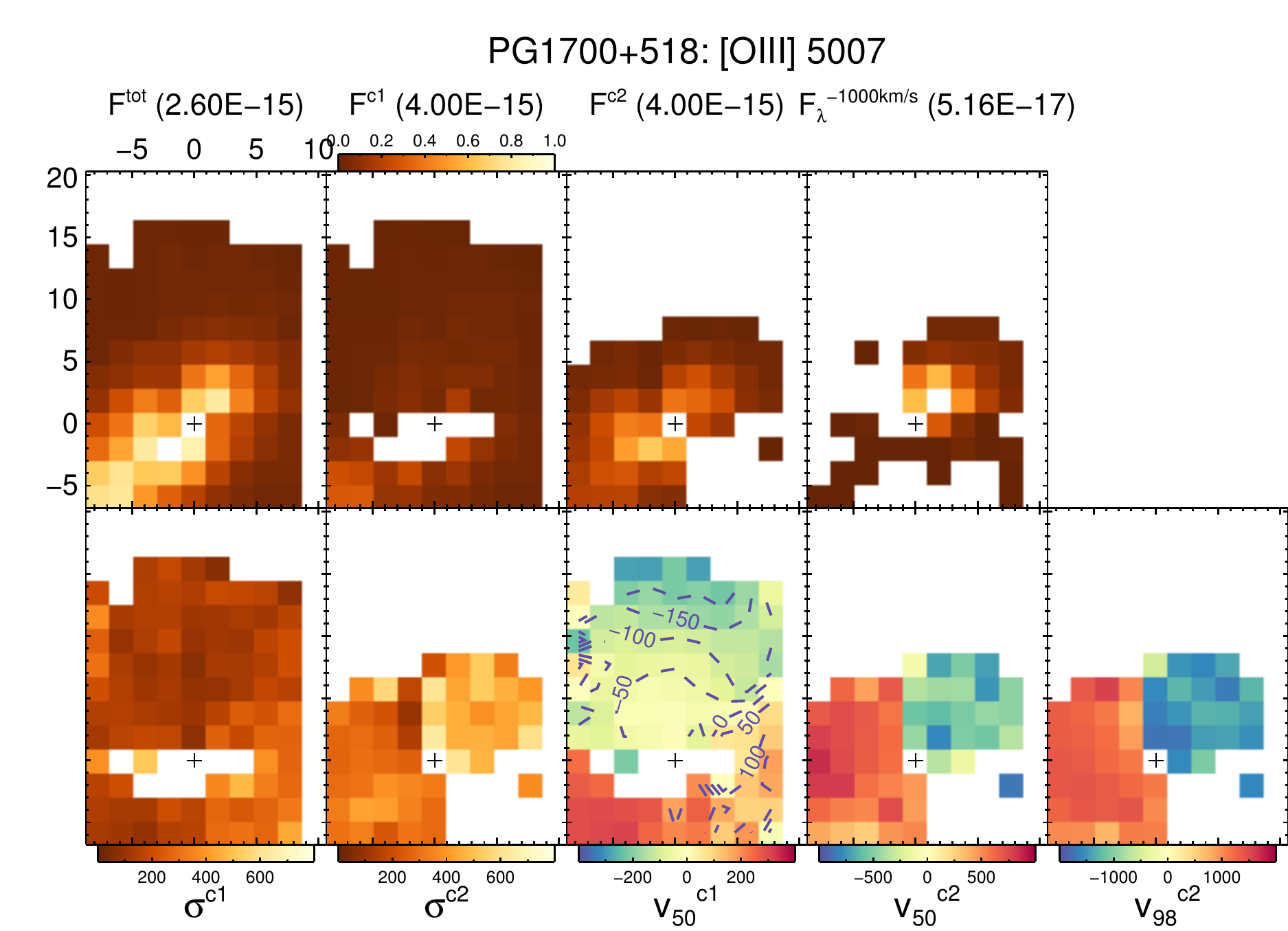}
  \caption{\it Continued.}
\end{figure*}

\setcounter{figure}{9}
\begin{figure*}
  \includegraphics[center]{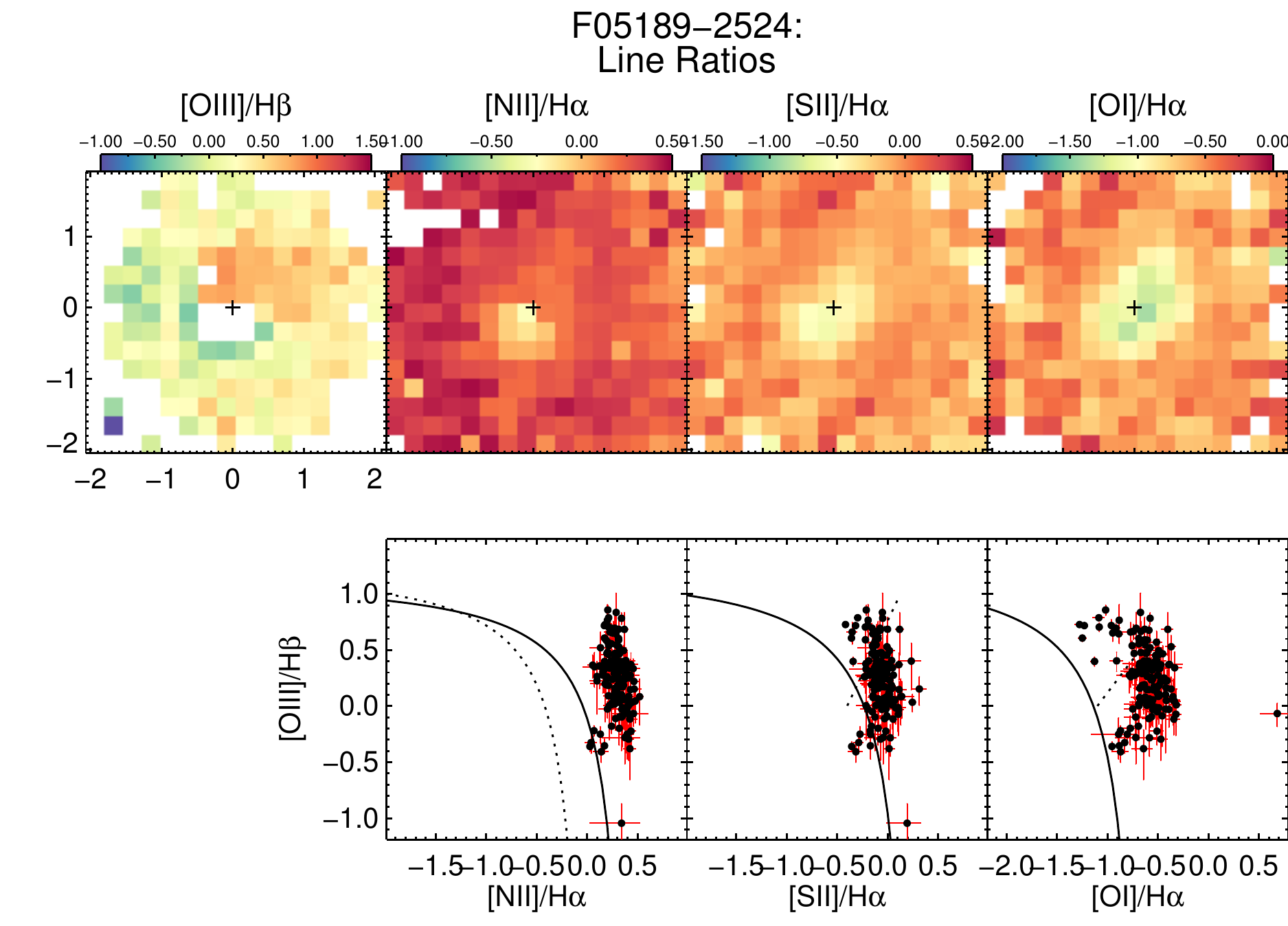}
  \caption{\it Continued.}
\end{figure*}
\setcounter{figure}{9}
\begin{figure*}
  \includegraphics[center]{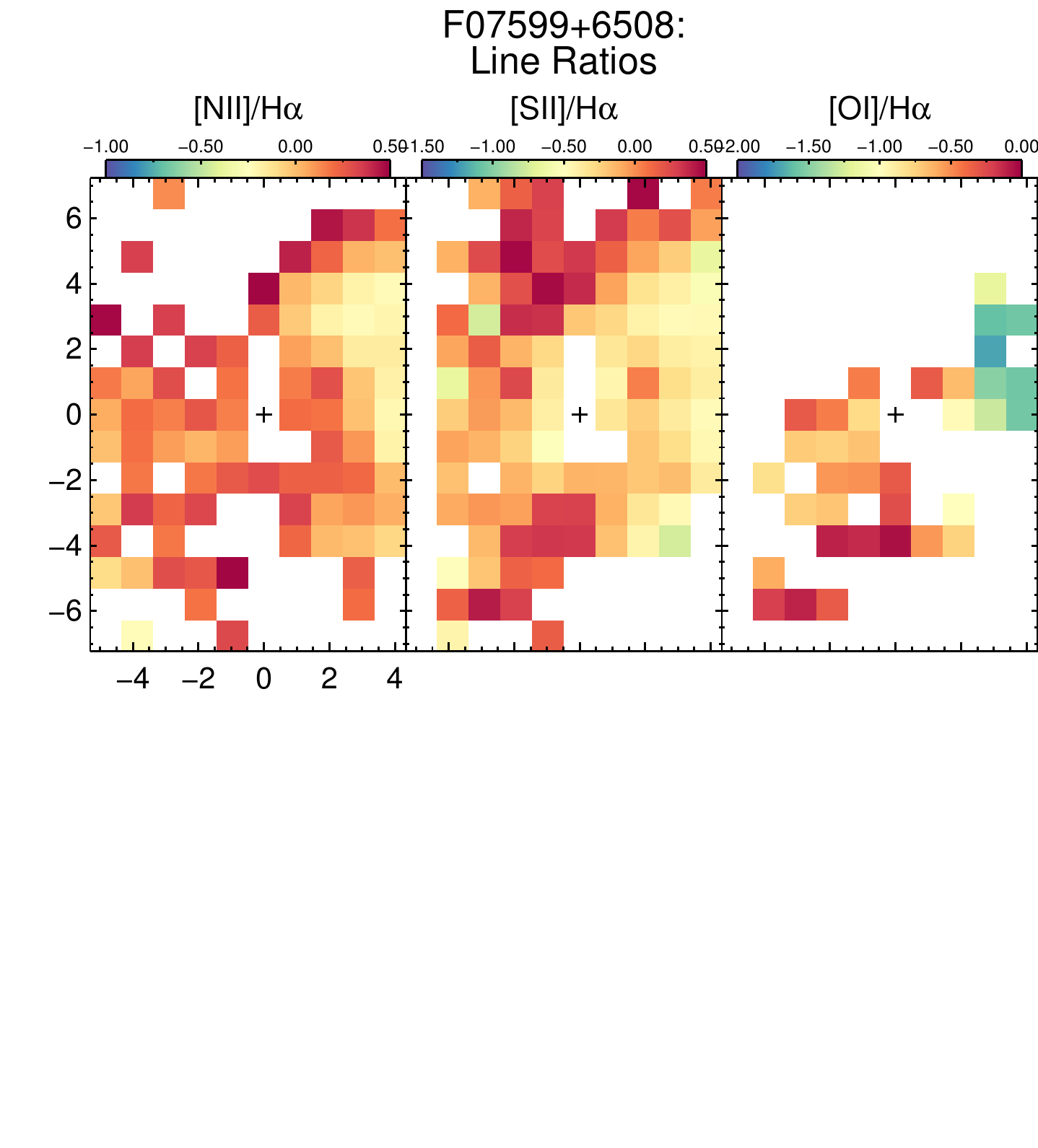}
  \caption{\it Continued.}
\end{figure*}
\setcounter{figure}{9}
\begin{figure*}
  \includegraphics[center]{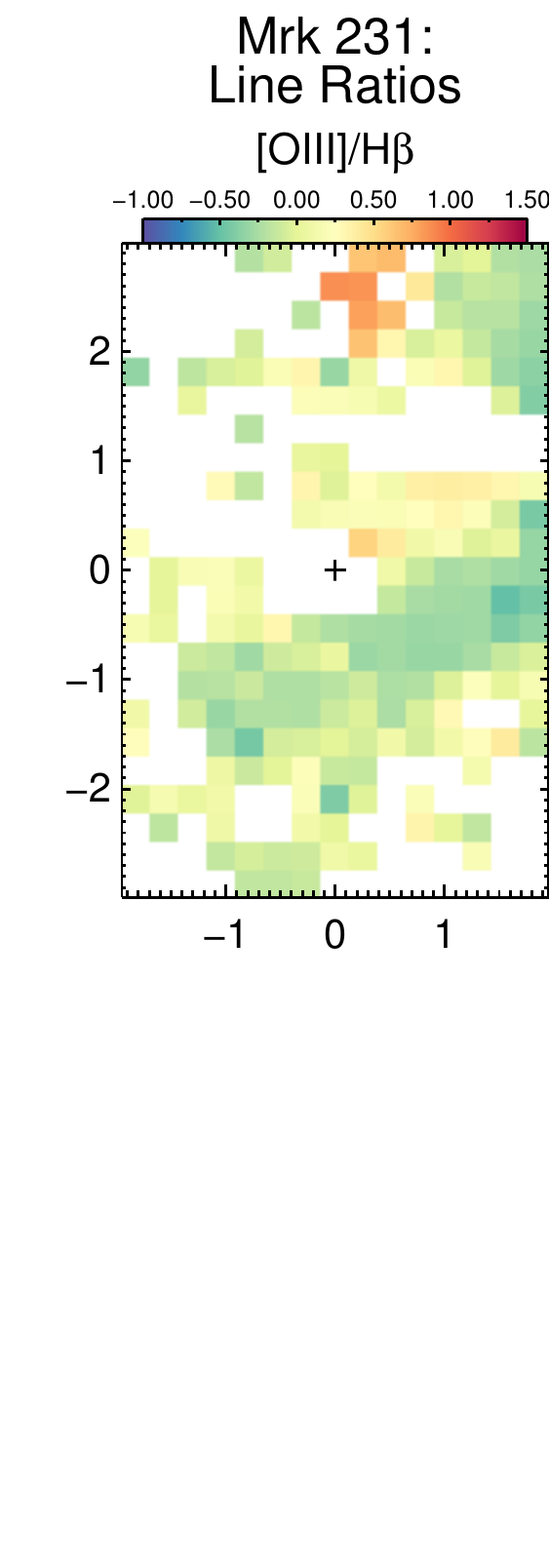}
  \caption{\it Continued.}
\end{figure*}
\setcounter{figure}{9}
\begin{figure*}
  \includegraphics[center]{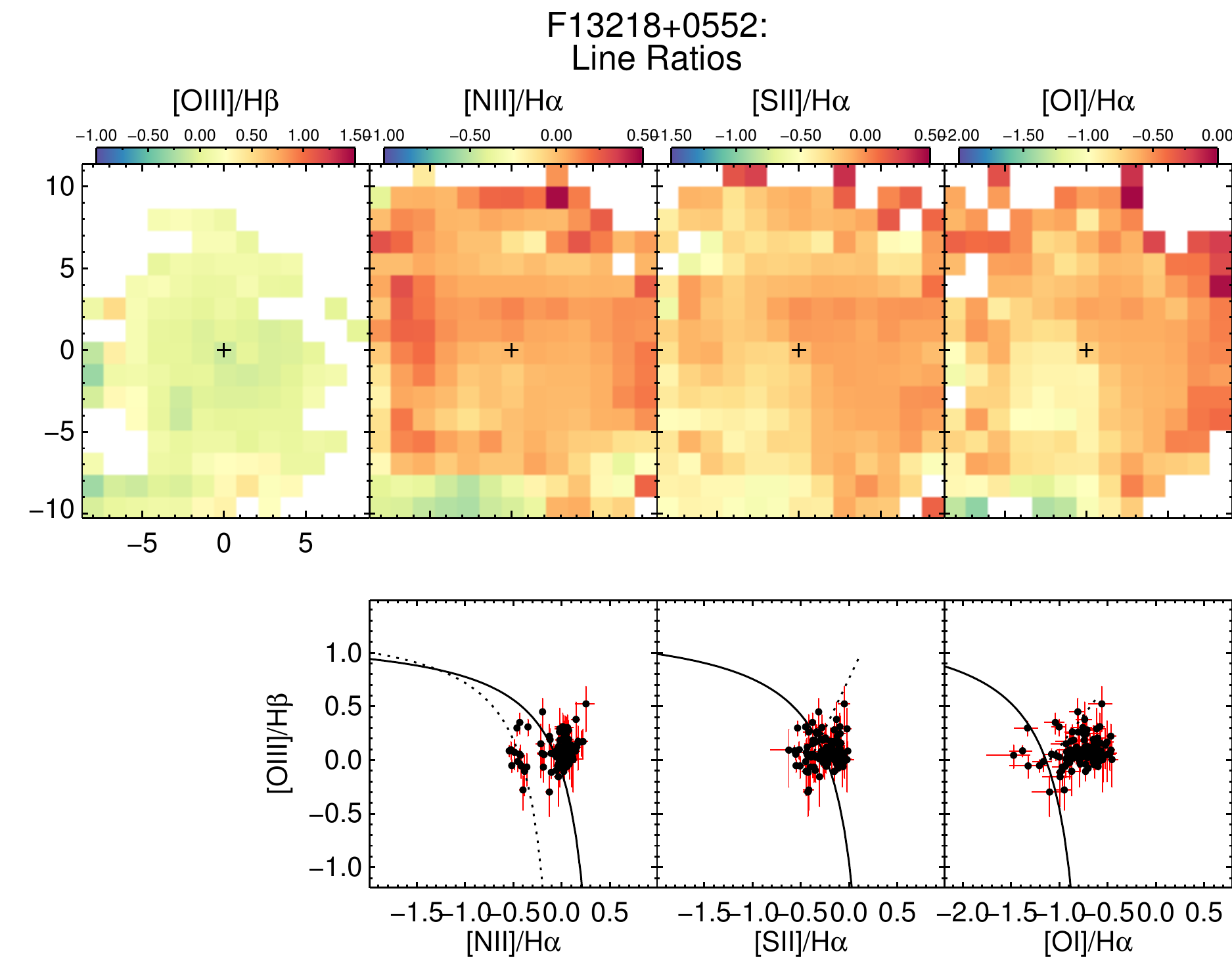}
  \caption{\it Continued.}
\end{figure*}
\setcounter{figure}{9}
\begin{figure*}
  \includegraphics[center]{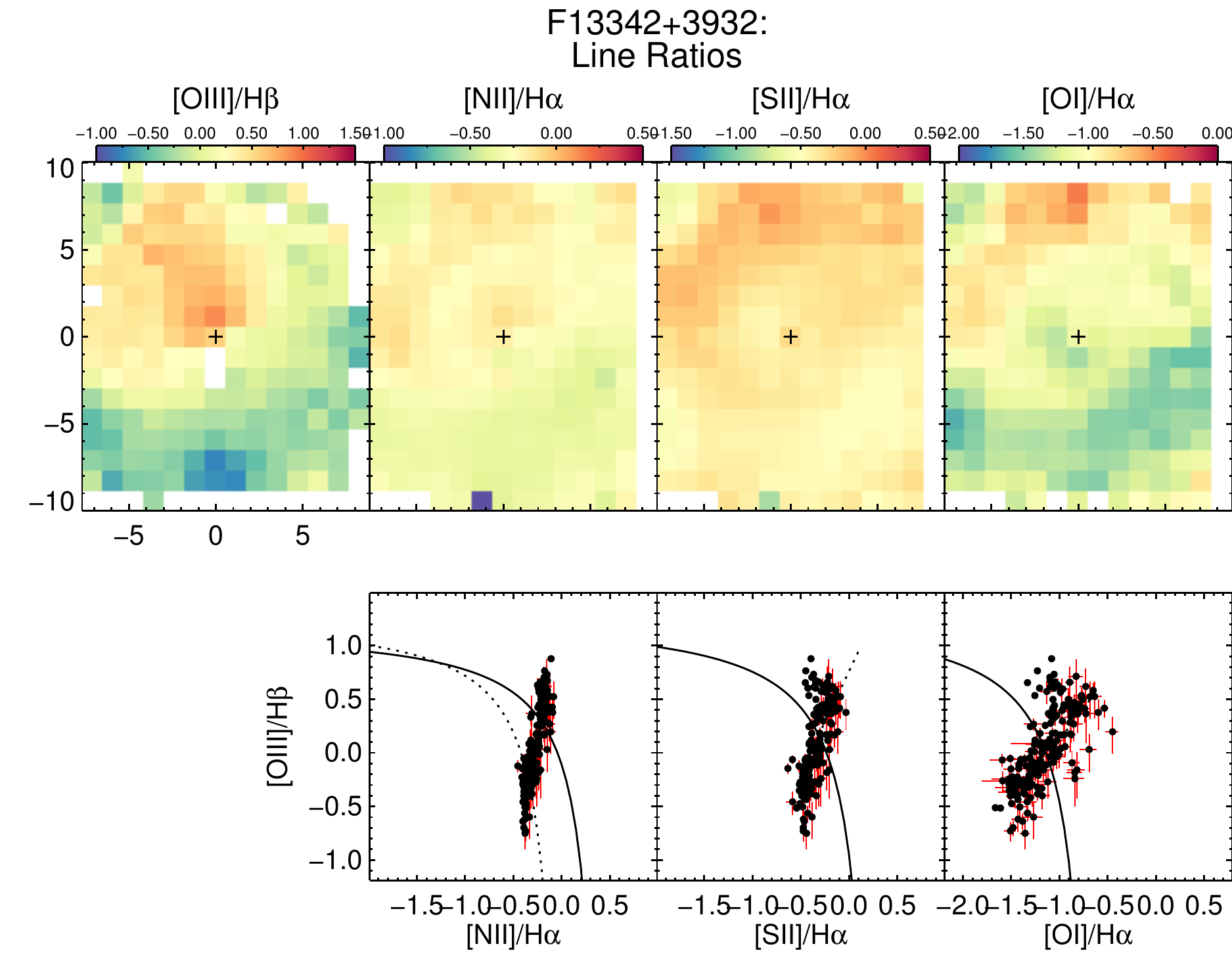}
  \caption{\it Continued.}
\end{figure*}
\setcounter{figure}{9}
\begin{figure*}
  \includegraphics[center]{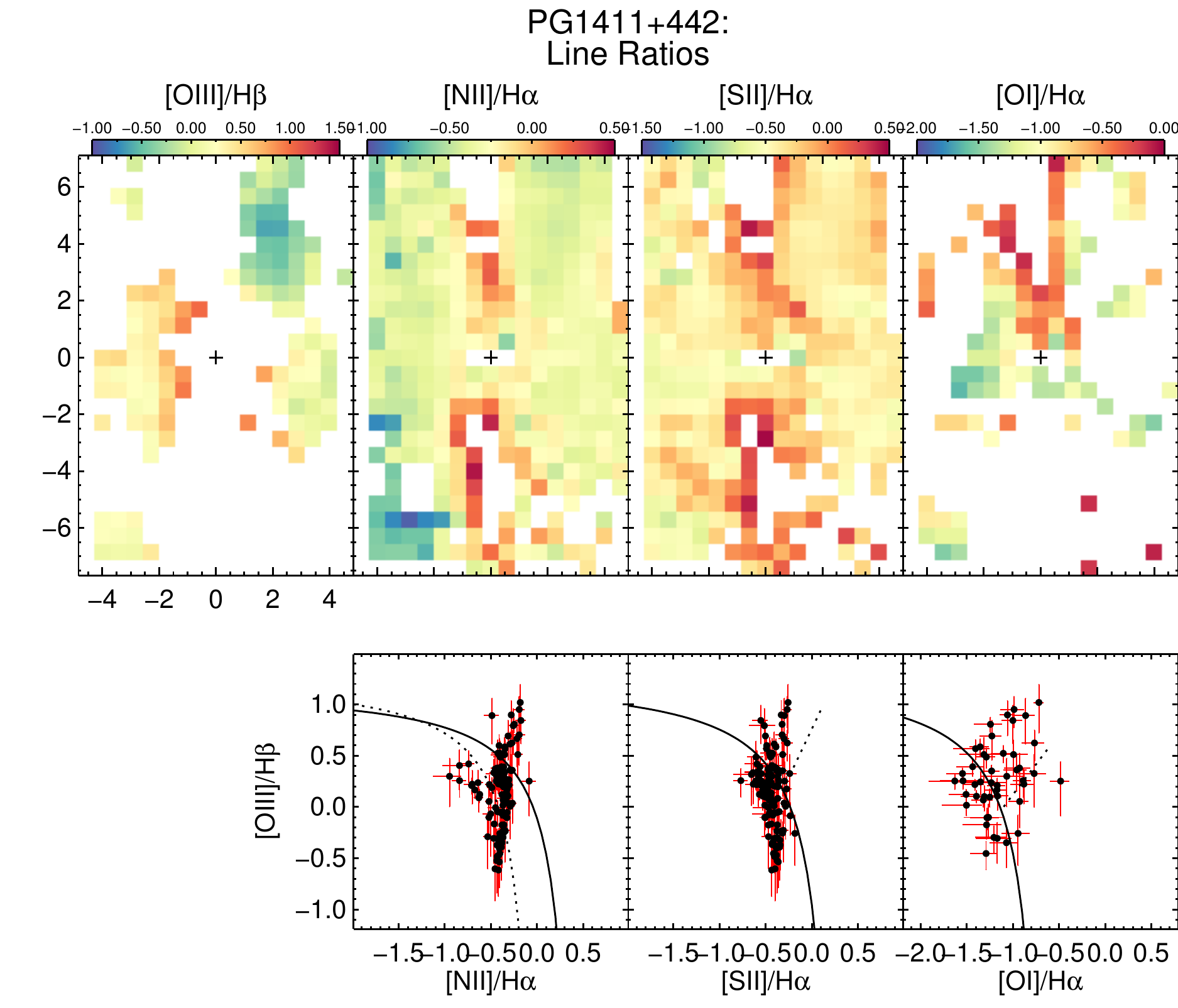}
  \caption{\it Continued.}
\end{figure*}
\setcounter{figure}{9}
\begin{figure*}
  \includegraphics[center]{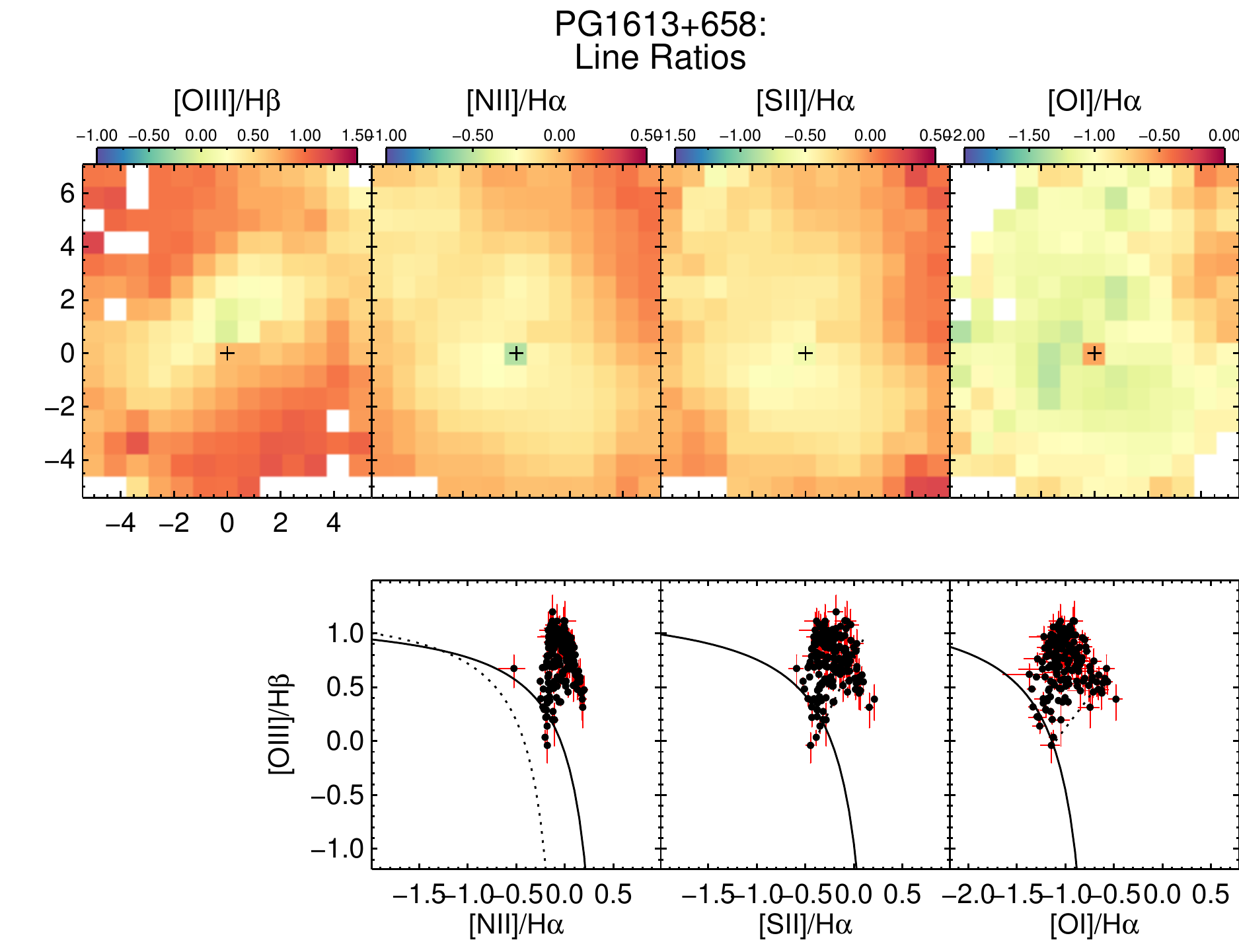}
  \caption{\it Continued.}
\end{figure*}
\setcounter{figure}{9}
\begin{figure*}
  \includegraphics[center]{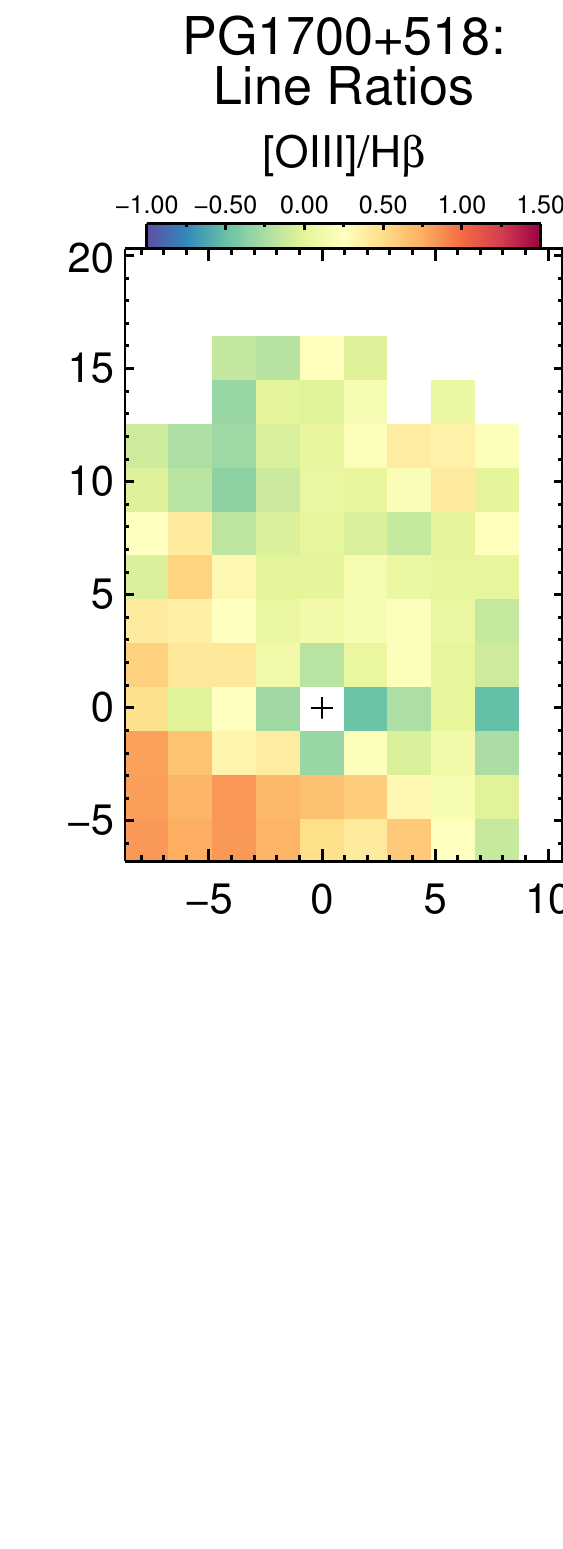}
  \caption{\it Continued.}
\end{figure*}

\clearpage

\setcounter{figure}{11}
\begin{figure*}
  \includegraphics[center]{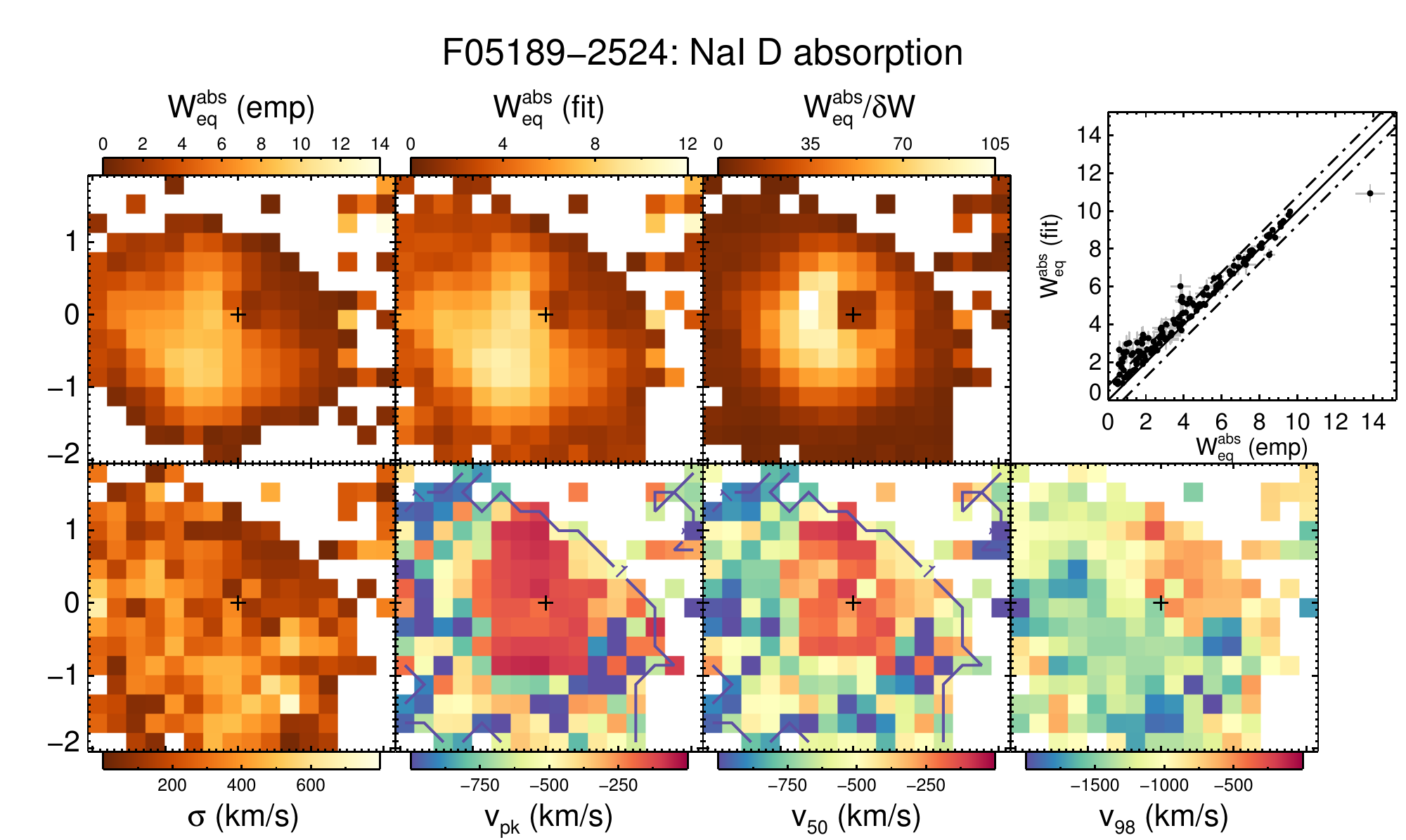}
  \caption{\it Continued.}
\end{figure*}
\setcounter{figure}{11}
\begin{figure*}
  \includegraphics[center]{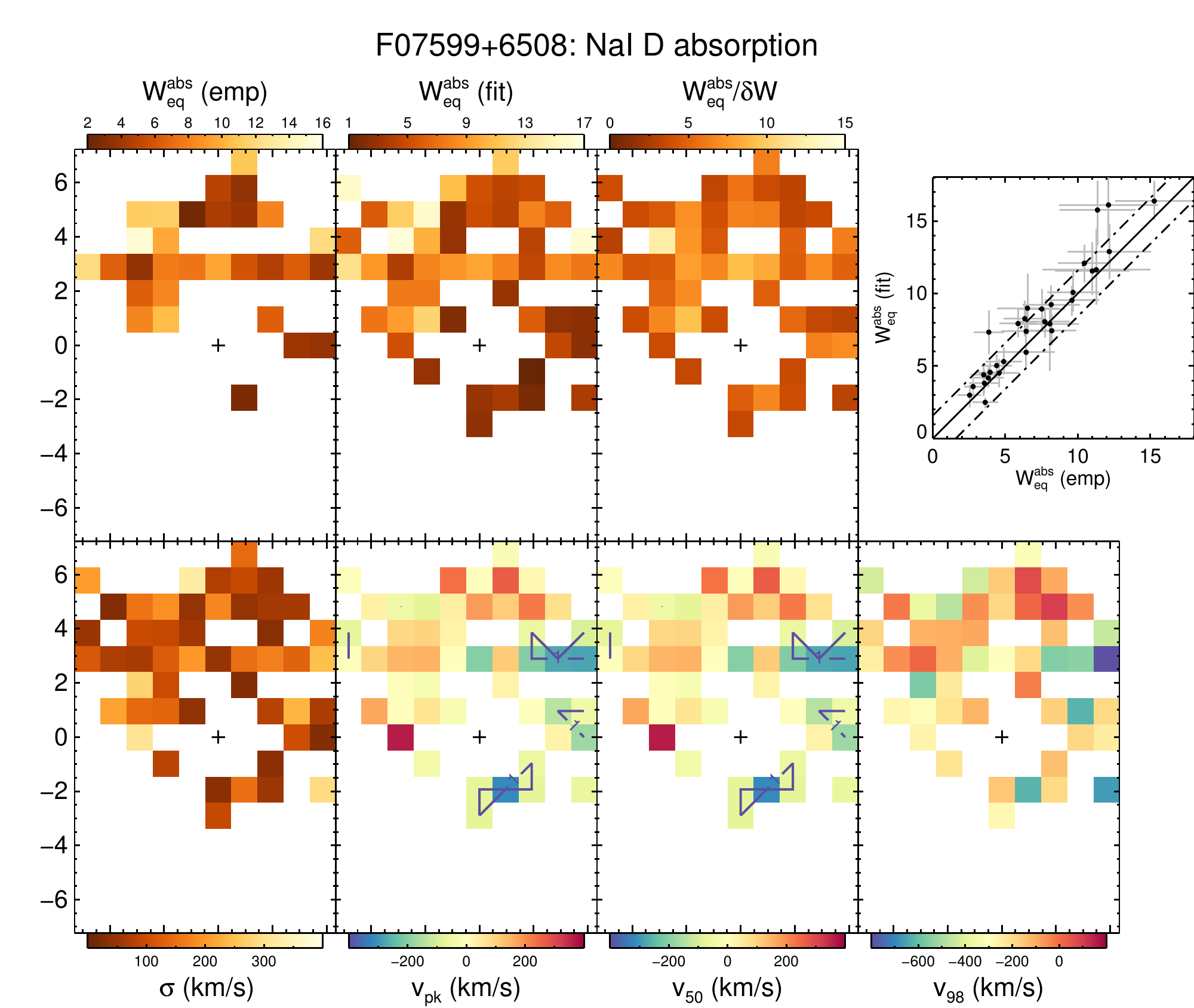}
  \caption{\it Continued.}
\end{figure*}
\setcounter{figure}{11}
\begin{figure*}
  \includegraphics[center]{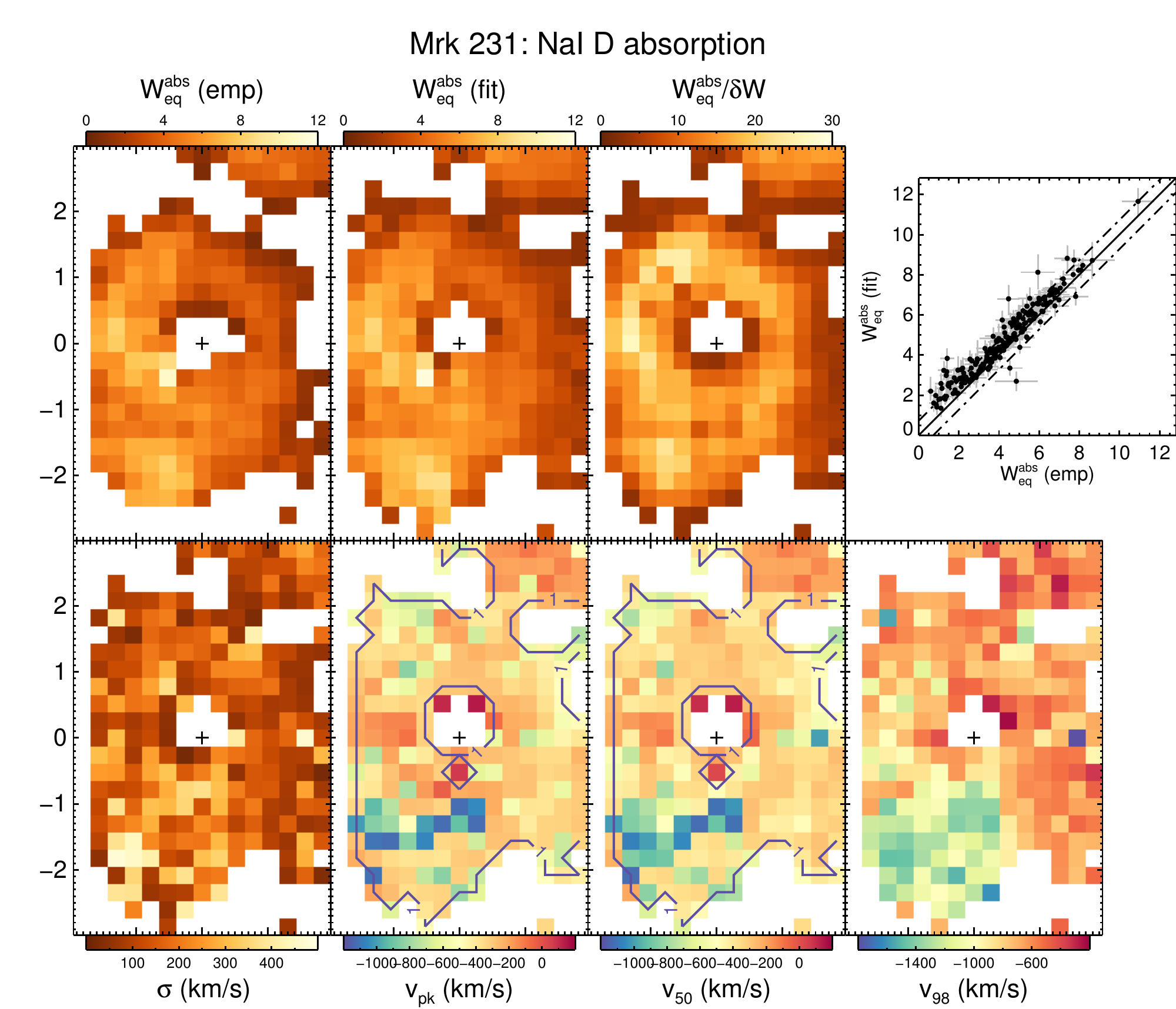}
  \caption{\it Continued.}
\end{figure*}
\setcounter{figure}{11}
\begin{figure*}
  \includegraphics[center]{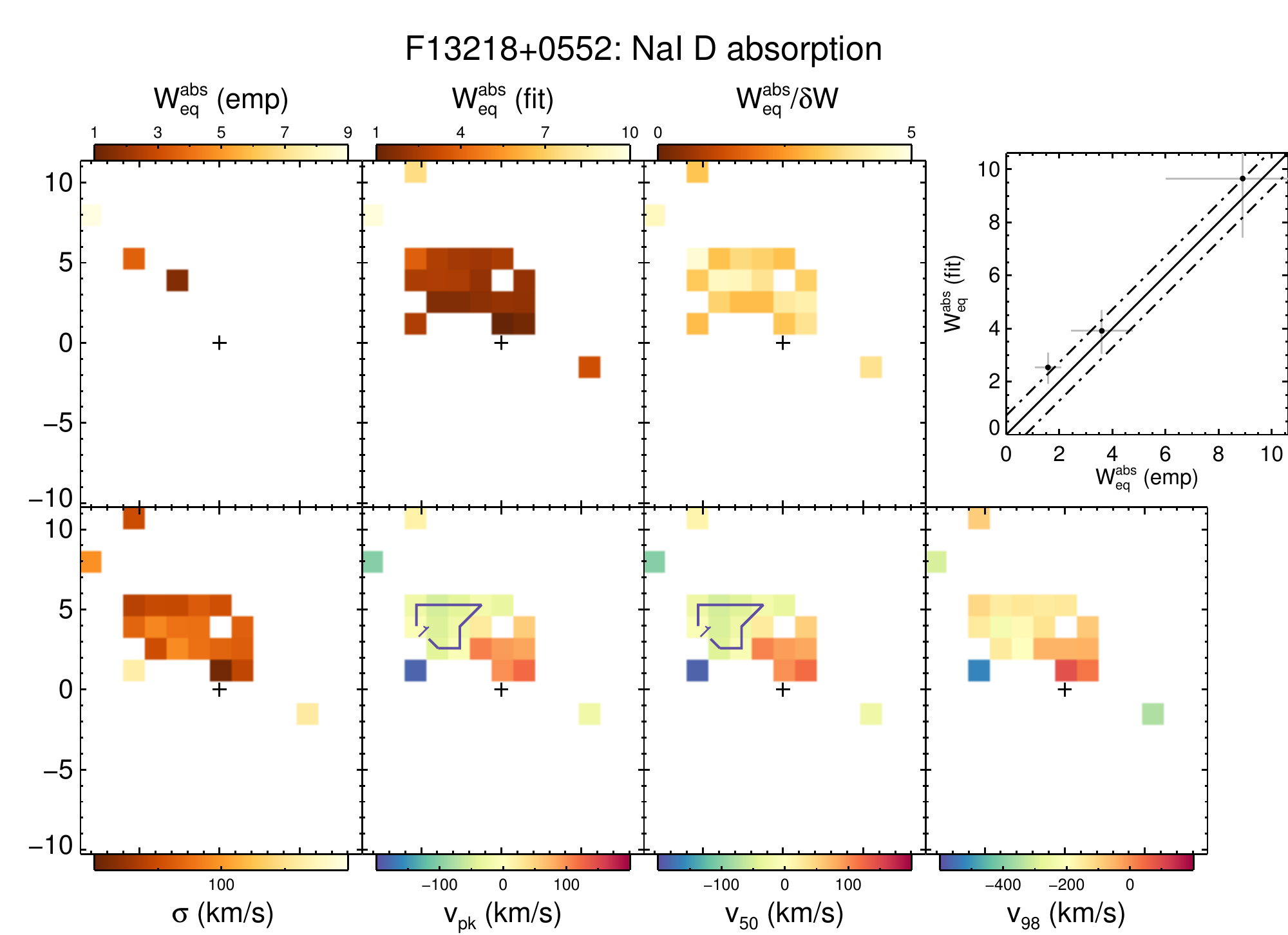}
  \caption{\it Continued.}
\end{figure*}
\setcounter{figure}{11}
\begin{figure*}
  \includegraphics[center]{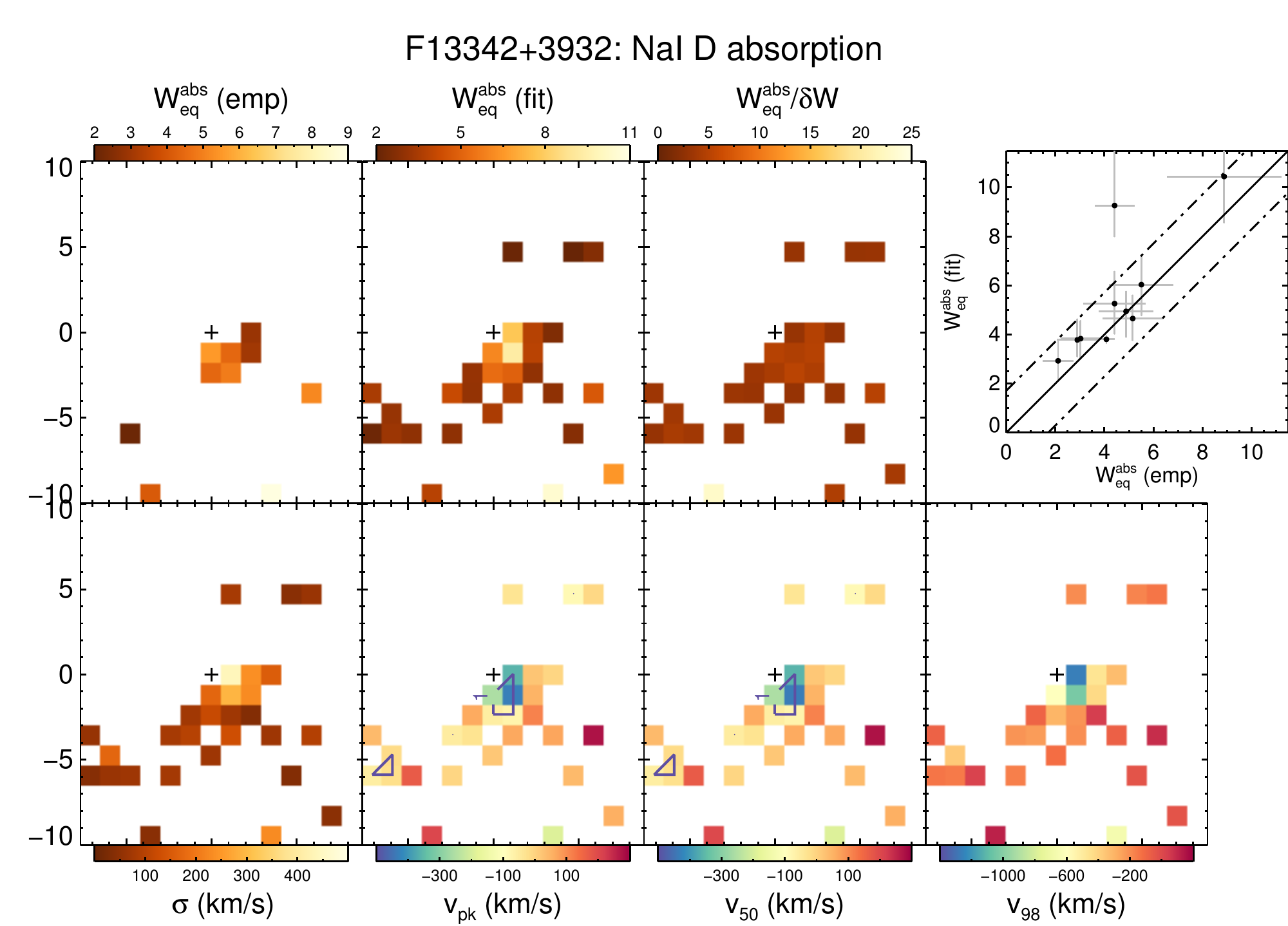}
  \caption{\it Continued.}
\end{figure*}
\setcounter{figure}{11}
\begin{figure*}
  \includegraphics[center]{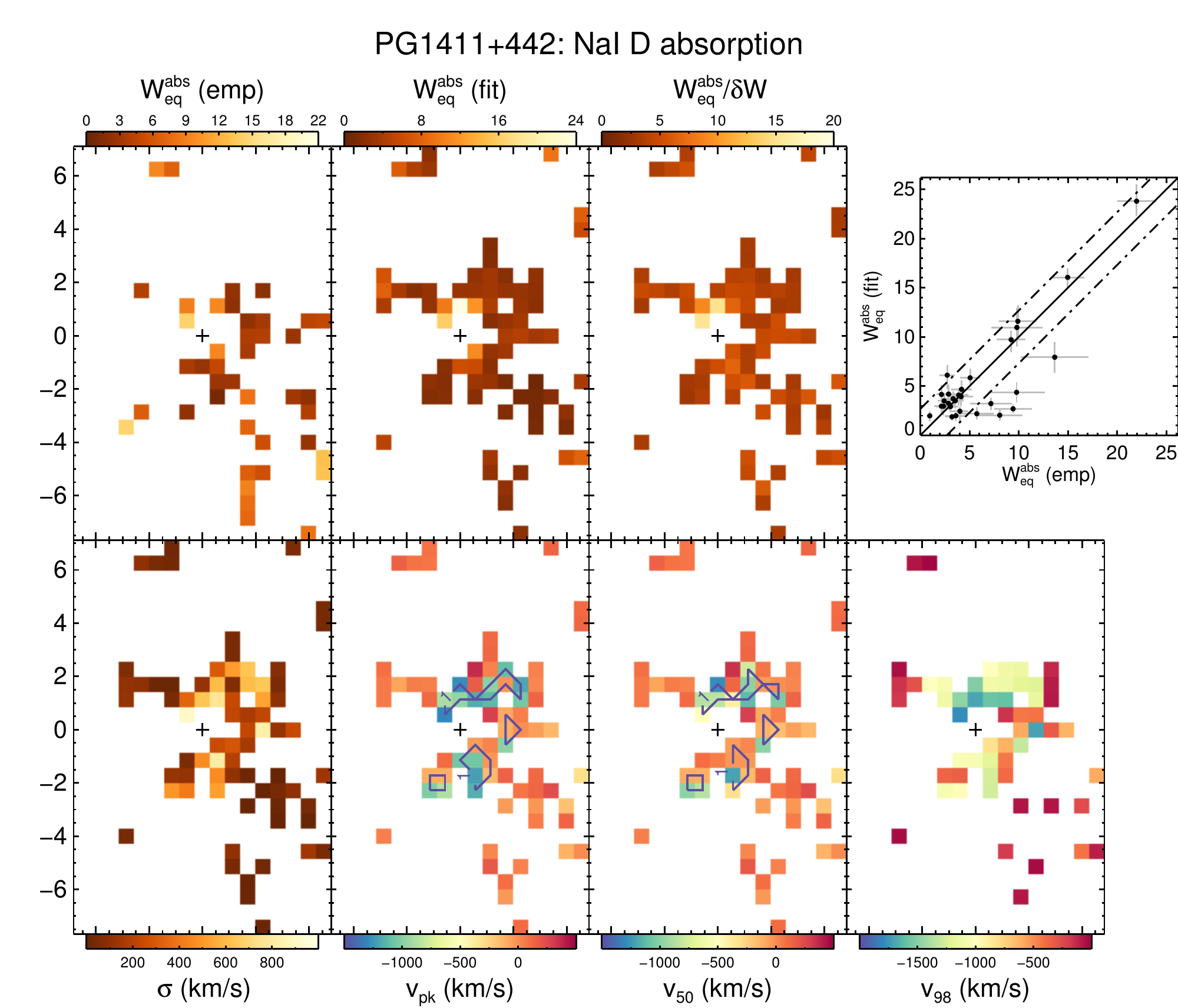}
  \caption{\it Continued.}
\end{figure*}
\setcounter{figure}{11}
\begin{figure*}
  \includegraphics[center]{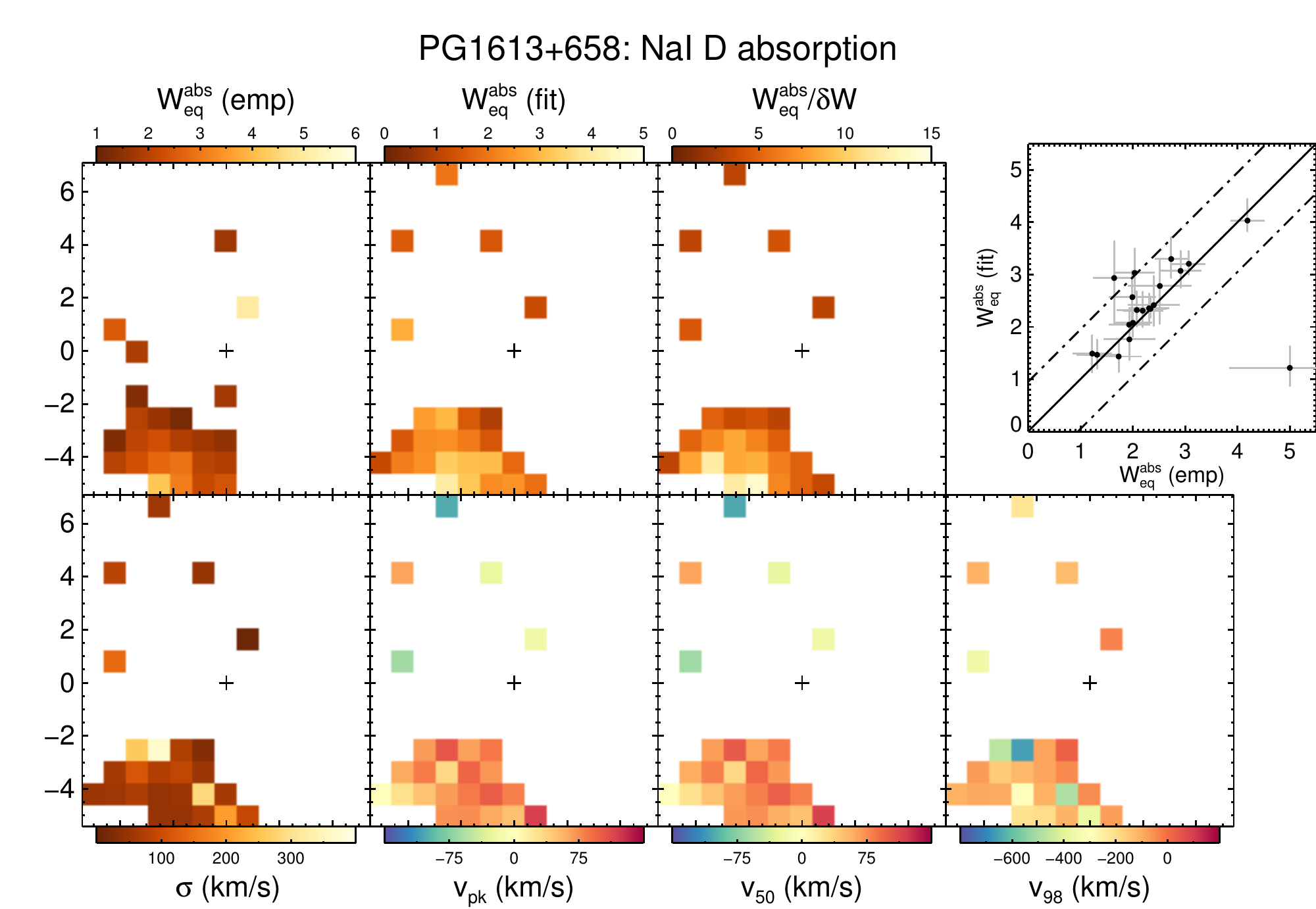}
  \caption{\it Continued.}
\end{figure*}
\setcounter{figure}{11}
\begin{figure*}
  \includegraphics[center]{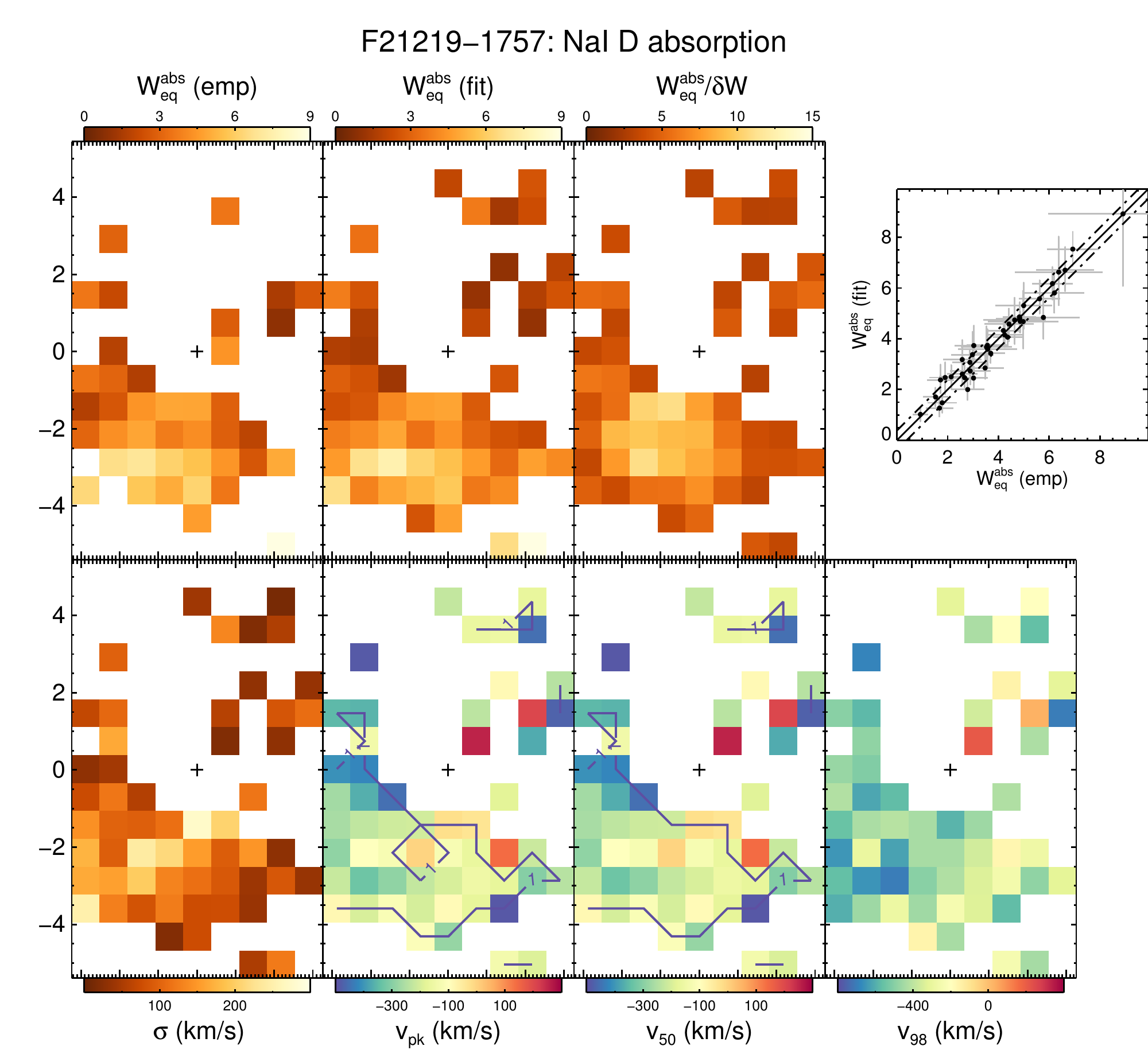}
  \caption{\it Continued.}
\end{figure*}

\end{document}